\documentclass[11pt,letter,notitlepage]{article}

\usepackage[left=1 in, top=1 in, right=1 in, bottom=1 in]{geometry}
\usepackage{graphicx}
\usepackage{amsmath}
\usepackage{amsfonts}
\usepackage{amssymb}
\usepackage{multicol}
\usepackage{fancyhdr}

\usepackage{enumitem}
\setlist[itemize]{noitemsep,topsep=0pt}
\usepackage{color}
\definecolor{soft_blue}{rgb}{0,0,0}
\newcommand{\hclr}[1]{\textcolor{soft_blue}{#1}}
\usepackage[labelfont={color=soft_blue, bf}, figurename=Figure, font = {small}, labelsep = period]{caption}

\definecolor{shiningblue}{rgb}{0.3,0.68,0.89}
\definecolor{brightgreen}{rgb}{0.3,0.8,0.0}

\usepackage[numbers,sort&compress]{natbib}
\parskip = \baselineskip

\newcommand{\beginsupplement}{
        \setcounter{table}{0}
        \renewcommand{\thetable}{S\arabic{table}}
        \setcounter{figure}{0}
        \renewcommand{\thefigure}{S\arabic{figure}}
        \setcounter{equation}{0}
        \renewcommand{\theequation}{S\arabic{equation}}
} 

\vspace{-3in}

\title{\hclr{\Large \textbf{Neither pulled nor pushed: Genetic drift and front wandering uncover a new class of reaction-diffusion waves}}}
\author{ Gabriel Birzu\\
         Department of Physics, Boston University, Boston, MA 02215, USA\\
		 Oskar Hallatschek\\
		 Departments of Physics and Integrative Biology,\\University of California, Berkeley, California 94720, USA\\
		 and \\
	     Kirill S. Korolev\\ 
         Department of Physics and Graduate Program in Bioinformatics,\\Boston University, Boston, MA 02215, USA\\ korolev@bu.edu\\}
\date{\today}

\begin{document}
\maketitle
\begin{abstract}
\noindent\textbf{Short Abstract: }
Traveling waves describe diverse natural phenomena from crystal growth in physics to range expansions in biology. Two classes of waves exist with very different properties: pulled and pushed. Pulled waves are driven by high growth rates at the expansion edge, where the number of organisms is small and fluctuations are large. In contrast, fluctuations are suppressed in pushed waves because the region of maximal growth is shifted towards the population bulk. Although it is commonly believed that expansions are either pulled or pushed, we found an intermediate class of waves with bulk-driven growth, but exceedingly large fluctuations. These waves are unusual because their properties are controlled by both the leading edge and the bulk of the front.

\noindent\textbf{Long Abstract: } 
Epidemics, flame propagation, and cardiac rhythms are classic examples of reaction-diffusion waves that describe a switch from one alternative state to another. Only two types of waves are known: pulled, driven by the leading edge, and pushed, driven by the bulk of the wave. Here, we report a distinct class of semi-pushed waves for which both the bulk and the leading edge contribute to the dynamics. These hybrid waves have the kinetics of pushed waves, but exhibit giant fluctuations similar to pulled waves. The transitions between pulled, semi-pushed, and fully-pushed waves occur at universal ratios of the wave velocity to the Fisher velocity. We derive these results in the context of a species invading a new habitat by examining front diffusion, rate of diversity loss, and fluctuation-induced corrections to the expansion velocity. All three quantities decrease as a power law of the population density with the same exponent. We analytically calculate this exponent taking into account the fluctuations in the shape of the wave front. For fully-pushed waves, the exponent is -1 consistent with the central limit theorem. In semi-pushed waves, however, the fluctuations average out much more slowly, and the exponent approaches 0 towards the transition to pulled waves. As a result, a rapid loss of genetic diversity and large fluctuations in the position of the front occur even for populations with cooperative growth and other forms of an Allee effect. The evolutionary outcome of spatial spreading in such populations could therefore be less predictable than previously thought.

\end{abstract}

\clearpage
\textbf{\large \hclr{Introduction}}

ave-like phenomena are ubiquitous in nature and have been extensively studied across many disciplines. In physics, traveling waves describe chemical reactions, kinetics of phase transitions, and fluid flow~\cite{saarloos:review, pattern_formation:rmp, gl_equation:review, sachdev:diffusive_waves, barenblatt:book, douglas:roughness, whitesides:flame, ramaswamy:wave}. In biology, traveling waves describe invasions, disease outbreaks, and spatial processes in physiology and development~\cite{murray:mathematical_biology, korolev:arrest, nelson:biological_physics, heart:waves, mitotic:wave, ishihara:explosive, development:wave, hastings:invasion_review, invasion:control, travis:projecting, hallatschek:noisy_fisher, network:wave}. Even non-spatial phenomena such as Darwinian evolution and dynamics on networks can be successfully modeled by waves propagating in more abstract spaces such as fitness~\cite{tsimring:wave, rouzine:wave, good:distribution, neher:genealogies, brunet:genealogies_transition, network:wave}. 

The wide range of applications stimulated substantial effort to develop a general theory of traveling waves that is now commonly used to understand, predict, and control spreading phenomena~\cite{saarloos:review, hastings:invasion_review, invasion:control, network:wave, murray:mathematical_biology, ishihara:explosive, lewis:pulled_pushed_commentary, nelson:gene_drive}. A major achievement of this theory was the division of traveling waves into two classes with very different properties~\cite{stokes:pulled_pushed, goldenfeld:structural_stability, murray:mathematical_biology, saarloos:review, kessler:velocity_selection, kessler:velocity_cutoff, douglas:roughness, lewis:pulled_pushed_commentary, gandhi:pulled_pushed, brunet:phenomenological_pulled}. The first class contains waves that are ``pulled'' forward by the dynamics at the leading edge. Kinetics of pulled waves are independent from the nonlinearities behind the front, but extremely sensitive to noise and external perturbations~\cite{saarloos:review, panja:review, goldenfeld:structural_stability}. In contrast, the waves in the second class are resilient to fluctuations and are ``pushed'' forward by the nonlinear dynamics behind the wave front. 

Fluctuations in traveling waves arise due to the randomness associated with discrete events such chemical reaction or birth and deaths. This microscopic stochasticity manifests in many macroscopic properties of the wave including its velocity, the diffusive wandering of the front position, and the loss of genetic diversity~\cite{brunet:phenomenological_pulled, hallatschek:diversity_wave, roques:allee_diversity,hallatschek:tuned_model, meerson:velocity_fluctuations}. For pulled waves, these quantities have been intensely studied because they show an apparent violation of the central limit theorem~\cite{brunet:velocity_cutoff, brunet:phenomenological_pulled,saarloos:review, panja:review, goldenfeld:structural_stability, kessler:velocity_selection, hallatschek:diversity_wave, roques:allee_diversity, brunet:genealogies_transition, hallatschek:tuned_model, hallatschek:tuned_moments}. Naively, one might expect that fluctuations self-average, and their variance is, therefore, inversely proportional to the population density. Instead, the strength of fluctuations in pulled waves has only a logarithmic dependence on the population density. This weak dependence is now completely understood and is explained by the extreme sensitivity of pulled waves to the dynamics at the front~\cite{ brunet:phenomenological_pulled,saarloos:review, hallatschek:tuned_model, hallatschek:tuned_moments}.

A complete understanding is however lacking for fluctuations in pushed waves~\cite{saarloos:review, panja:review, goldenfeld:structural_stability, kessler:velocity_selection, kessler:velocity_cutoff, hallatschek:diversity_wave, roques:allee_diversity, brunet:genealogies_transition}. Since pushed waves are driven by the dynamics at the bulk of the wave front, it is reasonable to expect that the central limit theorem holds, and fluctuations decrease as one over the population density~$N$. Consistent with this expectation, the~$1/N$ scaling was theoretically derived both for the effective diffusion constant of the front~\cite{meerson:velocity_fluctuations} and for the rate of diversity loss~\cite{hallatschek:diversity_wave}. Numerical simulations confirmed the~$1/N$ scaling for the diffusion constant~\cite{khain:wandering}, but showed a much weaker dependence for the rate of diversity loss~\cite{hallatschek:diversity_wave}. Ref.~\cite{khain:wandering}, however, considered only propagation into a metastable state, while Ref.~\cite{hallatschek:diversity_wave} analyzed only one particular choice of the nonlinear growth function. As a result, it is not clear whether the effective diffusion constant and the rate of diversity loss behave differently or if there are two distinct types of dynamics within the class of pushed waves.

The latter possibility was anticipated by the analysis of how the wave velocity changes if one sets the growth rate to zero below a certain population density~\cite{kessler:velocity_cutoff}. This study found that the velocity correction scales as a power law of the growth-rate cutoff with a continuously varying exponent. If the cutoff was a faithful approximation of fluctuations at the front, this result would suggest that the central limit theorem does not apply to pushed waves. Stochastic simulations, however, were not carried out in Ref.~\cite{kessler:velocity_cutoff} to test this prediction.

Taken together, previous findings highlight the need to characterize the dynamics of pushed waves more thoroughly. Here, we develop a unified theoretical approach to fluctuations in reaction-diffusion waves and show how to handle divergences and cutoffs that typically arise in analytical calculations. Theoretical predictions are tested against extensive numerical simulations. In simulations, we vary the model parameters to tune the propagation dynamics from pulled to pushed and determine how the front diffusion, diversity loss, and wave velocity depend on the population density. Our main result is that the simple pulled \textit{vs.} pushed classification does not hold. Instead, there are three distinct classes of traveling waves. Only one of these classes shows weak fluctuations consistent with the central limit theorem. The other two classes exhibit large fluctuations because they are very sensitive to the dynamics at the leading edge of the wave front.

\textbf{\large \hclr{Model}}

Traveling waves occur when a transport mechanism couples dynamics at different spatial locations. The nature of these wave-generating processes could be very different and ranges from reactions and diffusion in chemistry to growth and dispersal in ecology. The simplest and most widely-used model of a reaction-diffusion wave\footnote{Throughout the paper we use the term reaction-diffusion wave to describe propagating fronts that connect two states with different population densities. Reaction-diffusion models, especially with several components, also describe more intricate phenomena such as periodic waves, spatio-temporal chaos, and pulse propagation. While some of our results could be useful in these more general settings, our theory and numerical simulations are limited to regular fronts only.} is the generalized Fisher-Kolmogorov equation:

\begin{equation}
\frac{\partial n}{\partial t} = D\frac{\partial^2 n}{\partial x^2} + r(n) n + \sqrt{\gamma_{n}(n)n}\eta(t,x) ,
\label{eq:fisher}
\end{equation}

\noindent which, in the context of ecology, describes how a species colonizes a new habitat~\cite{saarloos:review, murray:mathematical_biology, fisher:wave, kolmogorov:wave, skellam:wave}. Here,~$n(t,x)$ is the population density of the species, $D$~is the dispersal rate, and~$r(n)$ is the density-dependent per capita growth rate. The last term accounts for demographic fluctuations: $\eta(t,x)$ is a Gaussian white noise, and~$\gamma_n(n)$ quantifies the strength of demographic fluctuations. In simple birth-death models,~$\gamma_n$ is a constant, but we allow for an arbitrary dependence on~$n$ provided that~$\gamma_n(0)>0$. The origin of the noise term and its effects on the wave dynamics are further discussed in Sec.~IV of the SI.

Pulled waves occur when~$r(n)$ is maximal at small~$n$; for example, when the growth is logistic:~$r(n)=r_0(1-n/N)$~\cite{saarloos:review, murray:mathematical_biology}. Here,~$r_0$ is the growth rate at low densities, and~$N$ is the carrying capacity that sets the population density behind the front. For pulled waves, the expansion dynamics are controlled by the very tip of the front, where the organisms not only grow at the fastest rate, but also have an unhindered access to the uncolonized territories. As a result, the expansion velocity is independent of the functional form of~$r(n)$ and is given by the celebrated result due to Fisher, Kolmogorov, and Skellam~\cite{fisher:wave, kolmogorov:wave, skellam:wave}:

\begin{equation}
v_{\mathrm{\textsc{f}}} = 2\sqrt{Dr(0)}.
\label{eq:v_f}
\end{equation} 

Expression~\ref{eq:v_f}, to which we refer as the Fisher velocity, can be defined for any model with~$r(0)>0$ even when the expansion is not pulled. We show below that~$v_{\mathrm{\textsc{f}}} $ provides a useful baseline for comparing different types of waves.

Pushed waves occur when a species grows best at intermediate population densities~\cite{saarloos:review, murray:mathematical_biology}. Such non-monotonic behavior of~$r(n)$ arises through a diverse set of mechanisms and is known as an Allee effect in ecology~\cite{lewis_kareiva:allee, veit:finch}. Most common causes of an Allee effect are cooperative feeding, collective defense against predators, and the difficulty in finding mates at low population densities~\cite{courchamp:allee_review, drake:allee_evidence, dai:science, tobin:moth_review}. The velocity of pushed waves is always greater than Fisher's prediction~$(v>v_{\mathrm{\textsc{f}}})$ and depends on all aspects of the functional form of~$r(n)$~\cite{saarloos:review, murray:mathematical_biology}.

\begin{figure}
\begin{center}
\includegraphics[width=17.8cm]{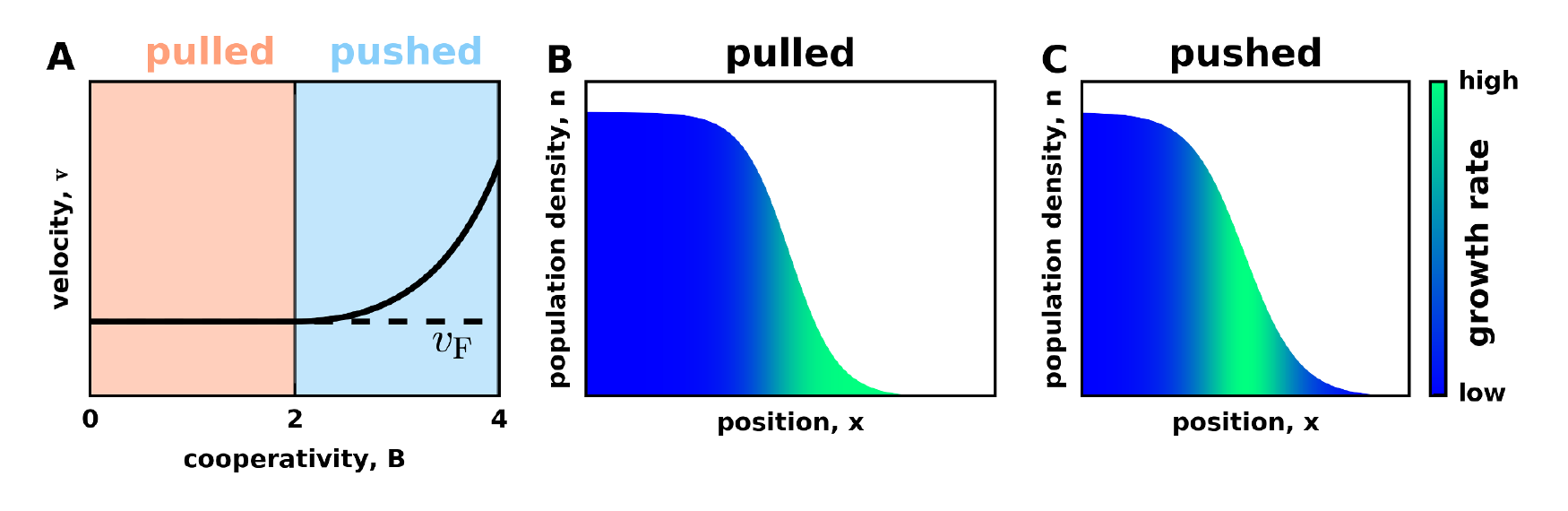}
\caption{\textbf{Waves transition from pulled to pushed as growth becomes more cooperative.} \textbf{(A)}~shows the expansion velocity as a function of cooperativity for the growth rate specified by Eq.~(\protect{\ref{eq:cooperative_growth}}). For low cooperativity, expansions are pulled and their velocity equals the Fisher velocity. Beyond the critical value of~$B=2$, expansions become pushed and their velocity exceeds~$v_{\mathrm{\textsc{f}}}$. Note that the region of high growth is at the leading edge of the front in pulled waves~\textbf{(B)}, but in the interior of the front in pushed waves~\textbf{(C)}. This difference is due to the dependence of the growth rate on the population density. For low cooperativity, the growth rate is maximal at low population densities, but, for high cooperativity, the growth rate is maximal at intermediate population densities. In all panels, the exact solution of Eq.~(\protect{\ref{eq:fisher}}) is plotted; $B=0$ in panel B, and~$B=4$ in panel C.} 
\label{fig:growth}
\end{center}  
\end{figure}

Allee effects are typically described by adding a cooperative term to the logistic equation:

\begin{equation}
r(n) = r_0 \left(1-\frac{n}{N}\right)\left(1+B\frac{n}{N}\right),
\label{eq:cooperative_growth}
\end{equation}

\noindent where~$B$ is the strength of cooperativity. For this model, the exact solutions are known for the expansion velocity and the population density profile; see SI(Sec.~II) and Ref.~\cite{murray:mathematical_biology, aronson:allee_wave, fife:allee_wave}. For~$B\le2$, expansions are pulled, and the expansion velocity equals~$v_{\mathrm{\textsc{f}}}$, which is independent of~$B$. That is cooperativity does not always increase the expansion velocity even though it always increases the growth rates at high densities. For~$B>2$, expansions are pushed, and~$v$ increases with~$B$. Figure~\ref{fig:growth}A illustrates this transition from pulled to pushed waves as cooperativity is increased. In Methods and SI(Sec.~II), we also present several alternative models of an Allee effect and show that our conclusions do not depend on a particular choice of~$r(n)$.

Increasing the value of cooperativity beyond~$B=2$ not only makes the expansion faster, but also shifts the region of high growth from the tip to the interior of the expansion front~(Fig.~\ref{fig:growth}BC). This shift is the most fundamental difference between pulled and pushed waves because it indicates the transition from a wave being ``pulled'' by its leading edge to a wave being ``pushed'' by its bulk growth.

The edge-dominated dynamics make pulled waves extremely sensitive to the vagaries of reproduction, death, and dispersal~\cite{goldenfeld:structural_stability, saarloos:review, panja:review, brunet:phenomenological_pulled}. Indeed, the number of organisms at the leading edge is always small, so strong number fluctuations are expected even in populations with a large carrying capacity,~$N$. These fluctuation affect both physical properties, such as the shape and position of the wave front, and evolutionary properties, such as the genetic diversity of the expanding population.\footnote{We refer to genetic drift and genetic diversity as an evolutionary property because they occurs only in systems where agents can be assigned heritable labels. In contrast, front wandering occurs in any physical system and can be quantified even when all agents are indistinguishable as is the case in chemical processes.} Consistent with these expectations, experiments with pulled waves reported an unusual roughness of the expansion front~\cite{douglas:roughness} and a rapid loss of genetic diversity~\cite{hallatschek:sectors, korolev:amnat}.

\begin{figure}
\begin{center}
\includegraphics[width=17.8cm]{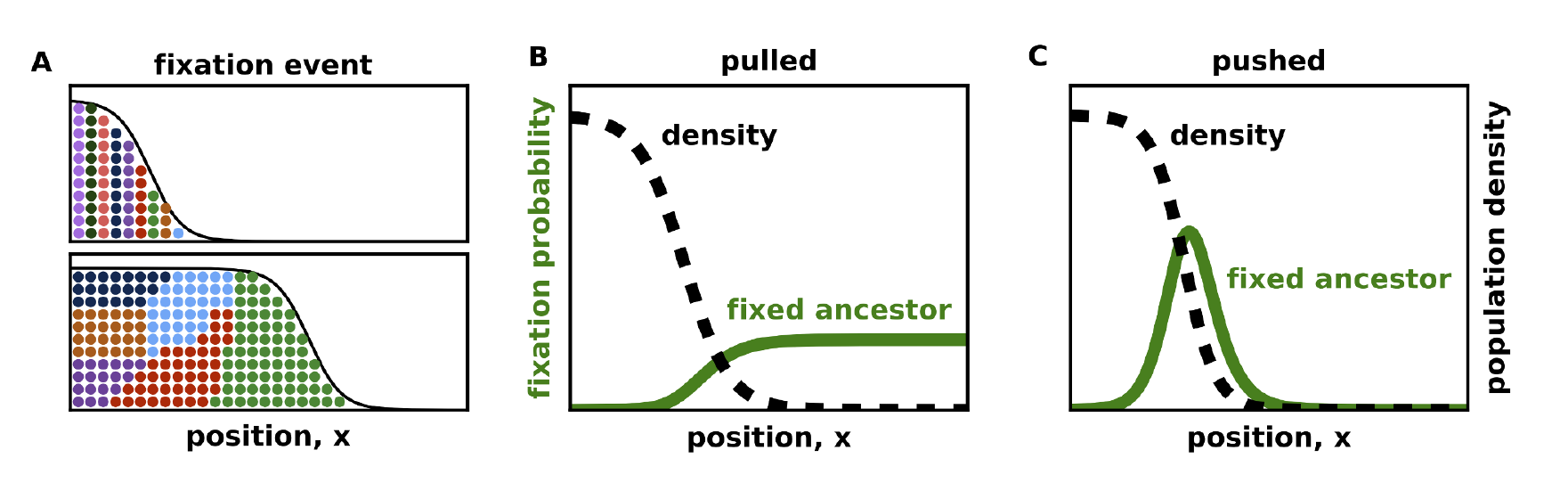}
\caption{\textbf{Ancestral lineages occupy distinct locations in pulled and pushed waves.} \textbf{(A)}~illustrates the fixation of a particular genotype. Initially, a unique and heritable color was assigned to every organism to visualize its ancestral lineage. There are no fitness differences in the population, so fixations are caused by genetic drift. \textbf{(B)}~and~\textbf{(C)} show the probability that the fixed genotype was initially present at a specific position in the reference frame comoving with the expansion. The transition from pulled to pushed waves is marked by a shift in the fixation probability from the tip to the interior of the expansion front. This shift indicates that most ancestral lineages are focused at the leading edge in pulled waves, but near the middle of the front in pushed waves. The fixation probabilities were computed analytically, following Refs.~\protect{\cite{hallatschek:diversity_wave, roques:allee_diversity}}, as described in the SI. We used~$B=0$ in panel B and~$B=4$ in panel C. }
\label{fig:fixation}
\end{center}  
\end{figure}

The transition from ``pulled'' to ``pushed'' dynamics is also evident in the number of organisms that trace their ancestry to the leading edge \textit{vs.} the bulk of the front. The expected number of descendants has been determined for any spatial position along the front for both pulled and pushed waves~\cite{vlad:hydrodynamic, hallatschek:diversity_wave, roques:allee_diversity, lewis:wave_fixation}. For pulled waves, only the very tip of the expansion contributes to future generations. On the contrary, the organisms at the leading edge leave few progeny in pushed waves, and the population descends primarily from the organisms in the region of high growth. This shift in the spatial patterns of ancestry has a profound effect on species evolution. In pulled waves, only mutations near the very edge of the expansion have an appreciable fixation probability, but the entire expansion front contributes to evolution in pushed waves~(Fig.~\ref{fig:fixation}).

\begin{figure}[!h]
\begin{center}
\includegraphics[width=8.7cm]{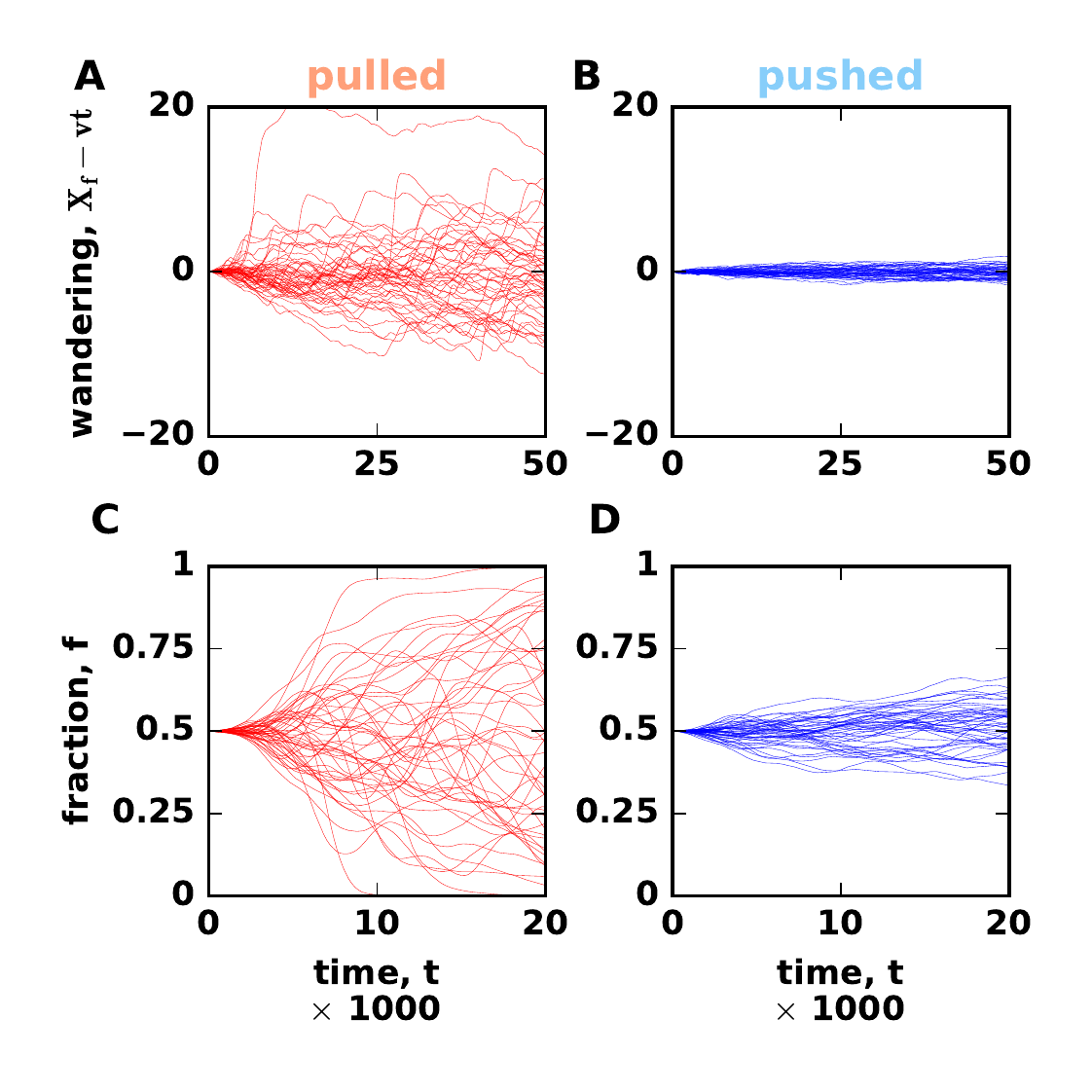}
\caption{\textbf{Fluctuations are much stronger in pulled than in pushed waves.} The top row compares front wandering between pulled~\textbf{(A)} and pushed~\textbf{(B)} expansions. Each line shows the position of the front~$X_{\mathrm{f}}(t)$ in a single simulation relative to the mean over all simulations in the plot. The bottom row compares the strength of genetic drift between pulled~\textbf{(C)} and pushed~\textbf{(D)} expansions. We started the simulations with two neutral genotypes equally distributed throughout the front and then tracked how the fraction of one of the genotypes changes with time. This fraction was computed from~300~patches centered on~$X_{\mathrm{f}}$ to exclude the fluctuations well behind the expansion front.}
\label{fig:fluctuations}
\end{center}  
\end{figure}

Fixation probabilities and, more generally, the dynamics of heritable markers provides an important window into the internal dynamics of a reaction-diffusion wave~\cite{vlad:hydrodynamic}. When the markers are neutral, i.e. they do not affect the growth and dispersal of the agents, the relative abundance of the markers changes only stochastically. In population genetics, such random changes in the genotype frequencies are known as genetic drift. To describe genetic drift mathematically, we introduce the relative fraction of one of the genotypes in the population~$f(t,x)$. The dynamics of~$f(t,x)$ follow from~Eq.~(\ref{eq:fisher}) and are derived in Sec.~III of the SI~(see also Ref.~\cite{mckane:genetic_drift_from_lv, korolev:review, vlad:hydrodynamic}). The result reads

\begin{equation}
\frac{\partial f}{\partial t} = D\frac{\partial^2 f}{\partial x^2} + 2\frac{\partial \ln n}{\partial x}\frac{\partial f}{\partial x} +  \sqrt{\frac{\gamma_f(n)}{n}f(1-f)}\eta_{f}(t),
\label{eq:drift}
\end{equation}

where~$\gamma_{f}(n)>0$ is the strength of genetic drift.

Equation~(\ref{eq:drift}) preserves the expectation value of~$f$, but the variance of~$f$ increases with time until one of the absorbing states is reached. The two absorbing states are~$f=0$ and~$f=1$, which correspond to the extinction and fixation of a particular genotype respectively. The fluctuations of~$f$ and front position are shown in Fig.~\ref{fig:fluctuations}. Both quantities show an order of magnitude differences between pulled and pushed waves even though the corresponding change in cooperativity is quite small.

Although the difference between pulled and pushed waves seems well-established, little is known about the transition between the two types of behavior. In particular, it is not clear how increasing the nonlinearity of~$r(n)$ transforms the patterns of fluctuations and other properties of a traveling wave. To answer this question, we solved Eqs.~(\ref{eq:fisher}) and~(\ref{eq:drift}) numerically. Specifically, our simulations described the dynamics of both the population density and the relative abundance of two neutral genotypes. The former was used to estimate the fluctuations in the position of the front, and the latter was used to quantify the decay rate of genetic diversity. In simulations, the species expanded in a one-dimensional array of habitable patches connected by dispersal between the nearest neighbors. Each time step consisted of a deterministic dispersal and growth followed by random sampling to simulate demographic fluctuations and genetic drift~(see Methods and Sec.~XIII in the SI). By increasing the cooperativity of the growth rate, we observed a clear transition from pulled~($v=v_{\mathrm{\textsc{f}}}$) to pushed~($v>v_{\mathrm{\textsc{f}}}$) waves accompanied by a dramatic reduction in fluctuations; see Fig.~\ref{fig:fluctuations}.

\textbf{\large \hclr{Results}}

Fluctuations provide an easy readout of the internal dynamics in a traveling wave, so we decided to determine how they change as a function of cooperativity. Because the magnitude of the fluctuations also depends on the population density, we looked for a qualitative change in this dependence while varying~$B$. In particular, we aimed to determine whether population dynamics change gradually or discontinuously at the transition between pulled and pushed waves.

\begin{figure}
\begin{center}
\includegraphics[width=17.8cm]{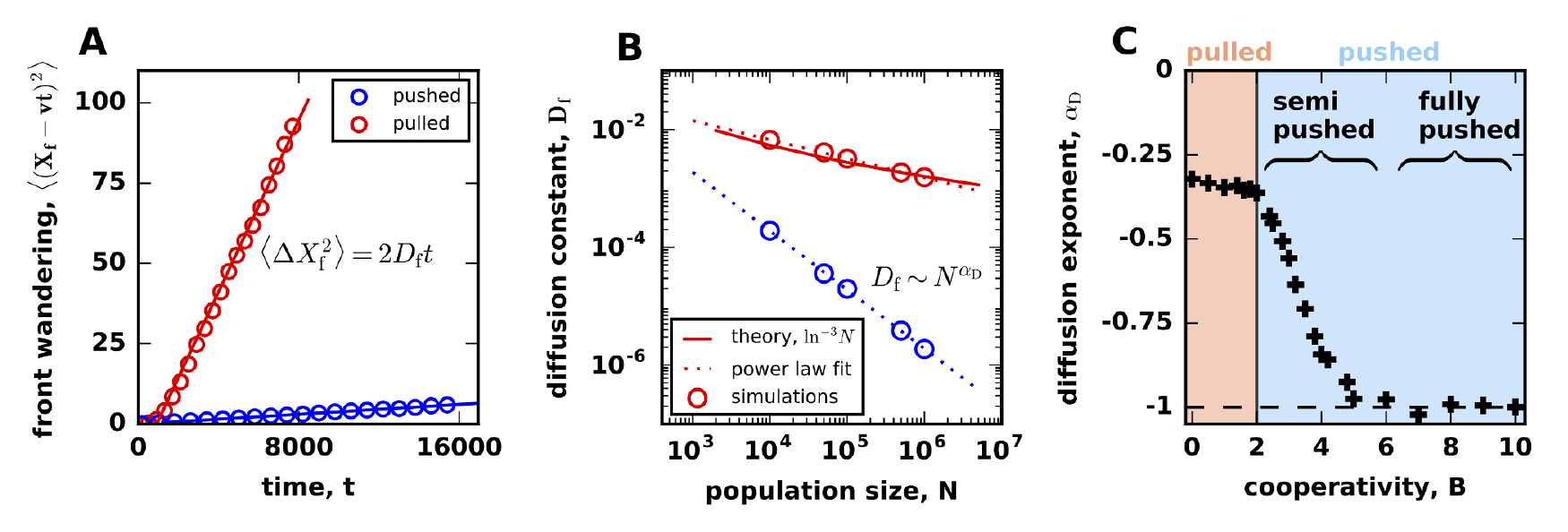}
\caption{\textbf{Front wandering identifies a new class of pushed waves.} \textbf{(A)}~Fluctuations in the front position can be described by simple diffusion for both pulled and pushed waves. \textbf{(B)}~The front diffusion is caused by the number fluctuations, so the effective diffusion constant,~$D_{\mathrm{f}}$, decreases with the carrying capacity,~$N$. For pulled waves,~$D_{\mathrm{f}}\sim\ln^{-3}N$~\protect{\cite{brunet:phenomenological_pulled}}, while, for pushed waves,~$D_{\mathrm{f}}$ can decrease much faster as~$N^{-1}$~\protect{\cite{meerson:velocity_fluctuations}}. We quantify the scaling of~$D_{\mathrm{f}}$ with~$N$ by the exponent~$\alpha_{\mathrm{\textsc{d}}}$ equal to the slope on the log-log plot shown. The overlap of the two red lines highlight the fact that, even though~$\alpha_{\mathrm{\textsc{d}}}$ should equal~$0$ for pulled waves, the limited range of~$N$ results in a different value of~$\alpha_{\mathrm{\textsc{d}}}\approx-0.33$. \textbf{(C)}~The dependence of the scaling exponent on cooperativity identifies two distinct classes of pushed waves.}
\label{fig:diffusion}
\end{center}  
\end{figure}

\textit{Spatial wandering of the front}\\
We first examined the fluctuations of the front position in the comoving reference frame. The position of the front~$X_{\mathrm{f}}$ was defined as the total population size in the colonized space normalized by the carrying capacity~$X_{\mathrm{f}}=\frac{1}{N}\int_{0}^{+\infty}n(t,x)dx$. As expected~\cite{saarloos:review, panja:review, brunet:phenomenological_pulled, meerson:velocity_fluctuations},~$X_{\mathrm{f}}$ performed a random walk due to demographic fluctuations in addition to the average motion with a constant velocity~(Fig.~\ref{fig:fluctuations}AB). For both pulled and pushed waves, the variance of~$X_{\mathrm{f}}$ grew linearly in time~(Fig.~\ref{fig:diffusion}A), i.e. the front wandering was diffusive and could be quantified by an effective diffusion constant~$D_{\mathrm{f}}$. 

The magnitude of the front wandering is expected to depend strongly on the type of the expansion~\cite{saarloos:review, panja:review, brunet:phenomenological_pulled, meerson:velocity_fluctuations}. For pulled waves, Ref.~\cite{brunet:phenomenological_pulled} found that~$D_{\mathrm{f}}\sim\ln^{-3}N$, but a very different scaling~$D_{\mathrm{f}} \sim N^{-1}$ was predicted for certain pushed waves~\cite{meerson:velocity_fluctuations}; see Fig.~\ref{fig:diffusion}B. Given that pulled and pushed waves belong to distinct universality classes, it is easy to assume that the transition between the two scaling regimens should be discontinuous~\cite{saarloos:review, panja:review, brunet:phenomenological_pulled, meerson:velocity_fluctuations, hallatschek:diversity_wave, roques:allee_diversity}. This assumption, however, has not been carefully investigated, and we hypothesized that there could be an intermediate regime with~$D_{\mathrm{f}}\sim N^{\alpha_{\mathrm{\textsc{d}}}}$. From simulations, we computed how~$\alpha_{\mathrm{\textsc{d}}}$ changes with~$B$ and indeed found that pushed waves have intermediate values of~$\alpha_{\mathrm{\textsc{d}}}$ between~$0$ and~$-1$ when~$B\in(2,4)$~(Fig.~S3).

The dependence of the scaling exponent on the value of cooperativity is shown in Fig.~\ref{fig:diffusion}C. For large~$B$, we found that~$\alpha_{\mathrm{\textsc{d}}}$ is constant and equal to~$-1$, which is consistent with the previous work~\cite{meerson:velocity_fluctuations}. Below a critical value of cooperativity, however, the exponent~$\alpha_{\mathrm{\textsc{d}}}$ continually changes with~$B$ towards~$0$. The critical cooperativity is much larger than the transition point between pulled and pushed waves, so the change in the scaling occurs within the class of pushed waves. This transition divides pushed waves into two subclasses, which we termed fully-pushed and semi-pushed waves. For pulled waves, we found that~$\alpha_{\mathrm{\textsc{d}}}$ is independent of~$B$, but our estimate of~$\alpha_{\mathrm{\textsc{d}}}$ deviated slightly from the expected value due to the finite range of~$N$ in the simulations~(compare~$N^{\alpha_{\mathrm{\textsc{d}}}}$ and~$\ln^{-3}N$ fits in Fig.~\ref{fig:diffusion}B). 

\textit{Loss of genetic diversity}\\
Our analysis of the front wandering showed that pushed waves consist of two classes with a very different response to demographic fluctuations. To determine whether this difference extends to other properties of expansions, we turned to genetic drift, a different process that describes fluctuations in the genetic composition of the front. Genetic drift occurs even in the absence of front wandering~(see Sec.~III in the SI and Ref.~\cite{hallatschek:constrained_expansions}), so these two properties are largely independent from each other and capture complementary aspects related to physical and evolutionary dynamics in traveling waves.\footnote{Front wandering and genetic drift are in general coupled because both arise due to the randomness of birth and death. The two processes are however not identical because the fluctuations in the total population density could differ from the fluctuations in the relative frequency of the genotypes. For example, in the standard Wright-Fisher model, only genetic drift is present since the total population size is fixed; see SI(Sec.~III) for further details.}

\begin{figure}
\begin{center}
\includegraphics[width=17.8cm]{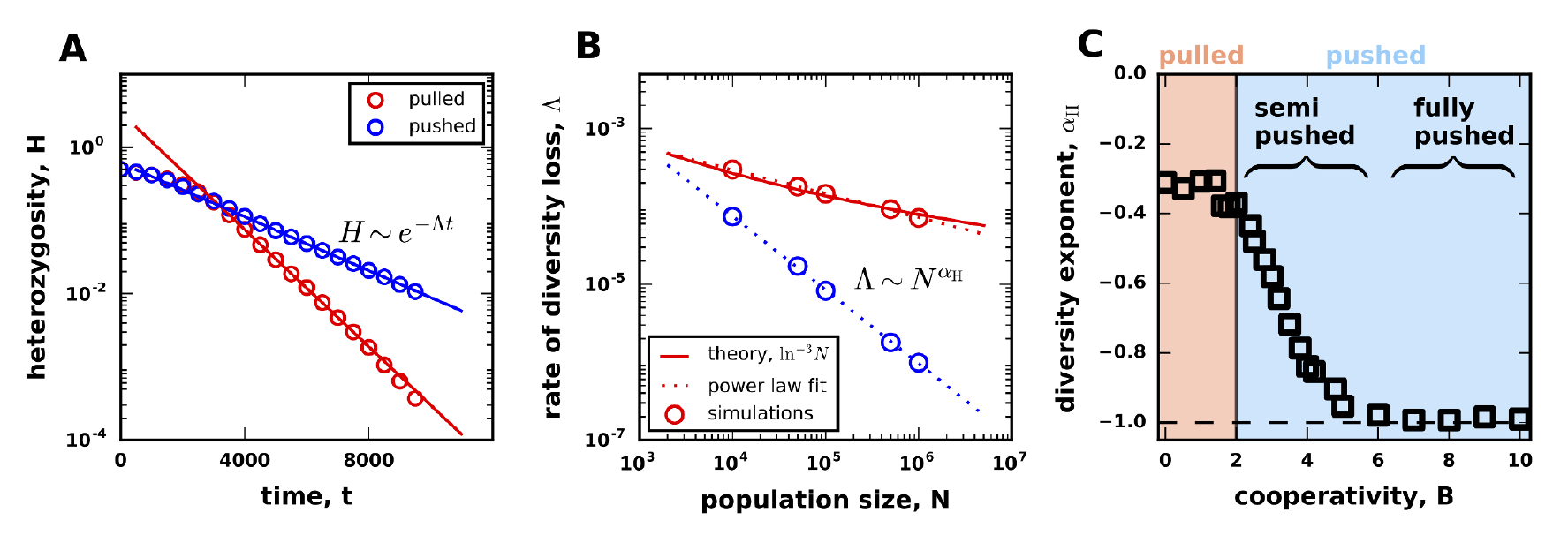}
\caption{\textbf{Genetic diversity is lost at different rates in pulled, semi-pushed waves, and fully-pushed waves.} \textbf{(A)}~The average heterozygosity,~$H$, is a measure of diversity equal to the probability to sample two distinct genotypes in the population. For both pulled and pushed expansions, the decay of genetic diversity is exponential in time:~$H\sim e^{-\Lambda t}$, so we used~$\Lambda$ to measure the strength of genetic drift. \textbf{(B)} Genetic drift decrease with~$N$. For pulled waves,~$\Lambda\sim\ln^{-3}N$~\protect{\cite{brunet:phenomenological_pulled}}, while, for fully-pushed waves, we predict that~$\Lambda\sim N^{-1}$; see Eq.~(\protect{\ref{eq:LD_result}}). To quantify the dependence of~$\Lambda$ on~$N$, we fit~$\Lambda \sim N^{\alpha_{\mathrm{\textsc{h}}}}$. The dashed red line shows that even though~$\alpha_{\mathrm{\textsc{h}}}$ should equal~$0$ for pulled waves, the limited range of~$N$ results in a different value of~$\alpha_{\mathrm{\textsc{h}}}\approx-0.33$. \textbf{(C)}~The dependence of the scaling exponent on cooperativity identifies the same three classes of waves as in Fig.~\protect{\ref{fig:diffusion}}C; the transitions between the classes occur at the same values of~$B$.}
\label{fig:diversity}
\end{center}  
\end{figure}

We quantified genetic fluctuations by the rate at which genetic diversity is lost during an expansion. The simulations were started in a diverse state with each habitable patch containing an equal number of two neutral genotypes. As the expansion proceeded, the relative fractions of the genotypes fluctuated and eventually one of them was lost from the expansion front~(Fig.~\ref{fig:fluctuations}C). To capture the loss of diversity, we computed the average heterozygosity~$H$, defined as the probability to sample two different genotypes at the front. Mathematically,~$H$ equals the average of $2f(1-f)$, where~$f$ is the fraction of one of the genotypes in an array of patches comoving with the front, and the averaging is done over independent realizations. Consistent with the previous work~\cite{brunet:phenomenological_pulled, hallatschek:diversity_wave}, we found that the heterozygosity decays exponentially at long times:~$H\sim e^{-\Lambda t}$ for both pulled and pushed waves~(Fig.~\ref{fig:diversity}A). Therefore, the rate~$\Lambda$ was used to measure the strength of genetic drift across all values of cooperativity. 

By analogy with the front wandering, we reasoned that~$\Lambda$ would scales as~$N^{\alpha_{\mathrm{\textsc{h}}}}$ for large~$N$, and~$\alpha_{\mathrm{\textsc{h}}}$ would serve as  an effective ``order parameter'' that distinguishes different classes of traveling waves. Indeed, Ref.~\cite{brunet:phenomenological_pulled} showed that~$\Lambda\sim\ln^{-3}N$ for pulled waves, i.e. the expected~$\alpha_{\mathrm{\textsc{h}}}$ is zero. Although no conclusive results have been reported for pushed waves, the work on adaptation waves in fitness space suggests~$\alpha_{\mathrm{\textsc{h}}}=-1$ for fully-pushed waves~\cite{brunet:genealogies_transition}. Our simulations confirmed both of these predictions~(Fig.~\ref{fig:diversity}B) and showed~$\Lambda\sim N^{\alpha_{\mathrm{\textsc{h}}}}$ scaling for all values of cooperativity. 

The dependence of~$\alpha_{\mathrm{\textsc{h}}}$ on~$B$ shows that genetic fluctuations follow exactly the same pattern as the front wandering~(Fig.~\ref{fig:diversity}C). In particular, both exponents undergo a simultaneous transition from~$\alpha_{\mathrm{\textsc{h}}}=\alpha_{\mathrm{\textsc{d}}}=-1$ to a continual dependence on~$B$ as cooperativity is decreased. Thus, genetic fluctuations also become large as waves switch from fully-pushed to semi-pushed. In the region of pulled waves,~$\alpha_{\mathrm{\textsc{d}}}$ and~$\alpha_{\mathrm{\textsc{h}}}$ are independent of~$B$, but their values deviate from the theoretical expectation due to the finite range of~$N$ explored in the simulations. Overall, the consistent behavior of the fluctuations in the position and composition of the front strongly suggests the existence of two classes of pushed waves, each with a distinct set of properties.

\textit{The origin of semi-pushed waves}\\
We next sought an analytical argument that can explain the origin of the giant fluctuations in semi-pushed waves. In the SI(Sec.~VI and VIII), we explain and extend the approaches from Refs.~\cite{hallatschek:diversity_wave} and~\cite{meerson:velocity_fluctuations} to compute~$D_{\mathrm{f}}$ and~$\Lambda$ using a perturbation expansion in~$1/N$. The main results are 

\begin{equation}
\begin{aligned}
 D_{\mathrm{f}} &= \frac{1}{N} \frac{\int_{-\infty}^{+\infty}\gamma_n(\rho)[\rho'(\zeta)]^2\rho(\zeta)e^{\frac{2v\zeta}{D}}d\zeta}{2\left(\int^{+\infty}_{-\infty}[\rho'(\zeta)]^2e^{\frac{v\zeta}{D}}d\zeta\right)^2}, \\
 \Lambda &= \frac{1}{N} \frac{\int_{-\infty}^{+\infty}\gamma_{f}(\rho)\rho^{3}(\zeta)e^{\frac{2v\zeta}{D}}d\zeta}{\left(\int^{+\infty}_{-\infty}\rho^2(\zeta)e^{\frac{v\zeta}{D}}d\zeta\right)^2}. \\
\end{aligned}
\label{eq:LD_result}
\end{equation}

\noindent Here, primes denote derivatives;~$\zeta=x-vt$ is the coordinate in the reference frame comoving with the expansion; $\rho(\zeta)=n(\zeta)/N$ is the normalized population density profile in the steady state; $v$~is the expansion velocity; $D$~is the dispersal rate as in Eq.~(\ref{eq:fisher}); and~$\gamma_n$ and~$\gamma_{f}$ are the strength of demographic fluctuations and genetic drift, which in general could be different~(see Sec.~III in the SI).

The~$N^{-1}$~scaling that we observed for fully-pushed waves is readily apparent from~Eqs.~(\ref{eq:LD_result}). The prefactors of~$1/N$ account for the dependence of microscopic fluctuations on the carrying capacity, and the ratios of the integrals describe the relative contribution of the different locations within the wave front.

For fully-pushed waves, the integrands in~Eqs.~(\ref{eq:LD_result}) vanish both in the bulk and at the leading edge, so~$\Lambda$ and~$D_{\mathrm{f}}$ are controlled by the number of organisms within the wave front. Hence, the~$N^{-1}$~scaling can be viewed as a manifestation of the central limit theorem, which predicts that the variance in the position and genetic diversity of the front should be inversely proportional to the effective population size of the front. To test this theory, we calculated the integrals in~Eqs.~(\ref{eq:LD_result}) analytically for the model specified by~Eq.~(\ref{eq:cooperative_growth}); see Sec.~VI and VIII in the SI. These exact results show excellent agreement with our simulations~(Fig.~S4) and thus confirm the validity of the perturbation approach for fully-pushed waves.

Why does the~$N^{-1}$~scaling break down in semi-pushed waves? We found that the integrals in the numerators in~Eqs.~(\ref{eq:LD_result}) become more and more dominated by large~$\zeta$ as cooperativity decreases, and, at a critical value of~$B$, they diverge. To pinpoint this transition, we determined the behavior of~$\rho(\zeta)$ for large~$\zeta$ by linearizing~Eq.~(\ref{eq:fisher}) for small population densities:

\begin{equation}
D\frac{d^2\rho}{d\zeta^2} + v\frac{d\rho}{d\zeta} + r(0)\rho = 0,
\label{eq:linearized_fisher}
\end{equation}

\noindent where we replaced~$n$ by~$\rho$ and shifted into the reference frame comoving with the front. Equation~(\ref{eq:linearized_fisher}) is linear, so the population density decreases exponentially at the front as~$\rho\sim e^{-k\zeta}$. The value of~$k$ is obtained by substituting this exponential form into Eq.~(\ref{eq:linearized_fisher}) and is given by~$k=\frac{v}{2D}\left(1+\sqrt{1-v_{\mathrm{\textsc{f}}}^2/v^2}\right)$ with~$v_{\mathrm{\textsc{f}}}$ as in Eq.~(\ref{eq:v_f}) (see Sec.~II and Sec.~IX in the SI). From the asymptotic behavior of~$\rho$, it is clear that the integrands in the numerators in~Eqs.~(\ref{eq:LD_result}) scale as~$e^{(2v/D-3k)\zeta}$, and the integrals diverge when~$v/D = 3k/2$. The integrals in the denominators converge for all pushed waves.

The divergence condition can be stated more clearly by expressing~$k$ in terms of~$v$ and then solving for the critical velocity~$v_{\mathrm{critical}}$. From this calculation, we found that the transition from fully-pushed to semi-pushed waves occurs at a universal ratio of the expansion velocity~$v$ to the linear spreading velocity~$v_{\mathrm{\textsc{f}}}$:

\begin{equation}
v_{\mathrm{critical}} = \frac{3}{2\sqrt{2}}v_{\mathrm{\textsc{f}}}.
\label{eq:v_critical}
\end{equation}

This result does not rely on Eq.~(\ref{eq:cooperative_growth}) and holds for any model of cooperative growth.

The ratio~$v/v_{\mathrm{\textsc{f}}}$ increases with cooperativity and serves as a model-independent metric of the extent to which a wave is pushed. Equation~(\ref{eq:v_critical}) and the results below further show that this metric is universal, i.e. different models with the same~$v/v_{\mathrm{\textsc{f}}}$ have the same patterns of fluctuations. We can then classify all reaction-diffusion waves using this metric. Pulled waves correspond to the special point of~$v/v_{\mathrm{\textsc{f}}}=1$. When~$1<v/v_{\mathrm{\textsc{f}}}<\frac{3}{2\sqrt{2}}$, waves are semi-pushed, and fully-pushed waves occur when~$v/v_{\mathrm{\textsc{f}}}\ge\frac{3}{2\sqrt{2}}$. Fully-pushed waves also occur when~$r(0)<0$; see Sec.~X in the SI. Such situations are called propagation into metastable state in physics~\cite{saarloos:review} and strong Allee effect in ecology~\cite{courchamp:allee_review}. Because the growth rate at the front is negative,~$v_{\mathrm{\textsc{f}}}$ does not exist, and the expansion proceeds only due to the growth in the bulk, where the fluctuations are small.

\textit{Properties of semi-pushed waves}\\
Although the perturbation theory breaks down for~$v<v_{\mathrm{critical}}$, we can nevertheless estimate the scaling exponents~$\alpha_{\mathrm{\textsc{d}}}$ and~$\alpha_{\mathrm{\textsc{h}}}$ by imposing an appropriate cutoff in the integrals in~Eqs.~(\ref{eq:LD_result}). One reasonable choice of the cutoff is~$\rho_{c} \sim 1/N$, which ensures that there is no growth in patches that have fewer than one organism. In Sec.~IX of the SI, we show that this cutoff is appropriate for deterministic fronts with~$\gamma_n=0$, but a different cutoff is needed for fluctuating fronts with~$\gamma_n>0$.

The need for a different cutoff had been recognized for a long time both from simulations~\cite{kessler:velocity_cutoff} and theoretical considerations~\cite{brunet:phenomenological_pulled}. However, a method to compute the cutoff has been developed only recently. For pulled waves, the correct value of the cutoff was obtained in Ref.~\cite{hallatschek:tuned_model} using a nonstandard moment-closure approximation for Eq.~(\ref{eq:fisher}). We extended this method to pushed waves and found that the integrals should be cut off when~$\rho$ falls below~$\rho_c \sim (1/N)^{\frac{1}{v/Dk-1}}$; see Sec.~IX in the SI. Note that the value of the cutoff depends not only on the absolute number of organisms, but also on the shape and velocity of the front. This dependence arises because population dynamics are much more sensitive to the rare excursions of the front ahead of its deterministic position than to the local fluctuations of the population density; see Sec.~IX in the SI and~\cite{brunet:phenomenological_pulled}. Since front excursions occur into typically unoccupied regions, we find that~$\rho_c<1/N$ and, therefore, genetic drift and front wandering are stronger than one would expect from~$\rho_c=1/N$.

Upon applying the correct cutoff to Eqs.~(\ref{eq:LD_result}), we find that the fluctuations in semi-pushed waves have a power-law dependence on~$N$ with a nontrivial exponent between~$0$ and~$-1$. The exponent is the same for both~$\Lambda$ and~$D_{\mathrm{f}}$ and depends only on~$v/v_{\mathrm{\textsc{f}}}$. Overall, our theoretical results can be summarized as follows 

\begin{equation}
\alpha_{\mathrm{\textsc{d}}} = \alpha_{\mathrm{\textsc{h}}} = \left\{
\begin{aligned}
& 0,\quad\quad v/v_{\mathrm{\textsc{f}}}=1,\\
& - 2\frac{\sqrt{1-v^2_{\mathrm{\textsc{f}}}/v^2}}{1-\sqrt{1-v^2_{\mathrm{\textsc{f}}}/v^2}} ,\quad\quad v/v_{\mathrm{\textsc{f}}}\in(1,\frac{3}{2\sqrt{2}}),\\
& -1,\quad\quad v/v_{\mathrm{\textsc{f}}}\ge\frac{3}{2\sqrt{2}}.\\
\end{aligned}
\right.
\label{eq:exponents_results}
\end{equation}

\noindent In the case of pulled waves, our cutoff-based calculation not only predicts the correct values of~$\alpha_{\mathrm{\textsc{d}}} = \alpha_{\mathrm{\textsc{h}}}=0$, but also reproduces the expected~$\ln^{-3}N$ scaling~(Sec.~X in the SI). 

\begin{figure}
\begin{center}
\includegraphics[width=8.7cm]{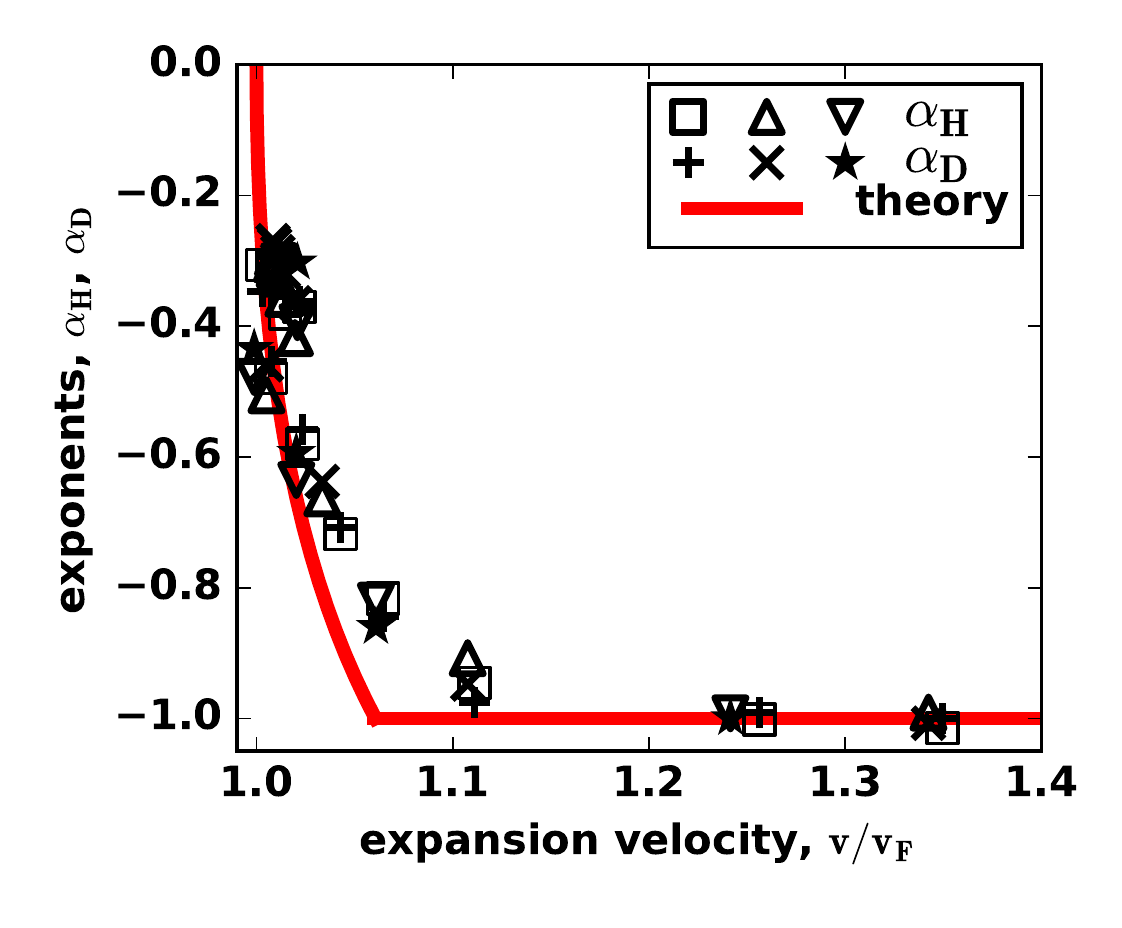}
\caption{\textbf{The universal transition from semi-pushed to fully-pushed waves.} For three different models on an Allee effect, the scaling exponents for the heterozygosity and front diffusion collapse on the same curve when plotted as a function of~$v/v_{\mathrm{\textsc{f}}}$. Thus,~$v/v_{\mathrm{\textsc{f}}}$ serves as a universal metric that quantifies the effects of cooperativity and separates semi-pushed from fully-pushed waves. We used used squares and plus signs for the model specified by Eq.~(\protect{\ref{eq:cooperative_growth}}), triangles and crosses for the model specified by Eq.~(\ref{eq:methods_allee}), inverted triangles and stars for the model specified by Eq.~(\ref{eq:methods_nonlinear}), and the red line for the theoretical prediction from~Eq.~(\ref{eq:LD_result}).}
\label{fig:collapse}
\end{center}  
\end{figure}

To test the validity of the cutoff approach, we compared its predictions to the simulations of Eq.~(\ref{eq:cooperative_growth}) and two other models of cooperative growth; see Fig.~\ref{fig:collapse}, Methods, and Fig. S5. The simulations confirm that the values of~$\alpha_{\mathrm{\textsc{d}}}$ and~$\alpha_{\mathrm{\textsc{h}}}$ are equal to each other and depend only on~$v/v_{\mathrm{\textsc{f}}}$. Moreover, there is a reasonable quantitative agreement between the theory and the data, given the errors in~$\alpha_{\mathrm{\textsc{d}}}$ and~$\alpha_{\mathrm{\textsc{h}}}$ due to the finite range of~$N$ in our simulations.

The success of the cutoff-based calculation leads to the following conclusion about the dynamics in semi-pushed waves: The fluctuations are controlled only by the very tip of the front while the growth and ancestry are controlled by the front bulk~(see Figs.~\ref{fig:growth}C and~\ref{fig:fixation}C). Thus, the counter-intuitive behavior of semi-pushed waves originates from the spatial segregation of different processes within a wave front. This segregation is not present in either pulled or fully-pushed waves and signifies a new state of the internal dynamics in a traveling wave.

\textit{Corrections to the expansion velocity due to demographic fluctuations}\\
Finally, we examined how the expansion velocity depends on the strength of demographic fluctuations. To quantify this dependence, we computed ~$\Delta v$, the difference between the actual wave velocity~$v$ and the deterministic wave velocity $v_{\mathrm{d}}$ obtained by setting~$\gamma_n=0$ in Eq.~(\ref{eq:fisher}). The perturbation theory in~$1/N$ shows that~$\Delta v\sim N^{\alpha_{\mathrm{\textsc{v}}}}$ with~$\alpha_{\mathrm{\textsc{v}}}$ equal to~$\alpha_{\mathrm{\textsc{d}}}=\alpha_{\mathrm{\textsc{h}}}$~(see Sec.~VII in the SI). Thus, we predict~$1/N$ scaling for fully-pushed waves and a weaker power-law dependence for semi-pushed waves with the exponent given by Eq.~(\ref{eq:exponents_results})\footnote{Note that, for pulled waves,~$v-v_{\mathrm{d}}\sim\ln^{-2}N$, which is different from the~$\ln^{-3}N$ scaling of~$D_{\mathrm{f}}$ and~$\Lambda$~\cite{brunet:phenomenological_pulled, hallatschek:tuned_model}. All three quantities, however, scale identically with~$N$ for semi- and fully-pushed waves}. Our simulations agreed with these results~(Fig.~S6) and, therefore, provided further support for the existence of two distinct classes of pushed waves.

Historically, corrections to wave velocity have been used to test the theories of fluctuating fronts~\cite{saarloos:review, panja:review, kessler:velocity_cutoff}. For pulled waves, the scaling~$\Delta v\sim\ln^{-2}N$ was first obtained using the~$1/N$ growth-rate cutoff~\cite{brunet:velocity_cutoff}. This calculation yielded the right answer because the correct value of the cutoff~$\rho_c \sim (1/N)^{\frac{1}{v/Dk-1}}$ reduces to~$1/N$ in the limit of pulled waves.\footnote{For pulled waves, $\Delta v$~depends on~$\ln\rho_c$, so any power-law dependence of~$\rho_c$ on~$N$ leads to the same scaling with~$N$. The coefficient of proportionality between $\Delta v$ and~$\ln^{-2}N$ is, however, also universal, and the correct value is obtained only for~$\rho_c\sim1/N$.} It is then natural to expect that the approach based on the~$1/N$ cutoff must fail for pushed waves. Indeed, Kessler~\textit{et al.}~\cite{kessler:velocity_cutoff} extended the cutoff-based approach to pushed wave and obtained results quite different from what we report here. They analyzed deterministic fronts and imposed a fixed growth-rate cutoff. Upon setting the value of this cutoff to~$1/N$, one obtains that~$\alpha_{\mathrm{\textsc{v}}}$ changes continuously from~$0$ to~$-2$ as cooperativity increases. Thus, for some values of cooperativity, the decrease with~$N$ is faster than would be expected from the central limit theorem. This clearly indicates that fluctuations rather than the modification of the growth rates play the dominant role. In Sec.~X of the SI, we show that the approach of Ref.~\cite{kessler:velocity_cutoff} supplemented with the correct value of the cutoff~$\rho_c \sim (1/N)^{\frac{1}{v/Dk-1}}$ captures the dependence of~$\Delta v$ on~$N$ for semi-pushed waves. We also explain why this approach does not apply to fully-pushed waves, in which~$\Delta v$ is not sensitive to the growth dynamics at the expansion edge, but is instead controlled by the fluctuations throughout the wave front. The SI also provides a detailed comparison of the rate of diversity loss in fluctuating \textit{vs.} deterministic fronts (Sec.~X and Figs. S7, S8 and S9).

\textbf{\large \hclr{Discussion}}

Spatially extended systems often change through a wave-like process. In reaction-diffusion systems, two types of waves have been known for a long time: pulled and pushed. Pulled waves are driven by the dynamics at the leading edge, and all their properties can be obtained by linearizing the equations of motion. In contrast, the kinetics of pushed waves are determined by nonlinear reaction processes. The distinction between pulled and pushed waves has been further supported by the recent work on the evolutionary dynamics during range expansions~\cite{hallatschek:diversity_wave, roques:allee_diversity}. In pulled waves, mutations spread only if they occur at the expansion edge, but the entire front contributes to adaptation in pushed waves. 

A natural conclusion from the previous work is that all aspects of the wave behavior are determined by whether the wave is pulled or pushed. Here, we challenged this view by reporting how fluctuation patterns change as the growth of a species becomes more nonlinear. Our main finding is that both front wandering~(a physical property) and genetic drift~(an evolutionary property) show identical behavior with increasing nonlinearity and undergo two phase transitions. The first phase transition is the classic transition between pulled and pushed waves. The second phase transition is novel and separates pushed waves into two distinct subclasses, which we termed fully-pushed and semi-pushed waves. 

The differences between the three wave classes can be understood from the spatial distribution of population dynamics. The transition from pulled to semi-pushed waves is marked by a shift of growth and ancestry from the edge to the bulk of the front (Fig.~S2). In pulled waves, the expansion velocity is determined only by the growth rate at the expansion edge, while the velocity of semi-pushed waves depends on the growth rates throughout the front. Similarly, all organisms descend from the individuals right at the edge of the front in pulled, but not in semi-pushed waves, where any organism at the front has a nonzero probability to become the sole ancestor of the future generations. The transition from semi- to fully-pushed waves is marked by an additional change in the spatial pattern of fluctuations. In fully-pushed waves, the wandering of the front arises due to the fluctuations in the shape of the entire wave front. Similarly, genetic drift at all regions of the wave front contributes to the overall fluctuations in genotype frequencies. The dynamics of semi-pushed wave are different: Both the bulk processes and rare excursions of the leading edge control the rate of diversity loss and front wandering. As a result, semi-pushed waves possess characteristics of both pulled and pushed expansions and require analysis that relies on neither linearization of the reaction-diffusion equation nor on an effective averaging within the wave front.

The shift of the fluctuations from the front to the bulk of the wave front explains the different scalings of fluctuations with the population density,~$N$. In fully-pushed waves, fluctuations obey the central limit theorem and decrease with the carrying capacity as~$N^{-1}$. This simple behavior arises because all processes are localized in a region behind the front. The number of organisms in this region grows linearly with~$N$, so the variance of the fluctuations scales as~$N^{-1}$. The central limit theorem seems not to apply to semi-pushed waves, for which we observed a nontrivial power-law scaling with variable exponents. The new scaling reflects the balance between the large fluctuations at the leading edge and the localization of the growth and ancestry processes behind the front. The departure from the~$N^{-1}$ scaling is the strongest in pulled waves, where all processes localize at the tip of the front. Since the number of organisms at the leading edge is always close to~$1$, the fluctuations are very large and decrease only logarithmically with the population size. 

The different scalings of fluctuations with population density may reflect the different structure of genealogies in pulled, semi-pushed, and fully-pushed waves. Although little is known about the structure of genealogies in the context of range expansions, we can nevertheless propose a conjecture based on an analogy with evolutionary waves in fitness space. Similar to range expansions, evolutionary waves are described by a one-dimensional reaction-diffusion equation, where the role of dispersal is assumed by mutations, which take populations to neighboring regions of the fitness space. The growth rate in evolutionary waves, however, depends not only on the local population density, but also on the location itself because the location of an organisms is its fitness. Despite this important difference, evolutionary waves and range expansions have striking similarities. Some evolutionary waves driven by frequent adaptive mutations are similar to pulled waves because their velocity is controlled by the dynamics at the wave edge, and their rate of diversity loss scales as~$\ln^{-3}N$~\cite{tsimring:wave, rouzine:wave, brunet:phenomenological_pulled, brunet:genealogies_transition, neher:genealogies, shraiman:coalescence, fisher:genealogies}. Approximately neutral evolution is in turn similar to pushed waves because its dynamics is controlled by the entire population, and the rate of diversity loss scales as~$N^{-1}$~\cite{brunet:genealogies_transition, good:diversity_selection}. The transition between these two regimes is not fully understood~\cite{brunet:genealogies_transition, good:diversity_selection}, and range expansions might provide a simpler context in which to approach this problem.

Based on the above similarity and the known structure of genealogies in evolutionary waves, it has been conjectured that genealogies in pulled waves are described by the Bolthausen-Sznitman coalescent with multiple mergers~\cite{brunet:selection_ancestry, neher:genealogies, good:diversity_selection, brunet:genealogies_transition, shraiman:coalescence, fisher:genealogies, kingman:coalescent, wakeley:textbook, berestycki:coalescent}. For fully-pushed waves we conjecture that their genealogies are described by the standard Kingman coalescent with pairwise merges. The Kingman coalescent was rigorously derived for well-mixed populations with arbitrary complex demographic structure~\cite{kingman:coalescent}, so it is natural to expect that it should apply to fully-pushed waves, where all of the dynamics occur in a well-defined region within the wave front. The structure of genealogies in semi-pushed waves is likely to be intermediate and could be similar to that of a~$\Lambda-$coalescent with multiple mergers~\cite{berestycki:coalescent, brunet:genealogies_transition}. Although these conjectures are in line with the results for evolutionary waves~\cite{brunet:genealogies_transition, good:diversity_selection, brunet:selection_ancestry, good:diversity_selection}, their applicability to range expansions requires further study, which we hope to carry out in the near future. Given that genealogies can be readily inferred from population sequencing, they could provide a convenient method to identify the class of a wave and characterize the pattern of fluctuations.

Our analysis of diversity loss and front wandering also revealed surprising universality in pushed waves. Because pushed waves are nonlinear, their velocity and front shape depend on all aspects of the growth rate, and it is natural to assume that there are as many types of pushed waves as there are nonlinear growth functions. Contrary to this expectation, we showed that many consequences of nonlinearities can be captured by a single dimensionless parameter~$v/v_{\mathrm{\textsc{f}}}$. This ratio was first used to distinguish pulled and pushed waves, but we found that~$v/v_{\mathrm{\textsc{f}}}$ also determines the transition from semi-pushed to fully-pushed waves and the magnitude of the fluctuations. We therefore suggest that~$v/v_{\mathrm{\textsc{f}}}$ could be a useful and possibly universal metric of the extent to which an expansion is pushed. Such a metric is needed to compare dynamics in different ecosystems and could play an important role in connecting the theory to empirical studies that can measure~$v/v_{\mathrm{\textsc{f}}}$ sufficiently accurately.

In most ecological studies, however, the measurements of both the observed and the Fisher velocities have substantial uncertainty. Our results caution against the common practice of using the approximate equality of~$v$ and~$v_{\mathrm{\textsc{f}}}$ to conclude that the invasion is pulled. The transition to fully-pushed waves occurs at~$v/v_{\mathrm{\textsc{f}}}=3/(2\sqrt{2})\approx1.06$, which is very close to the regime of pulled waves~$v/v_{\mathrm{\textsc{f}}}=1$. Therefore, expansions with velocities that are only a few percent greater than~$v_{\mathrm{\textsc{f}}}$ could behave very differently from pulled waves, e.g., have orders of magnitude lower rates of diversity loss. Given that Allee effects arise via a large number of mechanisms and are usually difficult to detect~\cite{courchamp:allee_review, drake:allee_evidence, berec:multiple_allee}, it is possible that many expansions thought to be pulled based on~$v\approx v_{\mathrm{\textsc{f}}}$ are actually semi- or even fully-pushed. The utility of~$v/v_{\mathrm{\textsc{f}}}$ for distinguishing pulled from semi-pushed waves could, therefore, be limited to systems where accurate measurements are possible such as waves in physical systems or in well-controlled experimental populations. Identifying fully-pushed waves based on the velocity ratio is, however, more straightforward because~$v$ substantially grater than~$v_{\mathrm{\textsc{f}}}$ unambiguously signals that the wave is fully-pushed and that the fluctuations are weak.

The somewhat narrow range of velocity ratios for semi-pushed waves,~$1<v/v_{\mathrm{\textsc{f}}}\lesssim 1.06$, does not imply that semi-pushed waves are rare. Indeed, the entire class of pulled waves is mapped to a single point~$v/v_{\mathrm{\textsc{f}}}=1$ even though a large number of growth functions lead to pulled expansions. For the growth function in Eq.~(\ref{eq:cooperative_growth}), pulled and semi-pushed waves occupy equally sized regions in the parameter space:~$B\in[0,2]$ for pulled and~$B\in(2,4)$ for semi-pushed waves. We examined several other models of cooperative growth in the SI(Sec.~XII and Fig. S1), including the one that describes the observed transition from pulled to pushed waves in an experimental yeast population~\cite{gandhi:pulled_pushed}. For all models, we found that pulled, semi-pushed, and fully-pushed waves occupy regions in the parameter space that have comparable size. Thus, all three classes of waves should be readily observable in cooperatively growing populations. 

\textbf{\large \hclr{Conclusions}}

Despite the critical role that evolution plays in biological invasions~\cite{travis:projecting, pateman:climate_invasion,nelson:gene_drive, roman:paradox_lost, lee:invasion_genetics, dlugosch:diversity_invasions, shine:spatial_sorting, korolev:arrest, phillips:toads, gray:rootworm}, only a handful of studies examined the link between genetic diversity and species ecology in this context~\cite{hallatschek:diversity_wave, roques:allee_diversity, hallatschek:constrained_expansions, lewis:wave_fixation}. The main result of the previous work is that Allee effects reduce genetic drift and preserve diversity. This conclusion, however, was reached without systematically varying the strength on the Allee effect in simulations and was often motivated by the behavior of the fixation probabilities rather than the diversity itself. Our findings not only provide firm analytical and numerical support for the previous results, but also demonstrate that the simple picture of reduced fluctuations in pushed waves does not accurately reflect the entire complexity of the eco-evolutionary feedback in traveling waves. In particular, we showed that the strength of genetic drift varies greatly between semi-pushed and fully-pushed waves. As a result, even a large Allee effect that makes the expansion pushed could be insufficient to substantially slow down the rate of diversity loss. 

Beyond specific applications in the evolution and ecology of expanding populations, our work provides an important conceptual advance in the theory of fluctuations in reaction-diffusion waves. We showed that there are three distinct classes of traveling waves and developed a unified approach to describe their fluctuations. In fully-pushed waves, fluctuations throughout the entire wave front contribute to the population dynamics. In contrast, the behavior of pulled and semi-pushed waves is largely controlled by rare front excursions, which can be captured by an effective cutoff at low population densities. Both the contribution of the dynamics at the leading edge and the value of the cutoff depend on the ratio of the wave velocity to the Fisher velocity. This dependence explains the transition from giant,~$\ln^{-3} N$, fluctuations in pulled waves to regular~$1/N$ fluctuations in fully-pushed waves. Extensions of our analytical approach could potentially be useful in other settings, where one needs to describe stochastic dynamics of non-linear waves.

\textbf{\large \hclr{Methods}}

The simulations in Figs.~\ref{fig:fluctuations}-\ref{fig:diversity} were carried out for the growth model defined by Eq.~(\ref{eq:cooperative_growth}). In Fig.~\ref{fig:collapse}, we also used two other growth models to demonstrate that our results do not depend on the choice of~$r(n)$. These growth models are specified by the following equations:

\begin{equation}
\label{eq:methods_allee}
r(n) = g_0\left(1-\frac{n}{N}\right) \left(\frac{n}{N} - \frac{n^*}{N}\right),
\end{equation}

\noindent and

\begin{equation}
\label{eq:methods_nonlinear}
r(n) = g_0\left[1-\left(\frac{n}{N}\right)^3\right] \left[\left(\frac{n}{N}\right)^3-\left(\frac{n^*}{N}\right)^3\right],
\end{equation} 

\noindent where~$N>0$ is the carrying capacity,~$g_0>0$ sets the time scale of growth, and~$c^*$ is the Allee threshold, which could assume both positive and negative values; see SI(Sec.~II). 

We simulated range expansions of two neutral genotypes in a one-dimensional habitat modeled by an array of patches separated by distance~$a$; the time was discretized in steps of duration~$\tau$. Thus, the abundance of each genotype was represented as~$n_i(t,x)$, where~$i\in\{1,2\}$ is the index of the genotype, and~$t$ and~$x$ are integer multiples of~$\tau$ and~$a$. Each time step, we updated the abundance of both genotypes simultaneously by drawing from a multinomial distribution with~$N$ trials and probability~$p_i$ to sample genotype~$i$. The values of~$p_i$ reflected the expected abundances of the genotypes following dispersal and growth:

\begin{equation}
p_i = \frac{\frac{m}{2}n_i(t,x-a) + (1-m)n_i(t,x) + \frac{m}{2}n_i(t,x+a)}{N(1-r(\tilde{n})\tau)},
\label{eq:p_i}
\end{equation}
 
\noindent where~$\tilde{n}=\frac{m}{2}n_1(t,x-a) + (1-m)n_1(t,x) + \frac{m}{2}n_1(t,x+a) + \frac{m}{2}n_2(t,x-a) + (1-m)n_2(t,x) + \frac{m}{2}n_2(t,x+a)$ is the total population density after dispersal. Note that~$p_1+p_2<1$ in patches, where the population density is less than the carrying capacity.

In the continuum limit, when~$r(n)\tau\ll1$ and~$ka\ll1$, our model becomes equivalent to Eq.~(\ref{eq:fisher}) for the population density and to Eq.~(\ref{eq:drift}) for the relative fraction of the two genotypes with~$D=ma^2/2$,~$\gamma_n = (1-n/N)/\tau$, and~$\gamma_f=1/(a\tau)$. For simplicity, we set both~$a$ and~$\tau$ to~$1$ in all of our simulations. We used~$r0=g0=0.01$ and ~$m=0.25$ for all simulations, unless noted otherwise. These values were chosen to minimize the effects of discreteness of space and time while preserving computational efficiency.

\textbf{\large \hclr{Acknowledgements}}
We thank Jeff Gore and Saurabh Gandhi for useful discussions. This work was supported by a grant from the Simons Foundation (\#409704, Kirill S. Korolev; \#327934, Oskar Hallatschek), by the startup fund from Boston University to Kirill S. Korolev, and by a National Science Foundation Career Award (\#1555330, Oskar Hallatschek). Simulations were carried out on the Boston University Shared Computing Cluster.

\clearpage
\beginsupplement

\title{\hclr{\Large \textbf{Supplemental Information}}}

Supplemental Information~(SI) provides additional results and explanations that further support the classification of reaction-diffusion waves into pulled, semi-pushed, and fully-pushed waves. SI can be roughly divided into two parts. The first part~(up to~``Cutoffs for deterministic and fluctuating fronts'') mostly reviews previous findings while the second part contains mostly new results. We describe the content of each part in more detail below.

The goal of the first half is to introduce common notation, clarify terminology, and state the results in a way that makes it easy to compare theoretical predictions to simulations, experiments, and field studies. The first two sections summarize the standard theory of deterministic reaction-diffusion waves and explain the terms that physicists and ecologists use for cooperative growth. The third section discusses the patterns of ancestry in reaction-diffusion waves. The fourth section introduces demographic fluctuations and genetic drift paying special attention to distinguishing fluctuations in population density from fluctuations in the genetic composition of the population. This distinction is not always drawn in the literature, but is important for applying the theory to specific populations. Sections~V-VIII develop a perturbation theory in~$1/N$ to compute~$\Delta v$,~$D_{\mathrm{f}}$, and~$\Lambda$. The only new results here are the second order perturbation theory for~$\Delta v$ and the expressions for~$D_{\mathrm{f}}$ and~$\Lambda$ for exactly solvable models. 

The second half of the SI begins with section~IX, which shows how to regularize the perturbation theory by introducing an effective cutoff at low population densities. The value of the cutoff differs between deterministic and fluctuating fronts and is a nontrivial result from our work. The following section contains our main arguments for the existence of the three distinct classes of reaction-diffusion waves. This section combines the results of the perturbation theory with an appropriate cutoff and provides the derivation of the scaling exponents~$\alpha_{\mathrm{\textsc{v}}}$,~$\alpha_{\mathrm{\textsc{d}}}$, and~$\alpha_{\mathrm{\textsc{h}}}$. The separation between foci of growth, ancestry, and diversity is discussed in Section~XI. Section~XII discusses the parameter range for pulled, semi-pushed and fully-pushed waves in different models. The details of computer simulations and data analysis are given in Section~XIII. The final section of the SI contains additional results from simulations. In particular, we show that (i)~the perturbation theory agrees with simulations for fully-pushed waves without any adjustable parameters; (ii)~the scaling of~$\Delta v$ with~$N$ is the same as for~$D_{\mathrm{f}}$ and~$\Lambda$; (iii)~the predicted scaling exponents match simulation results for both deterministic and fluctuating fronts. In this section, we also show that waves propagating into a metastable state are fully-pushed.

\textbf{\large {Table of Contents}}\vspace{0.25cm}\\
\textbf{\large{I. Classification of the growth rate~$r(n)$ in physics and ecology: metastability and Allee effects}}\dotfill 21\vspace{0.15cm}\\

\textbf{\large{II. Deterministic theory and classification of reaction-diffusion waves}}\dotfill 21\vspace{0.1cm}\\
\hspace*{2em}Reduced equation for traveling wave solutions\dotfill 22\vspace{0.1cm}\\
\hspace*{2em}Behavior near the boundaries\dotfill 22\vspace{0.1cm}\\
\hspace*{2em}Waves propagating into a metastable state are pushed\dotfill 23\vspace{0.1cm}\\
\hspace*{2em}Infinite number of solutions for propagation into an unstable state\dotfill 23\vspace{0.1cm}\\
\hspace*{2em}Fisher waves - a simple case of pulled waves\dotfill 23\vspace{0.1cm}\\
\hspace*{2em}Transition between pulled and pushed regimes of propagation into an unstable state\dotfill 25\vspace{0.1cm}\\
\hspace*{2em}Summary\dotfill 25\vspace{0.1cm}\\
\hspace*{2em}Example of pulled and pushed waves in exactly solvable models\dotfill 26\vspace{0.1cm}\\
\hspace*{2em}Connection to the model of cooperative growth in the main text\dotfill 26\vspace{0.1cm}\\
\hspace*{2em}Other exactly solvable models\dotfill 27\vspace{0.1cm}\\
\hspace*{2em}Comments on notation\dotfill 28\vspace{0.15cm}\\

\textbf{\large{III. Dynamics of neutral markers and fixation probabilities}}\dotfill 28\vspace{0.1cm}\\
\hspace*{2em}Forward-in-time dynamics\dotfill 29\vspace{0.1cm}\\
\hspace*{2em}Fixation probabilities\dotfill 30\vspace{0.1cm}\\
\hspace*{2em}Contribution to neutral evolution by different regions of the frontd\dotfill 31\vspace{0.1cm}\\
\hspace*{2em}Spatial distribution of ancestors\dotfill 31\vspace{0.1cm}\\
\hspace*{2em}Backward-in-time dynamics and the patterns of ancestry\dotfill 31\vspace{0.1cm}\\
\hspace*{2em}Fixation probabilities and ancestry in pulled vs. pushed waves\dotfill 33\vspace{0.1cm}\\
\hspace*{2em}Evaluation of integrals\dotfill 34\vspace{0.15cm}\\

\textbf{\large{IV. Demographic fluctuations and genetic drift}}\dotfill 35\vspace{0.1cm}\\
\hspace*{2em}Fluctuations in population size\dotfill 35\vspace{0.1cm}\\
\hspace*{2em}Fluctuations in population composition\dotfill 35\vspace{0.1cm}\\
\hspace*{2em}Relationship between demographic fluctuations and genetic drift\dotfill 36\vspace{0.1cm}\\
\hspace*{2em}Fluctuations in spatial models\dotfill 37\vspace{0.15cm}\\

\textbf{\large{V. Correction to the wave velocity,~$v$, due to a cutoff}}\dotfill 38 \vspace{0.15cm}\\

\textbf{\large{VI. Diffusion constant of the front,~$D_\mathrm{f}$}}\dotfill 41\vspace{0.1cm}\\
\hspace*{2em}Perturbation theory for demographic fluctuation\dotfill 41\vspace{0.1cm}\\
\hspace*{2em}Perturbation theory for migration fluctuations\dotfill 44\vspace{0.1cm}\\
\hspace*{2em}Results for exactly solvable models\dotfill 45\vspace{0.15cm}\\

\textbf{\large{VII. Correction to velocity due to demographic fluctuations}}\dotfill 46 \vspace{0.15cm}\\

\textbf{\large{VIII. Rate of diversity loss,~$\Lambda$}}\dotfill 50\vspace{0.1cm}\\
\hspace*{2em}Forward-in-time analysis of the decay of heterozygosity\dotfill 50\vspace{0.1cm}\\
\hspace*{2em}Backward-in-time analysis of lineage coalescence\dotfill 53\vspace{0.1cm}\\
\hspace*{2em}Explicit results for~$\Lambda$ in exactly solvable models and connection\dotfill 54\vspace{0.15cm}\\

\textbf{\large{IX. Cutoffs for deterministic and fluctuating fronts}}\dotfill 55 \vspace{0.1cm}\\
\hspace*{2em}Cutoff for deterministic fronts\dotfill 55\vspace{0.1cm}\\
\hspace*{2em}Cutoff for fluctuating fronts\dotfill 56\vspace{0.1cm}\\
\hspace*{2em}Cutoff for pushed waves expanding into a metastable state\dotfill 57\vspace{0.1cm}\\
\hspace*{2em}Cutoff for pushed waves expanding into an unstable state\dotfill 57\vspace{0.1cm}\\
\hspace*{2em}Cutoff for pulled waves\dotfill 58\vspace{0.15cm}\\

\textbf{\large{X. Scaling of~$\Delta v$,~$D_{\mathrm{f}}$, and~$\Lambda$ in pulled, semi-pushed, and fully-pushed waves}}\dotfill 58 \vspace{0.1cm}\\
\hspace*{2em}$1/N$ scaling in fully-pushed waves\dotfill 59\vspace{0.1cm}\\
\hspace*{2em}$N^{\alpha}$ scaling in semi-pushed waves\dotfill 60\vspace{0.1cm}\\
\hspace*{2em}Logarithmic scaling in pulled waves\dotfill 60\vspace{0.1cm}\\
\hspace*{2em}Scaling with N in deterministic fronts\dotfill 61\vspace{0.1cm}\\
\hspace*{2em}Comparison of~$\Delta v$ due to a cutoff in growth and due to demographic fluctuations\dotfill 61\vspace{0.15cm}\\

\textbf{\large{XI. Precise definition of the foci of growth, ancestry, and diversity}}\dotfill 63 \vspace{0.15cm}\\

\textbf{\large{XII. Prevalence of semi-pushed waves}} \dotfill 64 \vspace{0.15cm}\\

\textbf{\large{XIII. Computer simulations}}\dotfill 66 \vspace{0.1cm}\\
\hspace*{2em}Interpretation of the simulations as the Wright-Fisher model with vacancies\dotfill 66\vspace{0.1cm}\\
\hspace*{2em}Simulations of deterministic fronts\dotfill 66\vspace{0.1cm}\\
\hspace*{2em}Boundary and initial conditions\dotfill 66\vspace{0.1cm}\\
\hspace*{2em}Duration of simulations and data collection\dotfill 67\vspace{0.1cm}\\
\hspace*{2em}Computing the front velocity\dotfill 67\vspace{0.15cm}\\
\hspace*{2em}Computing the diffusion constant of the front\dotfill 67\vspace{0.15cm}\\
\hspace*{2em}Computing the heterozygosity and the rate of its decay\dotfill 67\vspace{0.15cm}\\
\hspace*{2em}Computing the scaling exponents for~$D_{\mathrm{f}}$,~$\Lambda$, and~$v-v_{\mathrm{d}}$\dotfill 68\vspace{0.15cm}\\

\textbf{\large{XIV. Supplemental results and figures}}\dotfill 68 \vspace{0.15cm}\\

\clearpage

\textbf{\large \hclr{I. Classification of the growth rate~$r(n)$ in physics and ecology: metastability and Allee effects}}

Here, we introduce the terminology used to characterize the growth term in Eq.~(5) in physics and ecology.

The physics literature typically distinguishes between propagation into an unstable state when~$r(0)>0$ and propagation into a metastable state when~$r(0)<0$. The main difference between these two cases is their response to small perturbations. For~$r(0)>0$, the introduction of any number of organisms into an uncolonized habitat results in a successful invasion, so~$n=0$ is an unstable state. In contrast, for~$r(0)<0$, invasions fail when the number of introduced organisms is sufficiently small. Large introductions, however, do result in an invasion, so~$n=0$ is a metastable state. Since~$n=N$ is stable against any perturbation, it is referred to as a stable state. When~$n=0$ and~$n=N$ are the only states stable against small perturbations,~$r(n)$ is often termed bistable. 

In ecology, populations with a metastable state at~$n=0$~($r(0)<0$) are said to exhibit a strong Allee effect. When~$r(0)>0$, the growth dynamics is further classified as exhibiting either a weak Allee effect or no Allee effect. A population exhibits no Allee effect if~$r(0)\ge r(n)$ for all~$n$; otherwise, it exhibits a weak Allee effect. A common example of~$r(n)$ without an Allee effect is the logistic growth, for which~$r(n)$ decays monotonically from~$r_0$ at~$n=0$ to~$0$ at~$n=N$.

\textbf{\large \hclr{II. Deterministic theory and classification of reaction-diffusion waves}}

The goal of this section is to explain the difference between pulled and pushed waves. Despite considerable work on the topic, it is easy to conflate related but distinct properties of traveling waves, and few concise and self-contained accounts are available in the literature. Here, we only provide minimal and mostly intuitive discussion following Ref.~\cite{saarloos:review}, which is one of the most lucid and comprehensive reviews on propagating reaction-diffusion fronts. This section contains no new results.

The distinction between pulled and pushed waves arises already at the level of deterministic reaction-diffusion equations, so stochastic effects are not considered in this section. Specifically, we are interested in the asymptotic behavior as~$t\to+\infty$ of the solutions of the following one-dimensional problem:

\begin{equation}
\frac{\partial n}{\partial t} = D\frac{\partial^2 n}{\partial x^2} + r(n) n,
\label{eq:deterministic_n}
\end{equation}

\noindent where~$r(n)$ is the per capita growth rate that is negative for large~$n$, but could be either positive or negative at small~$n$. Except for possibly~$n=0$, we assume that there is only one other stable fixed point of~$dn/dt=r(n)n$ at~$n=N$, where~$N$ is the carrying capacity. We further assume that the initial conditions are sufficiently localized, e.g.~$n(0,x)$ is strictly zero outside a finite domain. Under these assumptions, the long time behavior of~$n(t,x)$ is that of a traveling wave:

\begin{equation}
\begin{aligned}
&n(t,x) = n(\zeta), \\ 
&\zeta = x - vt, \\
&n(-\infty) = N,\\
&n(+\infty) = 0.
\end{aligned}
\label{eq:comoving}
\end{equation}

\noindent and our main task is to determine the wave velocity~$v$ and the shape of the front in the comoving reference frame~$n(\zeta)$. Throughout this paper, we focus on the right-moving part of the expansion; the behavior of the left-moving expansion is completely analogous.

\textit{Reduced equation for traveling wave solutions}\\
By substituting Eq.~(\ref{eq:comoving}) into Eq.~(\ref{eq:deterministic_n}), we obtain the necessary condition on~$v$ and~$n(\zeta)$:

\begin{equation}
Dn'' + vn' + r(n)n = 0,
\label{eq:reduced_ode}
\end{equation}

\noindent where primes denote derivatives with respect to~$\zeta$. The solutions of~Eq.~(\ref{eq:reduced_ode}) clearly have a translational degree of freedom, i.e., if~$n(\zeta)$ is a solution, then~$n(\zeta + \mathrm{const})$ is a solution. We can eliminate this degree of freedom by choosing the references frame such that~$n(0)=N/2$. The general solution of Eq.~(\ref{eq:reduced_ode}) then has only one remaining degree of freedom, which corresponds to the value of~$n'(0)$. For a given~$v$, the existence of the solution depends on whether the value of~$n'(0)$ can be adjusted to match the boundary conditions specified by Eq.~(\ref{eq:comoving}).

Depending on the behavior of the solution at~$\zeta\to\pm\infty$, each boundary condition may or may not provide a constraint on the solution and, therefore, remove either one or zero degrees of freedom. To determine the number constraints, we linearize Eq.~(\ref{eq:reduced_ode}) near each of the boundaries.

\textit{Behavior near the boundaries}\\
For~$\zeta\to-\infty$, we let~$u=N-n$ and obtain:

\begin{equation}
Du'' + vu' + \left.\frac{dr}{dn}\right|_{n=N}Nu = 0,
\label{eq:reduced_ode_bulk}
\end{equation}

\noindent which has the following solution:

\begin{equation}
\label{eq:solution_bulk}
u = A_b e^{k_b \zeta} + C_b e^{q_b \zeta},
\end{equation}

\noindent where

\begin{equation}
\label{eq:k_q_bulk}
\begin{aligned}
&k_b = \frac{\sqrt{v^2 - 4D\left.\frac{dr}{dn}\right|_{n=N}N}-v}{2D},\\
&q_b = \frac{-\sqrt{v^2 - 4D\left.\frac{dr}{dn}\right|_{n=N}N}-v}{2D}.\\
\end{aligned}
\end{equation}

\noindent Here, we used subscript~$b$ to indicate that we refer to the behavior of the population bulk. Because the carrying capacity is an attractive fixed point,~$dr/dn$ is negative at~$n=N$, and therefore~$k>0$ and~$q<0$. We then conclude that the boundary condition at~$\zeta \to -\infty$ requires that~$C_b=0$ and thus selects a unique value of~$n'(0)$. 

Next, we analyze the behavior at the front and linearize Eq.~(\ref{eq:reduced_ode}) for small~$n$:

\begin{equation}
Dn'' + vn' + r(0)n = 0,
\label{eq:reduced_ode_front}
\end{equation}

\noindent which has the following solution:

\begin{equation}
\label{eq:solution_front}
n = A e^{-k \zeta} + C e^{-q \zeta},
\end{equation}

\noindent where

\begin{equation}
\label{eq:k_q_front}
\begin{aligned}
&k = \frac{v+\sqrt{v^2 - 4Dr(0)}}{2D},\\
&q = \frac{v-\sqrt{v^2 - 4Dr(0)}}{2D}.\\
\end{aligned}
\end{equation}

\textit{Waves propagating into a metastable state are pushed}\\
The implications of~Eq.~(\ref{eq:k_q_front}) depend on the sign of~$r(0)$. For a strong Allee effect~($r(0)<0$), that is when the wave propagates into a metastable state, we find that~$k>0$ and~$q<0$. In consequence, the boundary condition requires that~$C=0$ and imposes an additional constraint. For an arbitrary value of~$v$ this constraint cannot be satisfied because the value of~$n'(0)$ is already determined by the behavior in the bulk. However, there could be a value of~$v$ for which the constraints in the bulk and at the front are satisfied simultaneously. This special~$v$ is then the velocity of the wave. Because the value of~$v$ depends population dynamics throughout the wave front, the propagation into a metastable state is classified as a pushed wave.

\textit{Infinite number of solutions for propagation into an unstable state}\\
For~$r(0)>0$, i.e. when the wave propagates into an unstable state, the analysis is more subtle because the behavior at the front does not fully constrain the velocity of the wave. To demonstrate this, we draw the following two conclusions from Eq.~(\ref{eq:k_q_front}). First,~$v$ must be greater or equal to $v_{\mathrm{\textsc{f}}}=2\sqrt{Dr(0)}$; otherwise, the solutions are oscillating around zero as~$\zeta\to+\infty$ and violate the biological constraint that~$n\ge0$. Second, both~$k$ and~$q$ are positive for~$v\ge v_{\mathrm{\textsc{f}}}$, so the solution decays to~$0$ for arbitrary~$A$ and~$C$. Thus, the boundary condition at the front does not impose an additional constraint, and one can find a solution of Eq.~(\ref{eq:reduced_ode}) satisfying Eq.~(\ref{eq:comoving}) for arbitrary~$v\ge v_{\mathrm{\textsc{f}}}$.

\textit{Fisher waves---a simple case of pulled waves}\\
The multiplicity of solutions for~$r(0)>0$ posed a great challenge for applied mathematics, statistical physics, and chemistry, and its resolution greatly stimulated the development of the theory of front propagation. The simplest context in which one can show that the wave velocity is unique is when there is no Allee effect, i.e.~$r(0)\ge r(n)$ for all~$n\in(0,N)$. This condition guarantees that the expansion velocity for Eq.~(\ref{eq:deterministic_n}) is less or equal than the expansion velocity for the dynamical equation linearized around~$n=0$:

\begin{equation}
\frac{\partial n}{\partial t} = D\frac{\partial^2 n}{\partial x^2} + r(0) n.
\label{eq:deterministic_n_linear}
\end{equation}

\noindent Moreover, we show below that the upper bound on~$v$ from the linearized dynamics coincides with the lower bound imposed by Eq.~(\ref{eq:k_q_front}). Thus, the velocity of the wave is~$2\sqrt{Dr(0)}$. This analysis was first carried out in Ref.~\cite{kolmogorov:wave} for an equation, which is now commonly known as the Fisher, Fisher-Kolmogorov, or Fisher-Kolmogorov-Petrovskii-Piskunov equation~\cite{kolmogorov:wave, fisher:wave}. Therefore, we refer to waves with~$r(0)\ge r(n)$ as Fisher or Fisher-like waves. Since the expansion of the population is determined by the linearized dynamics, Fisher waves are pulled.

The upper bound on the velocity from the linearized equation can be obtained for an arbitrary initial condition via the standard technique of Fourier and Laplace transforms~\cite{saarloos:review}. However, it is much simpler to use an initial condition for which a closed form solution is available. The results of such an analysis are completely general because the velocity of the wave should not depend on the exact shape of~$n(0,x)$ at least when the population density is zero outside a finite domain. So, we make a convenient choice of~$n(0,x)=\delta(x)$, where~$\delta(x)$ is the Dirac delta function. Then, the solution of Eq.~(\ref{eq:deterministic_n_linear}) is the product of an exponential growth term and a diffusively widening Gaussian:

\begin{equation}
\label{eq:linear_solution}
n(t,x) = \frac{1}{\sqrt{4\pi D t}}e^{r(0)t}e^{-\frac{x^2}{4Dt}}.
\end{equation}

To determine the long-time behavior, it is convenient to shift into a comoving reference frame~(see Eq.~(\ref{eq:comoving})):

\begin{equation}
\label{eq:linear_solution}
n(t,\zeta) = \frac{1}{\sqrt{4\pi D t}}e^{\left(r(0)-\frac{v^2}{4D}\right)t}e^{-\frac{\zeta^2}{4Dt}}e^{-\frac{v}{4D}\zeta}.
\end{equation}

\noindent From the first exponential term, it immediately clear that the wave velocity must be equal to~$2\sqrt{Dr(0)}$; otherwise the population size will either exponentially grow or decline as~$t\to+\infty$. As we said earlier, this result demonstrates that the upper and the lower bounds on~$v$ coincide with each other and uniquely specify the expansion velocity.

The last exponential term in Eq.~(\ref{eq:linear_solution}) further shows that the wave profile decays exponentially with~$\zeta$ as~$e^{-k_{\mathrm{\textsc{f}}}\zeta}$, where

\begin{equation}
k_{\mathrm{\textsc{f}}} = \sqrt{\frac{r(0)}{D}}
\label{eq:k_f}
\end{equation}

consistent with~Eq.~(\ref{eq:k_q_front}) for~$v=v_{\mathrm{\textsc{f}}}$. 

Finally, the middle exponential term in Eq.~(\ref{eq:linear_solution}) describes the transition from a sharp density profile at~$t=0$ to the asymptotic exponential decay of~$n(\zeta)$. For~$\zeta < 2\sqrt{Dt}$, this term is order one, and the front is well approximated by the asymptotic exponential profile. For~$\zeta > 2\sqrt{Dt}$, the Gaussian term becomes important, and the shape of the front is primarily dictated by the diffusive spreading from the initial conditions. Thus, the linear spreading dynamics builds up a gradual decay of the population density starting from a much sharper population front due to localized initial conditions. 

The steepness of the front created by the linearized dynamics is very important for the transition from pulled to pushed waves, so we emphasize that Eq.~(\ref{eq:k_f}) specifies the lower bound on the rate of exponential decay of the population density. Indeed, Eq.~(\ref{eq:linear_solution}) indicates that~$n$ decays no slower than~$e^{-k_{\mathrm{\textsc{f}}}\zeta}$ at all times. 

\textit{Transition between pulled and pushed regimes of propagation into an unstable state}\\
Up to here, we have shown that waves are pushed when the Allee effect is strong, but the waves are pulled when there is no Allee effect. Now, we shift the focus to the remaining case of the weak Allee effect and show that the transition from pulled to pushed waves occurs when the growth at intermediate~$n$ is sufficiently high to support an expansion velocity greater than the linear spreading velocity~$v_{\mathrm{\textsc{f}}}$.

To understand the origin of pushed waves, we need to consider the behavior of the population density near the front for~$v>v_{\mathrm{\textsc{f}}}$. Equations~(\ref{eq:solution_front}) and (\ref{eq:k_q_front}) predict that~$n(\zeta)$ is a sum of two exponentially decaying terms with different decay rates: One term decays with the rate~$k>k_{\mathrm{\textsc{f}}}$, but the other with the rate~$q<k_{\mathrm{\textsc{f}}}$. This behavior is in general inconsistent with the solution of the linearized dynamical equation~(\ref{eq:deterministic_n_linear}). Indeed, our analysis of a localized initial condition suggests and more rigorous analysis proves~\cite{saarloos:review} that~$n(t,x)$ decays to zero at least as fast as~$k_{\mathrm{\textsc{f}}}$. Therefore, we must require that~$C=0$ just as in the case of propagation into a metastable state. 

The extra requirement makes the problem of satisfying the boundary conditions overdetermined. As a result, there are two alternatives. First, there could be no solutions for any~$v>v_{\mathrm{\textsc{f}}}$. In this case, the expansion must be pulled because it is the only feasible solution. Second, a special value of~$v$ exists for which the boundary conditions at~$\zeta\to\pm\infty$ can be satisfied simultaneously. In this case, the wave is pushed because the pulled expansion at low~$n$ is quickly overtaken by the faster expansion from the bulk. 

For completeness, we also mention that initial conditions that are not localized and decay asymptotically as~$e^{-\tilde{k}x}$ with~$\tilde{k}<k_{\mathrm{\textsc{f}}}$ lead to pulled expansions with~$v=1/2v_{\mathrm{\textsc{f}}}(\tilde{k}/k_{\mathrm{\textsc{f}}}+k_{\mathrm{\textsc{f}}}/\tilde{k})>v_{\mathrm{\textsc{f}}}$. This result immediately follows from the solution of~Eq.~(\ref{eq:deterministic_n_linear}) using either Laplace transforms or a substitution of an exponential ansatz~$n=e^{-\tilde{k}(x-vt)}$. Discreteness of molecules or individuals, however, make such initial conditions impossible, so the wave propagation with~$v>v_{\mathrm{\textsc{f}}}$ can only describe transient behavior for initial conditions with a slow decay at large~$x$. 

\textit{Summary}\\
To summarize, we have shown that all waves propagating into a metastable state are pushed, and their profile decays as~$e^{-k\zeta}$ at the front with~$k$ given by~Eq.~(\ref{eq:k_q_front}). Propagation into an unstable state could be either pulled or pushed depending on the relative strength of the growth behind vs. at the front. Pushed waves expand with~$v>v_{\mathrm{\textsc{f}}}$, and their profile decays as~$e^{-k\zeta}$. In contrast, pulled waves expand with~$v=v_{\mathrm{\textsc{f}}}$, and their profile decays as~$e^{-k_\mathrm{\textsc{f}}\zeta}$. In general, the class of a wave propagating into an unstable state cannot be determined without solving the full nonlinear problem~(Eq.~(\ref{eq:deterministic_n}) or~Eq.~(\ref{eq:reduced_ode})). However, there are a few rigorous results that determine the class of the wave from simple properties of~$r(n)$. To the best of our knowledge, the most general result of this type is that waves are always pulled when there is no Allee effect. 

\textit{Example of pulled and pushed waves in exactly solvable models}\\
We conclude this section by illustrating the transition from pulled to pushed waves in an exactly solvable model. Our main goal is to provide an concrete example for the abstract concepts discussed so far. In addition, the models presented below are used in simulations and to explicitly calculate the diffusion constant of the front and the rate of diversity loss in the regime of fully-pushed waves.

Consider~$r(n)$ specified by

\begin{equation}
\label{eq:allee_quadratic}
r(n) = g_0\left(1-\frac{n}{N}\right) \left(\frac{n}{N} - \frac{n^*}{N}\right),
\end{equation}

\noindent where~$g_0$ sets the time scale of growth,~$N$ is the carrying capacity, and~$n^*$ is a parameter that controls the strength of an Allee effect. For every value of~$n^*$, the growth function~$r(n)$ can be classified in one of three types: no Allee effect, weak Allee effect, or strong Allee effect. The growth function does not exhibit an Allee effect when~$r(0)\ge r(n)$ for all~$n\in(0,N)$. For the model defined above, the region of no Allee effect corresponds to~$n^*/N\le-1$. When an Allee effect is present, one distinguished between a strong Allee effect,~$r(0)<0$, and a weak Allee effect,~$r(0)\ge0$. Thus, the Allee effect is weak for~$n^*/N\in(-1,0]$ and strong for~$n^*>0$. In the latter case,~$n^*$ represents the minimal population density required for net growth and is known as the Allee threshold. In the following, we will refer to~$n^*$ as the Allee threshold regardless of its sign.

With~$r(n)$ specified by Eq.~(\ref{eq:allee_quadratic}), Eq.~(\ref{eq:reduced_ode}) admits an exact solution~\cite{fife:allee_wave, aronson:allee_wave, murray:mathematical_biology, korolev:wave_splitting}:

\begin{equation}
\label{eq:quadratic_solution}
\begin{aligned}
& n(\zeta) = \frac{N}{1 + e^{\sqrt{\frac{g_0}{2D}} \zeta}}\\
& v  = \sqrt{\frac{g_0D}{2}}\left(1 - 2\frac{n^*}{N} \right).
\end{aligned}
\end{equation} 

For~$n^*>0$, we expect a unique pushed wave, so Eq.~(\ref{eq:quadratic_solution}) must provide the desired solution. For~$n^*/N<-1$, we know that the wave must be pulled with~$v=v_{\mathrm{\textsc{f}}}=2\sqrt{-Dg_0n^*/N}$ and~$k=k_{\mathrm{\textsc{f}}}=\sqrt{-g_0n^*/(ND)}$. To identify the transition between pulled and pushed waves, we equate the two expressions for the velocity and obtain that the critical value of the Allee threshold is given by~$n^*/N=-1/2$. Thus, the transition between pulled and pushed waves occurs within the region of a weak Allee effect in agreement with the general theory developed above. The behavior of this exactly solvable model is summarized in Table~\ref{tab:quadratic}.

\begin{table}
\begin{tabular}{ | l | *{4}{ c |} }
\hline
	Allee threshold, $n^*/N$ & $(-\infty, -1]$ & $(-1, -0.5]$ & (-0.5, 0] & (0, 0.5) \\
\hline
Allee effect & none & \multicolumn{2}{| c |}{weak} & strong \\
\hline
Stability of invaded state & \multicolumn{3}{| c |}{unstable} & metastable \\
\hline
Type of expansion & \multicolumn{2}{| c |}{pulled} & \multicolumn{2}{| c |}{pushed} \\
\hline
\end{tabular}
	\caption{\textbf{Comparison of wave type, state stability, and Allee effect for an exactly solvable model of range expansions given by Eq.~(\protect{\ref{eq:allee_quadratic}}).} Note that the transition from pulled to pushed waves does not coincide with a change in the type of growth. In particular, the pulled-pushed transition is distinct from the transition between no Allee effect and a weak Allee effect; it is also distinct from the transition between propagation into unstable and metastable state. For~$c^*/N>1/2$, the relative stability of the populated and unpopulated states changes, and the expansion wave propagates from~$n=0$ state into~$n=N$ state.}
\label{tab:quadratic}
\end{table}

\textit{Connection to the model of cooperative growth in the main text}\\
The model of cooperative growth that we defined in the main text~(Eq.~(3)) is a simple re-parameterization of~Eq.~(\ref{eq:allee_quadratic}) with~$r_0=-g_0n^*/N$ and~$B=-N/n^*$. In consequence,~$r(n)$~exhibits no Allee effect for~$B\le1$ and a weak Allee effect for~$B>1$. A strong Allee effect is not possible for any~$B$ because~$r(0)>0$. Hence, the wave always propagates into an unstable state. This model choice was convenient for us because it ensures that the transition between semi-pushed and fully-pushed waves is unambiguously distinct from the transition between propagation into unstable and metastable states. 

The transition between pulled and pushed waves occurs at~$B=2$. For pushed waves~($B>2$), the velocity and front shape are given by

\begin{equation}
\label{eq:cooperative_growth_v_n_pushed}
\begin{aligned}
&n(\zeta) = \frac{N}{1 + e^{\sqrt{\frac{r_0B}{2D}} \zeta}}\\
&v = \sqrt{\frac{r_0DB}{2}}\left(1 + \frac{2}{B} \right),
 \end{aligned}
\end{equation}

\noindent while for pulled waves~($B\le 2$) the corresponding results are

\begin{equation}
\label{eq:cooperative_growth_v_n_pulled}
\begin{aligned}
&n(\zeta) \sim  e^{-\sqrt{\frac{r_0}{D}}\zeta},\quad \zeta\to+\infty,\\
&v = 2\sqrt{Dr_0}.
 \end{aligned}
\end{equation}

Although Eq.~(3) is equivalent to Eq.~(\ref{eq:allee_quadratic}), our computer simulations of these models reveal complementary information because they explore different cuts through the parameter space. In particular, when we vary~$B$ in one model, we change both~$g_0$ and~$n^*/N$ in the other model. Similarly, changes in~$n^*/N$ modify both~$r_0$ and~$B$. Concordant results for the two parameterizations indicate that the transitions from pulled to semi-pushed and from semi-pushed to fully-pushed waves are universal and do not depend on the precise definition of cooperativity or the strength of an Allee effect. 

\textit{Other exactly solvable models}\\
For completeness, we also mention that exact solutions for pushed waves are known for a slightly more general class of~$r(n)$ than the quadratic growth function discussed so far. The following results are from Ref.~\cite{petrovskii:exact_models}, which is an excellent resource for exactly solvable models of traveling waves. 

For~$r(n)$ defined by

\begin{equation}
\label{eq:nonlinear_growth}
r(n) = g_0\left[1-\left(\frac{n}{N}\right)^b\right] \left[\left(\frac{n}{N}\right)^b-\left(\frac{n^*}{N}\right)^b\right],
\end{equation}

\noindent pushed waves occurs for~$n^*/N>-(b+1)^{-1/b}$, and their velocity and profile shape are given by

\begin{equation}
\label{eq:nonlinear_growth_v_n_pushed}
\begin{aligned}
& v = \sqrt{(b + 1)g_{0}D} \left[ \frac{1}{b + 1} - \left(\frac{n^{*}}{N}\right)^b \right],\\
& n(\zeta) = \frac{N}{\left(1 + e^{b\sqrt{\frac{g_0}{(b+1)D}} \zeta} \right)^{1/b}}. 
 \end{aligned}
\end{equation}

The transition point from semi-pushed to fully-pushed waves follows from these results and the condition that~$v=\sqrt{9/8}v_\mathrm{\textsc{f}}$. The value of this critical Allee threshold is given by~$n^*/N=-[2(b+1)]^{-1/b}$. Finally, the transition from weak to no Allee effect occurs at~$n^*=-N$ and from weak to strong Allee effect at~$n^*=0$.

\textit{Comments on notation}\\
For all models of~$r(n)$, it is sometimes convenient to use the normalized population density~$\rho = n/N$. For the exactly solvable models introduced above, it is also convenient to define~$\rho^*=n^*/N$. This notation is used in the main text and in the following sections.

We can also now be more precise about the definitions of population bulk, front, interior regions of the front, and the leading edge, which we use throughout the paper. The population bulk is defined the region where~$n$ is close to~$N$ and Eq.~(\ref{eq:solution_bulk}) holds. Similarly, the leading edge, the tip of the front, front edge,~\textit{etc.} refer to the region of~$n\ll N$, where Eq.~(\ref{eq:solution_front}) holds. The region with intermediate~$n$ is termed as the front or more precisely as the interior region of the front or the bulk of the front. We tried to avoid using the generic term front whenever that can cause confusion between the leading edge and the interior region of the front.

\textbf{\large \hclr{III. Dynamics of neutral markers and fixation probabilities}}

Heritable neutral markers provide a window in the internal dynamics of an expanding population. These dynamics can be studied either forward in time or backward in time. The former approach describes how the spatial distribution of neutral markers changes over time and provides an easy way to compute the fixation probabilities of neutral mutations. The latter approach describes the patterns of ancestry that emerge during a range expansion and provides a natural way to infer population parameters from genetic data. The main goal of this section is to demonstrate that both forward-in-time and backward-in-time dynamics fundamentally change at the transition from pulled to pushed waves. In pulled waves, all organisms trace their ancestry to the very tip of the expansion front, which is also the only source of successful mutations. In contrast, the entire expansion front contributes to the evolutionary dynamics in pushed waves. This section contains no new results except for the analytical calculation of the fixation probabilities in the exactly solvable models. The discussion largely follows that in Refs.~\cite{hallatschek:diversity_wave} and~\cite{roques:allee_diversity}. Our main goal here is to introduce the notation to be used in the following sections and explain how the patterns of ancestry depend on cooperativity.

\textit{Forward-in-time dynamics}\\
Let us consider a subpopulation carrying a neutral marker~$i$ and describe how its population density~$n_i(t,x)$ changes forward in time. Since the growth and migration rates are the same for all markers,~$n_i$ obey the following equation:

\begin{equation}
\frac{\partial n_i}{\partial t} = D\frac{\partial^2 n_i}{\partial x^2} + r(n)n_i,
\label{eq:deterministic_n_i}
\end{equation}

\noindent where, as before,~$n$ is the total population density, which is given by~$n=\sum_i n_i$ if all individuals are labeled by some marker.

To isolate the behavior of neutral markers from the overall population growth, it is convenient to define their relative frequency in the population:~$f_i=n_i/n$. From Eqs.~(\ref{eq:deterministic_n}) and~(\ref{eq:deterministic_n_i}), it follows that~\cite{vlad:hydrodynamic, hallatschek:diversity_wave}  

\begin{equation}
\frac{\partial f_i}{\partial t} = D\frac{\partial^2 f_i}{\partial x^2} + 2D\frac{\partial \ln n}{\partial x}\frac{\partial f_i}{\partial x}.
\label{eq:deterministic_f}
\end{equation}

The new advection-like term arises from the nonlinear change of variables and accounts for a larger change in~$f_i$ due to immigration from regions with high population density compared to regions with low population density. The main effect of the new term is to establish a ``flow'' of~$f_i$ from the posterior to the anterior of the front. 

It is also convenient to shift into the comoving reference frame~(see Eq.~(\ref{eq:comoving})) in order to focus on the dynamics that occurs at a fixed position within the front region rather than at a fixed position in the stationary reference frame. The result reads

\begin{equation}
\frac{\partial f_i}{\partial t} = D\frac{\partial^2 f_i}{\partial \zeta^2} + v\frac{\partial f_i}{\partial \zeta} + 2D\frac{\partial \ln n}{\partial \zeta}\frac{\partial f_i}{\partial \zeta}.
\label{eq:deterministic_f_comoving}
\end{equation}

We now drop the index~$i$ because, for the rest of this section, we focus on the frequency of a single marker, which we denote simply by~$f(t,\zeta)$. In the following, we also assume that~$n(t,\zeta)$ has reached the steady-state given by~Eq.~(\ref{eq:reduced_ode}). The evolutionary dynamics are typically much slower than ecological dynamics, so the initial transient in the dynamics of~$n$ could be neglected.

The analysis of Eq.~(\ref{eq:deterministic_f_comoving}) is greatly simplified by the existence of a time-invariant quantity:

\begin{equation}
\label{eq:pi}
	\pi = \frac{\int_{-\infty}^{+\infty}f(t,\zeta)n^{2}(\zeta)e^{v\zeta/D}d\zeta}{\int_{-\infty}^{+\infty}n^{2}(\zeta)e^{v\zeta/D}d\zeta},
\end{equation}

\noindent which was first demonstrated in Refs.~\cite{hallatschek:diversity_wave} and~\cite{roques:allee_diversity}.

\noindent To show that~$\pi$ is conserved, we evaluate its time derivative:

\begin{equation}
\label{eq:pi_conserved}
	\frac{d\pi}{dt} = \frac{\int_{-\infty}^{+\infty}\frac{\partial f}{\partial t}n^{2}e^{v\zeta/D}d\zeta}{\int_{-\infty}^{+\infty}n^{2}e^{v\zeta/D}d\zeta},
\end{equation}

\noindent and replace~$\partial f/\partial t$ by the right hand side of~Eq.~(\ref{eq:deterministic_f_comoving}). The numerator can then be simplified via the integration by parts to eliminate the derivatives of~$f$ with respect to~$\zeta$ in favor of~$f$. This leads to the cancellation of all the terms and thereby proves that~$\pi$ does not depend on time. 

The conservation on~$\pi$ makes it quite straightforward to determine the fixation probabilities, the spatial distribution of ancestors, and the contribution of different parts of the front to the neutral evolution. We now discuss each of these results separately.

\textit{Fixation probabilities}\\
Since only spatial derivatives of~$f$ enter Eq.~(\ref{eq:deterministic_f_comoving}), we conclude that~$f=\mathrm{const}$ is a solution that describes the steady state after migration has smoothed out the spatial variations in the initial conditions. The value of the constant is given by~$\pi$ because

\begin{equation}
\label{eq:f_limit}
	\pi = \lim_{t\to+\infty} \frac{\int_{-\infty}^{+\infty}f(t,\zeta)n^{2}(\zeta)e^{v\zeta/D}d\zeta}{\int_{-\infty}^{+\infty}n^{2}(\zeta)e^{v\zeta/D}d\zeta} =  \frac{\int_{-\infty}^{+\infty}\left(\lim_{t\to+\infty}f\right)n^{2}e^{v\zeta/D}d\zeta}{\int_{-\infty}^{+\infty}n^{2}e^{v\zeta/D}d\zeta} =  \lim_{t\to+\infty}f(t,\zeta).
\end{equation}

At the level of deterministic dynamics, this result captures how the final fraction of a genotype depends on its initial distribution in the population. Genetic drift, however, leads to the extinction of all but one genotype, so~$f(t,\zeta)$ should be interpreted as the average over the stochastic dynamics~(one can take the expectation value of both sides of Eq.~(\ref{eq:stochastic_f_space}) below). Since the expected value of~$f$ is~$1$ times the probability of fixation plus~$0$ times the probability of extinction, we immediately conclude that the fixation probability equals~$\pi$. Thus, given the initial distribution of a neutral marker~$f(0,\zeta)$, we can obtain its fixation probability by evaluating the integrals in Eq.~(\ref{eq:pi}) at~$t=0$.

We can also express this result in terms of the absolute abundance of the neutral marker~$n_i$~(here we keep the subscript to distinguish~$n_i$ from the total population density). Since~$f_i = n_i/n$, the fixation probability is given by

\begin{equation}
\label{eq:pi_n_i}
	\pi = \frac{\int_{-\infty}^{+\infty}n_i(t,\zeta)n(\zeta)e^{v\zeta/D}d\zeta}{\int_{-\infty}^{+\infty}n^{2}(\zeta)e^{v\zeta/D}d\zeta}.
\end{equation}

For a single organism present at location~$\zeta_0$, we can approximate~$n_i$ as~$\delta(\zeta-\zeta_0)$ and thus obtain the fixation probability of a single mutant:

\begin{equation}
\label{eq:pi_mutant}
	u(\zeta_0) = \frac{n(\zeta_0)e^{v\zeta_0/D}}{\int_{-\infty}^{+\infty}n^{2}(\zeta)e^{v\zeta/D}d\zeta}.
\end{equation}

\textit{Contribution to neutral evolution by different regions of the front}\\
From Eq.~(\ref{eq:pi_mutant}), we can determine how different regions contribute to the neutral evolution during a range expansion. Neutral evolution proceeds through two steps: first a random mutations appears somewhere in the population and second the frequency of the mutation fluctuations until the mutation either reaches fixation or becomes extinct. For a given mutation, the probability that it first occurred at location~$\zeta$ is proportional to~$n(\zeta)$, and its fixation probability is given by~$u(\zeta)$. Thus, the fraction of fixed mutations that first originated at~$\zeta$ is given by~$n^{2}(\zeta)e^{v\zeta/D}/\int_{-\infty}^{+\infty}n^{2}(\zeta)e^{v\zeta/D}d\zeta$.

\textit{Spatial distribution of ancestors}\\
Finally, we note that the last result also represents the probability that the ancestor of a randomly sampled individual from the population used to live at location~$\zeta$ sufficiently long ago. To demonstrate this, we label all individuals between~$\zeta$ and~$\zeta + d\zeta$ at a long time in the past and note that the probability of a random individual to have its ancestor at~$\zeta$ equals the expected number of labeled descendants. Since the long time limit of~$f$ is given by~$\pi$, we immediately conclude that the probability distribution of ancestor locations is given by

\begin{equation}
\label{eq:S_forward_in_time_stationary}
S(\zeta) = \frac{n^{2}(\zeta)e^{v\zeta/D}}{\int_{-\infty}^{+\infty}n^{2}(\zeta)e^{v\zeta/D}d\zeta}.
\end{equation}

\textit{Backward-in-time dynamics and the patterns of ancestry}\\
To characterize the patterns of ancestry in a population, it is convenient to describe the dynamics of ancestral lineages backward in time. Following the approach of Ref.~\cite{hallatschek:diversity_wave}, we show below that the probability~$S(\tau,\zeta)$ that an ancestor of a given individual lived at position~$\zeta$ time~$\tau$ ago is governed by the following equation~\cite{hallatschek:diversity_wave}:

\begin{equation}
\label{eq:S_backward_in_time}
\frac{\partial S}{\partial \tau} = D\frac{\partial^2 S}{\partial \zeta^2} - v\frac{\partial S}{\partial \zeta} - 2D\frac{\partial}{\partial \zeta}\left(\frac{\partial \ln n}{\partial \zeta}S\right).
\end{equation}

Here, the diffusion term randomizes the position of the ancestor; the term proportional to~$v$ pushes the ancestor towards the tip of the wave and reflects the change into the comoving reference frame; the last term pushes the ancestor towards the population bulk and reflects the fact that the ancestor is more likely to have emigrated from the region where the population density is higher. For large~$\tau$, it is easy to check that these forces balance and result in a stationary distribution~$S(\zeta)$ given by Eq.~(\ref{eq:S_forward_in_time_stationary}). Note that, unlike in Eq.~(\ref{eq:deterministic_f_comoving}) for the dynamics of~$f$, the right hand side of Eq.~(\ref{eq:S_backward_in_time}) contains a divergence of a flux and, therefore, preserves the normalization of~$S$, i.e.~$\int S(\tau,\zeta) d\zeta = \mathrm{const}$. Indeed, the probability that an ancestor was present somewhere in the population must always equal to one. 

The derivation of Eq.~(\ref{eq:S_backward_in_time}) from Eq.~(\ref{eq:deterministic_f_comoving}) follows the standard procedure for changing from the forward-in-time to the backward-in-time description~\cite{risken:fpe, gardiner:handbook} and consists of three steps. The first step is to define the ``propagator'' function that can describe both forward-in-time and backward-in-time processes. We denote this function as~$G(t_d,\zeta_d;t_a,\zeta_a)$ and define it as the probability that a descendant located at~$\zeta_d$ at time~$t_d$ originated from an ancestor who lived at time~$t_a$ at position~$\zeta_a$. On the one hand, with~$t_d$ and~$\zeta_d$ fixed,~$G$ can be viewed as a function of~$t_a$ and~$\zeta_a$ that specifies the probability distribution of ancestor location at a specific time. On the other hand, with~$t_a$ and~$\zeta_a$ fixed,~$G$ can be viewed as a function of~$t_d$ and~$\zeta_d$ that describes the spatial and temporal dynamics of the expected frequency of the descendants from all organisms that were present at~$\zeta_a$ at time~$t_a$. This two-way interpretation follows from the labeling thought-experiment that we used to derive~$S(\zeta)$ using the forward-in-time formulation.

In the second step, we claim that~$G$ obeys the same equation as~$f(t,\zeta)$, i.e. Eq.~(\ref{eq:deterministic_f_comoving}). This statement immediately follows from the forward-in-time interpretation of~$G$. We formally state this result as

\begin{equation}
\label{eq:Lf}
\frac{\partial G}{\partial t_d} = \mathcal{L}_{\zeta_d}G,
\end{equation}

\noindent where~$\mathcal{L}_{\zeta_d}$ is the linear operator from the right hand side of Eq.~(\ref{eq:deterministic_f_comoving}), and we used the subscript~$\zeta_{d}$ to indicate variable on which the operator acts.

The third step is to derive an equation for~$G$ that involves only the ancestor-related variables. To this purpose, we consider an infinitesimal change in~$t_a$,  

\begin{equation}
\begin{aligned}
G(t_d, \zeta_d; t_a-dt, \zeta_a) &= \int_{-\infty}^{+\infty} G(t_d, \zeta_d; t_a, \zeta') G(t_a, \zeta'; t_a-dt, \zeta_a) d\zeta' \\ &= \int_{-\infty}^{+\infty} G(t_d, \zeta_d; t_a, \zeta') [G(t_a-dt, \zeta'; t_a-dt, \zeta_a) + dt \mathcal{L}_{\zeta'}G(t_a-dt, \zeta'; t_a-dt, \zeta_a)] d\zeta' \\ & = \int_{-\infty}^{+\infty} G(t_d, \zeta_d; t_a, \zeta') [\delta(\zeta'-\zeta_a)+ dt \mathcal{L}_{\zeta'}\delta(\zeta'-\zeta_a)] d\zeta' 
\end{aligned}
\label{eq:bit_derivation_one}
\end{equation}

\noindent where we used the Markov property of the forward-in-time dynamics, then Eq.~(\ref{eq:Lf}), and finally the fact that~$G(t, \zeta_a; t, \zeta_d) = \delta(\zeta_a-\zeta_d)$, which immediately follows from the definition of~$G$. Equation~(\ref{eq:bit_derivation_one}) can be further simplified by expanding the left hand side in~$dt$:

\begin{equation}
-\frac{\partial G}{\partial t_a} = \int_{-\infty}^{+\infty}  G(t_d, \zeta_d; t_a, \zeta')\mathcal{L}_{\zeta'}\delta(\zeta'-\zeta_a) d\zeta'. 
\label{eq:bit_derivation_two}
\end{equation}

\noindent Finally, we integrate by parts to transfer the derivatives in~$\mathcal{L}_{\zeta_a}$ from the delta function to~$G(t_d, \zeta_d; t_a, \zeta')$:

\begin{equation}
-\frac{\partial G}{\partial t_a} = \int_{-\infty}^{+\infty} \delta(\zeta'-\zeta_a) \mathcal{L}^{+}_{\zeta'} G(t_d, \zeta_d; t_a, \zeta') d\zeta' = \mathcal{L}^{+}_{\zeta_a} G(t_d, \zeta_d; t_a, \zeta_a), 
\label{eq:bit_derivation_two} 
\end{equation}

\noindent where,~$\mathcal{L}^{+}_{\zeta_a}$ is the operator that results from the integration by parts and is known as the adjoint operator of~$\mathcal{L}_{\zeta_a}$. 

Equation~(\ref{eq:bit_derivation_two}) is the desired backward-in-time formulation that involves only the ancestor-related variables. To see that it is equivalent to Eq.~(\ref{eq:S_backward_in_time}), one needs to explicitly compute~$\mathcal{L}^{+}_{\zeta_a}$, substitute the definition of~$\tau = t_d - t_a$, and change the notation from~$G$ to~$S$.

\textit{Fixation probabilities and ancestry in pulled vs. pushed waves}\\
We conclude this section by comparing the fixation probabilities and patterns of ancestry in pulled and pushed waves. We focus on~$S(\zeta)$ as a typical example; other quantities, e.g.,~$u(\zeta)$ can be analyzed in the same fashion. 

Up to a constant factor,~$S(\zeta)$ is given by~$n^2(\zeta)e^{v\zeta/D}$. For large negative~$\zeta$, the exponential factor tends rapidly to zero indicating that the bulk of the wave contributes little to the neutral evolution and is unlikely to contain the ancestor of future generations. This conclusion applies to both pulled and pushed waves. The behavior at the front is more subtle because~$n(\zeta)\to0$ and~$e^{v\zeta/D}\to+\infty$ as~$\zeta\to+\infty$. To determine the scaling of~$S(\zeta)$ at the front, we replace~$n$ by its asymptotic form~$e^{-k\zeta}$ and obtain that

\begin{equation}
\label{eq:S_scaling}
S(\zeta) \propto e^{-\zeta(2k - v/D)} \propto e^{-\zeta(k-q)},
\end{equation}

\noindent where the last expression follows from the fact that~$v = D(k+q)$; see Eq.~(\ref{eq:k_q_front}).

For pushed waves,~$k>q$ so the tip of the front makes a vanishing contribution to the neutral evolution. Therefore, the main contribution to~$S(\zeta)$ must come from the interior regions of the front. In fact, for the exactly solvable model specified by Eq.~(\ref{eq:allee_quadratic}), one can express~$\zeta$ in terms of~$n$ and show that

\begin{equation}
S(\zeta) = \frac{2\pi\rho^*}{\sin(2\pi\rho^*)}\rho^{1+2\rho^*}(1-\rho)^{1-2\rho^*},
\label{eq:S_exact}
\end{equation}

\noindent where~$\rho=n/N$, and~$\rho^*=n^*/N$. This result clearly demonstrates that~$S(\zeta)$ is peaked at intermediate population densities~(specifically at~$\rho = (1+2\rho^*)/2$). 

For pulled waves,~$k=q$, and~$n^2(\zeta)e^{v\zeta/D}\to\mathrm{const}$ as~$\zeta\to+\infty$. Thus, every point arbitrarily far ahead of the front contributes equally to the neutral evolution. Since the region ahead of the front is infinite, the relative contribution of the bulk and the interior of the front must be negligible compared to that of the leading edge. A more careful analysis requires one to impose a cutoff on~$\zeta$ at sufficiently low densities so that~$S(\zeta)$~can be normalized. We show how to introduce such a cutoff in section~IX.  

In summary, we determined fixation probabilities and patterns of ancestry, which are plotted in Fig.~2 of the main text. We also compared the dynamics in pulled and pushed waves and found that they are driven by distinct spatial regions of the front. In pulled waves, the very tip of the front not only ``pulls'' the wave forward, but also acts as the focus of ancestry and the sole source of successful mutations. In pushed waves, however, the entire front contributes to both the expansion dynamics and evolutionary processes. We refer the readers to Ref.~\cite{hallatschek:diversity_wave, roques:allee_diversity, hallatschek:tuned_model} for the original derivations of these results and further discussion. 

\textit{Evaluation of integrals}\\
Let us briefly explain how one can evaluate the integrals that appear in Eq.~(\ref{eq:S_forward_in_time_stationary}) and in similar equations for the diffusion constant of the front and the rate of diversity loss. The main insight is to change the independent variable from~$\zeta$ to~$\rho$ using equation Eq.~(\ref{eq:quadratic_solution}). The following formulas are useful for this purpose:

\begin{equation}
	\label{eq:variable_change}
	\begin{aligned}
		& \zeta = \sqrt{\frac{2D}{g_0}}\ln\left(\frac{1-\rho}{\rho}\right), \\
		& d\zeta =  -\sqrt{\frac{2D}{g_0}}\frac{d\rho}{\rho(1-\rho)}, \\
		& e^{\zeta v/D} = \rho^{-(1-2\rho^*)}(1-\rho)^{1-2\rho^*}, \\
		& \frac{d\rho}{d\zeta} = - \sqrt{\frac{g_0}{2D}}\rho(1-\rho).
	\end{aligned}
\end{equation}

\noindent After this change of variable, all integrals become beta functions~\cite{gradshteyn:tables}. For example,

\begin{equation}
	\label{eq:int_example_beta}
	\int^{+\infty}_{-\infty}n^2(\zeta)e^{\frac{v\zeta}{D}}d\zeta = \sqrt{\frac{2D}{g_0}} \int_{0}^{1} \rho^{2\rho^*}(1-\rho)^{-2\rho^*} = \mathrm{Be}(1+2\rho^*,1-2\rho^*).
\end{equation}

\noindent The integrals of this type can be evaluated in the complex plane. Specifically, one can equate the integral around the branch cut from~$0$ to~$1$ to the residue at~$\zeta=\infty$. For the integral above, this results in

\begin{equation}
	\label{eq:int_example_final}
	\sqrt{\frac{2D}{g_0}} \int_{0}^{1} \rho^{2\rho^*}(1-\rho)^{-2\rho^*} = \frac{2\pi\rho^*}{\sin(2\pi\rho^*)}.
\end{equation}

An alternative method to evaluate the integrals is to use the following properties of the gamma and beta functions~\cite{gradshteyn:tables}: 
\begin{equation}
	\label{eq:beta_functions}
	\begin{aligned}
		& \mathrm{Be}(x,y) = \frac{\Gamma(x)\Gamma(y)}{\Gamma(x+y)}, \\
		& \Gamma(z)\Gamma(1 - z) = \frac{\pi}{\sin{\pi z}}. \\
	\end{aligned}
\end{equation}
Using these formulas, we can evaluate all beta functions of the type~\(\mathrm{Be}(n + z, m - z)\), where~$n$ and~$m$ are positive integers and~$z \in (0,1)$. The general expressions are
\begin{equation}
	\label{eq:beta_general_expression}
	\begin{aligned}
	\mathrm{Be}(n+z, 1 - z) & = \frac{\prod_{k = 0}^{n - 1}(z + k)}{n!}\frac{\pi}{\sin{\pi z}}, &\\
	\mathrm{Be}(n+z, m - z) & = \frac{\prod_{k = 0}^{n - 1}(z + k)\prod_{l = 1}^{m - 1}(l - z)}{(n + m - 1)!}\frac{\pi}{\sin{\pi z}}, & m > 1.
	\end{aligned}
\end{equation}

\textbf{\large \hclr{IV. Demographic fluctuations and genetic drift}}

In this section, we describe how to move beyond the deterministic approximation in Eqs.~(\ref{eq:deterministic_n}) and~(\ref{eq:deterministic_f}) and account for the effects of demographic fluctuations and genetic drift. Because the magnitude of the fluctuations depends on the details of the reproductive process, we need to introduce two additional functions of the population density~$\gamma_{n}(n)$ and~$\gamma_{f}(n)$ that describe the strength of fluctuations in the population size and composition respectively. This section contains no new results, and its main purpose is to specify the relevant notation and carefully discuss how stochastic dynamics should be added to deterministic reaction-diffusion equations.

For simplicity, we will consider well-mixed populations first and limit the discussion to neutral markers, e.g. genotypes that do not differ in fitness.

\textit{Fluctuations in population size}\\
Demographic fluctuations in the size~$n(t)$ of a well-mixed population arise due to the randomness of births and deaths. The simplest and most commonly used assumption is that an independent decision is made for each organism on whether it dies or to reproduces~\cite{mckane:genetic_drift_from_lv, nelson:demographic_noise_2015}. In a short time interval~$\Delta t$, the number of births and deaths are therefore two independent Poisson random variables with mean and variance equal to~$\mu_b(n)n\Delta t$ for births and to~$\mu_d(n)n\Delta t$ for deaths. Here,~$\mu_b(n)$ and~$\mu_d(n)$ are the per capita rates of birth and death respectively. Since the change in the population size is the difference between these two independent random variables, we conclude that the mean change of~$n$ is~$[\mu_b(n)-\mu_d(n)]n\Delta t$ and the variance is~$[\mu_b(n)+\mu_d(n)]n\Delta t$. In the continuum limit, the dynamics of the population size is then described by the following stochastic differential equation 

\begin{equation}
\frac{dn}{dt} = [b(n) - d(n)]n + \sqrt{[b(n) + d(n)]n}\eta(t),
\label{eq:birth_death}
\end{equation}

\noindent where~$\eta(t)$ is the It\^o white noise, i.e.~$\langle \eta(t_1) \eta(t_2) \rangle = \delta(t_1-t_2)$. In the following, we denote ensemble averages by angular brackets, and use~$\delta(t)$ for the Dirac delta function. 

The assumption that births and death events are independent random variables is however too restrictive. For example, the number of birth always equals the number of deaths in the classic Wright-Fisher model, which exhibits no fluctuations in~$n$ as a result. In addition, the number of births or deaths could deviate from the Poisson distribution and therefore have the variance not equal to the mean. To account for such scenarios, we need to generalize Eq.~(\ref{eq:birth_death}) as follows

\begin{equation}
\frac{dn}{dt} = r(n) n + \sqrt{\gamma_{n}(n)n}\eta(t),
\label{eq:demographic_fluctuations_wm}
\end{equation}

\noindent where~$r(n)$ is the difference between the birth and death rates, and $\gamma_{n}$ characterizes the strength of the demographic fluctuations. The value of~$\gamma_{n}$ can be easily determined from model parameters because~$\gamma_{n}n\Delta t$ is the sum of the variances of births and deaths during~$\Delta t$ minus twice their covariance.  

\textit{Fluctuations in population composition}\\
Genetic drift arises because the choice of the genotype that is affected by a specific birth or death event is random. This randomness does not lead to a change in the average abundance of neutral genotypes, but induces a random walk in the space of population compositions described by the species fractions~$\{f_i\}$. It is easy to show that both births and deaths contribute equally to the increase in the variance of~$f_{i}$ in a short time interval~$\Delta t$~\cite{mckane:genetic_drift_review, mckane:genetic_drift_from_lv, nelson:demographic_noise_2015}, so the strength of the genetic drift depends only on the total number of updates due to both births and deaths:~$\gamma_f(n)n\Delta t$, where~$\gamma_f=\mu_b(n)+\mu_d(n)$. Since probability to choose genotype~$i$ for an update is proportional to its current fraction in the population~$f_{i}(t)$, the number of updates for each genotype will be given by a multinomial distribution with~$\gamma_f(n)n\Delta t$ trials and outcome probabilities given by~$\{f_i\}$. This leads to the following continuum limit~\cite{mckane:genetic_drift_review, mckane:genetic_drift_from_lv, nelson:demographic_noise_2015, korolev:review}:

\begin{equation}
\label{eq:genetic_drift_wm}
\frac{df_{i}}{dt} = \sqrt{\frac{\gamma_f(n)}{n}f_{i}(1-f_{i})}\eta_{i}(t), 
\end{equation}

\noindent with the covariance structure of the noises~$\eta_{i}$ specified by

\begin{equation}
\label{eq:drift_noise}
\langle \eta_{i}(t_1)\eta_{j}(t_2) \rangle =  \delta(t_1-t_2) 
\left \{
\begin{aligned}
& 1,\quad i=j,\\
& -\sqrt{\frac{f_i f_j}{(1-f_i)(1-f_j)}}, \quad i\ne j. 
\end{aligned}
\right.
\end{equation} 

\noindent The factor of~$1/n$ under the square root in Eq.~(\ref{eq:genetic_drift_wm}) arises because a single birth or death event changes the frequency of the genotype at most by~$1/n$. The dependence on~$f_i$ reflects the properties of the multinomial distribution and ensures that the sum of~$f_i$ does not fluctuate and remains equal to~$1$.

Since genetic drift and demographic fluctuations are independent from each other, i.e.~$\langle \eta(t_1)\eta_{i}(t_2)\rangle =0$, and Eqs.~(\ref{eq:demographic_fluctuations_wm}),~(\ref{eq:genetic_drift_wm}), and~(\ref{eq:drift_noise}) completely specify population dynamics. In particular, one can easily obtain the dynamical equations for the genotype abundances~$n_{i}$ by differentiating~$n_i=f_in$.

An alternative, but completely equivalent, formulation of Eqs.~(\ref{eq:genetic_drift_wm}) and~(\ref{eq:drift_noise}) reads

\begin{equation}
\label{eq:genetic_drift_wm_a}
\frac{df_{i}}{dt} = \sqrt{\frac{\gamma_f(n)}{n}f_{i}}\left(\tilde{\eta}_{i}(t)-\sqrt{f_i}\sum_j\tilde{\eta}_j\right), 
\end{equation}

\noindent where~$\langle \tilde{\eta}_i(t_1)\tilde{\eta}_{j}(t_2)\rangle=\delta_{ij}\delta(t_1-t_2)$; we use~$\delta_{ij}$ to denote the Kronecker delta, i.e. the identity matrix. This alternative definition arises naturally when one derives the equations for~$f_i$ starting from the dynamical equations for species abundances~$n_i$ and shows that~$\sum_j f_j$ is constant more clearly. We provide this formulation only for completeness and do not use in the following.

\textit{Relationships between demographic fluctuations and genetic drift}\\
For the simple processes of uncorrelated births and deaths described by Eq.~(\ref{eq:birth_death}), one can show that~$\gamma_f(n)=\gamma_n(n)$~\cite{mckane:genetic_drift_from_lv, nelson:demographic_noise_2015}, but this is equality does not hold in general. For example, in the simulations that we describe below~$\gamma_f$ is independent of~$n$, but~$\gamma_n$ monotonically decreases to zero as the population size approaches the carrying capacity. Nevertheless, in a wide set of models,~$\gamma_f(0)=\gamma_n(0)$ because the dynamics of different genotypes becomes uncorrelated at low population densities and their fluctuations are determined by~$\gamma_0 = \mu_b(0) + \mu_d(0)$. As a result, the fluctuations in pulled and semi-pushed waves depends only on~$\gamma_0$ when the carrying capacity is large enough to justify the asymptotic limit.

Exceptions to~$\gamma_f(0)=\gamma_n(0)$ are in principle possible, for example, when many cycles of birth and death occur without an appreciable change in the total population. Such dynamics could arise when a slow and quasi-deterministic niche construction is required to increase the current limit on the population size.  

For completeness, we also mention that the dynamical equations for~$n_{i}$ take a particularly simple form when~$\gamma_n(n) = \gamma_f(n) = \gamma(n)$:

\begin{equation}
\label{eq:ni_independent}
\frac{dn_{i}}{dt} = g(n)n_i + \sqrt{\gamma(n)n_{i}}\tilde{\eta}_{i}(t),
\end{equation}

\noindent where~$\langle \tilde{\eta}_i(t_1)\tilde{\eta}_{j}(t_2)\rangle=\delta_{ij}\delta(t_1-t_2)$. Thus, for simple birth-death models, the fluctuations in genotype abundances are independent from each other as expected. Although Eq.~(\ref{eq:ni_independent}) is often used as a starting point for the analysis~\cite{hallatschek:tuned_moments}, it does not capture the full complexity of possible eco-evolutionary dynamics.

\textit{Fluctuations in spatial models}\\
It is straightforward to extend the above discussion to spatial populations where~$n$ and~$f_i$ depend on both~$t$ and~$x$. The net result is that Eqs.~(\ref{eq:deterministic_n}) and~(\ref{eq:deterministic_f}) acquire stochastic terms specified by Eqs.~(\ref{eq:demographic_fluctuations_wm}) and~(\ref{eq:genetic_drift_wm}). The results read

\begin{equation}
\frac{\partial n}{\partial t} = D\frac{\partial^2 n}{\partial x^2} + r(n) n + \sqrt{\gamma_{n}(n)n}\eta(t),
\label{eq:stochastic_n_space}
\end{equation}

\noindent and

\begin{equation}
\frac{\partial f_{i}}{\partial t} = D\frac{\partial^2 f_{i}}{\partial x^2} + 2\frac{\partial \ln n}{\partial x}\frac{\partial f_{i}}{\partial x} +  \sqrt{\frac{\gamma_f(n)}{n}f_{i}(1-f_{i})}\eta_{i}(t).
\label{eq:stochastic_f_space}
\end{equation}

\noindent The noise-noise correlations are specified by the following equations

\begin{equation}
\langle \eta(t_1,x_1)\eta_i(t_2,x_2) \rangle = 0,
\end{equation}

\begin{equation}
\label{eq:demographic_noise}
\langle \eta(t_1,x_1)\eta(t_2,x_2) \rangle = \delta(t_1-t_2)\delta(x_1-x_2),
\end{equation}

and

\begin{equation}
\label{eq:drift_noise_space}
\langle \eta_{i}(t_1,x_1)\eta_{j}(t_2,x_2) \rangle =  \delta(t_1-t_2)\delta(x_1-x_2) \left[\delta_{ij}- (1-\delta_{ij})\sqrt{\frac{f_i f_j}{(1-f_i)(1-f_j)}}\right].
\end{equation}

Note that, in Eqs.~(\ref{eq:stochastic_n_space}) and~(\ref{eq:stochastic_f_space}), we omitted a noise term that accounts for the randomness of migration~(or diffusion in the context of chemical reactions). Such noise inevitably arises when each organisms makes an independent decision on whether to migrate to a particular nearby site. Because this noise conserves the number of individual it appears as a derivative of the flux in the dynamical equation for~$n_{i}$. The general form of this noise is~$\partial_x[\sqrt{\gamma_m n_i} \chi_i]$ with~$\langle \chi_i(t_1,x_1)\chi_{j}(t_2,x_2)\rangle=\delta_{ij}\delta(t_1-t_2)\delta(x_1-x_2)$, and~$\chi_i$ are uncorrelated with~$\eta$ and~$\eta_i$~\cite{meerson:velocity_fluctuations}. 

Migration noise does not typically lead to any new qualitative dynamics, and we will show below it leads to the same scaling of the fluctuations with the bulk population density~$N$. Moreover, migration noise is often negligible compared to genetic drift. For example, it can be neglected when the migration rate is small or the number of organisms at the dispersal stage is much larger then the number of reproducing adults~(compare the number of seeds vs. the number of trees). We do not consider migration noise further because it is absent in our computer simulation. For the sake of simplicity and greater computational speed, we chose to perform the migration update deterministically.

\textbf{\large \hclr{V. Correction to the wave velocity,~$v$, due to a cutoff}}

How do demographic fluctuations modify the dynamics of wave propagation? This question is central to our paper and has generated significant interest in nonequilibrium statistical physics. Most early studies explored how demographic stochasticity modifies the expansion velocity~\cite{brunet:velocity_cutoff, kessler:velocity_cutoff, tsimring:wave, goldenfeld:structural_stability, panja:review}. While velocity corrections are small and likely negligible in the context of range expansions, they are essential for the description of evolving populations, which are often modeled as traveling waves in fitness space~\cite{tsimring:wave, rouzine:wave, hallatschek:tuned_model}. More importantly, wave velocity serves a salient and easy to measure observable that has been frequently used to test the theories of fluctuating fronts. This section shows how to compute the corrections to wave velocity using perturbation theory. All results in this section have been derived previously in Refs.~\cite{mikhailov:diffusion,goldenfeld:structural_stability, armero:corrections, rocco:diffusion, meerson:velocity_fluctuations, saarloos:review}. Our main goal here is to introduce the relevant notation and to explain the perturbation theory in the simplest context.  

Because non-linear stochastic equations are notoriously difficult to analyze, a direct calculation of~$v$ is challenging, and several approximate approaches were developed instead~\cite{saarloos:review}. In this section, we describe the simplest of these approaches that imposes a cutoff on the growth rate below a certain population density~$n_{c}$:

\begin{equation}
r_{\mathrm{cutoff}}(n) = r(n)\theta(n-n_c),
\label{eq:growth_cutoff}
\end{equation}

\noindent where~$\theta(n)$ is the Heaviside step function, which equals one for positive arguments and zero for negative arguments.

Although the value of~$n_c$ must reflect the strength of the demographic fluctuations, it is not entirely clear how to determine~$n_c$ a priori. A natural guess is to set~$n_c$ to one over the size of the patch size in simulations so that no growth occurs in regions where the expected number of individuals is less than one. However, this choice does not capture the full complexity of demographic fluctuations as shown in section~IX. For now, we keep~$n_c$ as an unspecified parameter and focus on the corrections to~$v$ due to the change in the growth rate specified by Eq.~(\ref{eq:growth_cutoff}). The position of the cuttoff where the deterministic profile reaches~$n_c$ is denoted as~$\zeta_c$, i.e.~$n(\zeta_c)=n_c$.

The corrections to~$v$ can be computed using a perturbation expansion in~$\Delta r(n) = r_{\mathrm{cutoff}}(n) - r(n)$. This approach has been developed by different groups either for computing the corrections due to a cutoff or for computing the diffusion constant of the front~\cite{mikhailov:diffusion, goldenfeld:structural_stability, armero:corrections, rocco:diffusion, meerson:velocity_fluctuations, saarloos:review}. 

Let us first introduce a convenient notation for the perturbation expansion that is also used in the following sections, where the perturbation is a stochastic variable rather than a deterministic cutoff. All quantities that solve the deterministic, unperturbed problem~(Eq.~(\ref{eq:deterministic_n}) or Eq.~(\ref{eq:reduced_ode})) are denoted with subscript~$d$. All quantities that solve the full, perturbed problem are denoted without a subscript. And, the differences between the two types of quantities are denotes with~$\Delta$. 

With this notation, the perturbed equation reads

\begin{equation}
\frac{\partial n}{\partial t} = D\frac{\partial^2 n}{\partial x^2} + n r_{\mathrm{cutoff}}(n), 
\end{equation}

\noindent or equivalently 

\begin{equation}
\frac{\partial n}{\partial t} = D\frac{\partial^2 n}{\partial x^2} + n r(n) + n \Delta r(n).
\label{eq:perturbed_problem}
\end{equation}

\noindent We seek the solution correct to the first order in~$\Delta r$ via the following ansatz

\begin{equation}
n(t,x) = n_d(x-v_dt-\Delta vt) + \Delta n (x-v_dt-\Delta vt),
\label{eq:perturbed_profile}
\end{equation}

\noindent where~$\Delta n$ is the correction to the shape of the stationary density profile, and~$\Delta v$ is the correction to the expansion velocity.

The zeroth order in perturbation theory yields the unperturbed equation:

\begin{equation}
Dn''_d + v_d n'_d + r n_d = 0,
\label{eq:zeroth_order}
\end{equation}

\noindent which is automatically satisfied by our choice of~$n_d$.

To obtain the equations for the next order, we expand~$\partial n/\partial t$ as

\begin{equation}
\frac{\partial n}{\partial t} \approx - v_d n'_d -  n'_d \Delta v- v_d \Delta n',
\end{equation}

\noindent the diffusion term as

\begin{equation}
D\frac{\partial^2 n}{\partial x^2} \approx D n''_d + \Delta n'',
\end{equation}

\noindent and the growth term as

\begin{equation}
r_{\mathrm{cutoff}}(n)n \approx n_dr(n_d) + r(n_d)\Delta n + r'(n_d)n_d\Delta n + n_d\Delta r(n_d).
\end{equation}

\noindent As before, we use primes to denote derivatives of functions of a single argument.

The resulting equation for the first order in perturbation theory reads

\begin{equation}
\label{eq:perturbation_first_order}
\mathfrak{L}_p \Delta n = - n_d\Delta v - n_d\Delta r,
\end{equation}

\noindent where

\begin{equation}
\label{eq:perturbation_L}
\mathfrak{L}_p  = D\frac{d^2}{d\zeta^2} + v_d\frac{d}{d\zeta} + r(n_d) + r'(n_d)n_d,
\end{equation}

\noindent is the linear operator that acts on the comoving spatial variable~$\zeta=x-vt=x-(v_d+\Delta v)t$.

Although Eq.~(\ref{eq:perturbation_first_order}) has two unknowns~$\Delta n$ and~$\Delta v$, both quantities can be determined simultaneously because the solution for~$\Delta n$ exists only for a specific value of~$\Delta v$. The constraint on~$\Delta v$ comes from the fact that~$\mathfrak{L}_p$ has an eigenvalue equal to zero and, therefore, its image does not span the entire space of functions possible on the right hand side of Eq.~(\ref{eq:perturbation_first_order}). As a result,~$\Delta v$ must be chosen to make~$-n_d\Delta v - n_d\Delta r$ lie in the image of~$\mathfrak{L}_p$. 

The zero mode of~$\mathfrak{L}_p$ originates from the translational invariance of the unperturbed problem, for which both~$n_d(\zeta)$ and~$n_d(\zeta+\mathrm{const})$ are solutions. Therefore, there should be no restoring force from the dynamical equation for~$\Delta n$ that effectively translates the front by an infinitesimal distance~$\delta\zeta$. Since~$n_d(\zeta +\delta \zeta) \approx n_d(\zeta) + n'_d(\zeta) \delta\zeta$, we expect that~$\Delta n \propto n_d'(\zeta)$ should not alter the left hand side of Eq.~(\ref{eq:perturbation_first_order}). Consistent with reasoning, the differentiation of Eq.~(\ref{eq:zeroth_order}) with respect to~$\zeta$ shows that~$\mathfrak{L}_pn'_d=0$. Thus,~$\mathfrak{L}_p$ indeed has a zero mode with~$n'_d$ being the right eigenvector. The corresponding left eigenvector can be obtained by solving~$\mathfrak{L}^{+}_{p}L(\zeta) = 0$ and is given by

\begin{equation}
\label{eq:eigen_peturbation_L}
L(\zeta) = n'_d(\zeta)e^{v_d\zeta/D}.
\end{equation}

To compute~$\Delta v$, we multiply both sides of~Eq.~(\ref{eq:perturbation_first_order}) by~$L(\zeta)$ and integrate over~$\zeta$. Since~$L\mathfrak{L}_p$ is equivalent to zero, the terms on the left hand side cancel, and we obtain that

\begin{equation}
\label{eq:delta_v_cutoff}
\Delta v  =-\frac{\int_{-\infty}^{+\infty} e^{v_d \zeta/D} n_d'(\zeta) n_d(\zeta) \Delta r(n_d(\zeta)) d\zeta}{\int_{-\infty}^{+\infty}  e^{v_d \zeta/D} [n_d'(\zeta)]^2 d\zeta},
\end{equation}

\noindent which is the same result as in Refs.~\cite{goldenfeld:structural_stability, gottwald:melnikov}. For the specific form of~$\Delta r$ due to a cutoff, this formula simplifies to

\begin{equation}
\label{eq:delta_v_cutoff}
\Delta v  = \frac{\int_{\zeta_c}^{+\infty} e^{v_d \zeta/D} n_d'(\zeta) n_d(\zeta) r(n_d(\zeta)) d\zeta}{\int_{-\infty}^{+\infty}  e^{v_d \zeta/D} [n_d'(\zeta)]^2 d\zeta},
\end{equation}

\noindent which is the main result of this section. 

The solvability condition that we used to compute~$\Delta v$ has a simple interpretation: All perturbations that act along the zero eigenmode of~$\mathfrak{L}_p$ accumulate unattenuated and contribute to the translation of the front, i.e. to~$\Delta v$ rather than to~$\Delta n$. This fact can be seen more clearly from the time-dependent perturbation theory that we use in sections~VI and~VII to compute the diffusion constant of the front and the corrections to the wave velocity due to demographic noise rather than a cutoff.

\textbf{\large \hclr{VI. Diffusion constant of the front,~$D_{\mathrm{f}}$}}

While a cutoff can account for changes in the velocity due to demographic fluctuations, it cannot capture the fluctuations in the front shape and position. In this section, we describe the stochastic properties of the front using an extension of the perturbation theory developed above. Originally developed in Refs.~\cite{mikhailov:diffusion} and~\cite{rocco:diffusion, meerson:velocity_fluctuations}, this approach shows that the position of the front performs a random walk that can be described by an effective diffusion constant. Following Ref.~\cite{meerson:velocity_fluctuations}, we derive the general formula for~$D_{\mathrm{f}}$ given by Eq.~(5) of the main text and evaluate it explicitly for the exactly solvable models introduced in the beginning of the SI. The calculations for the exactly solvable models are the only new results in this section.

\textit{Perturbation theory for demographic fluctuations}\\
The calculation follows exactly the same steps as in section~V. We begin by restating Eq.~(\ref{eq:stochastic_n_space}) in a more convenient form:

\begin{equation}
\frac{\partial \rho}{\partial t} = D\frac{\partial^2 \rho}{\partial x^2} + r(\rho)\rho +\frac{1}{\sqrt{N}}\Gamma(\rho)\eta
\label{eq:diffusion_problem_formulation}
\end{equation}

\noindent where~$\Gamma$ denotes the strength of the noise term

\begin{equation}
\label{eq:Gamma}
\Gamma(\rho) = \sqrt{\gamma_n(\rho)\rho}.
\end{equation}

\noindent In this section, we use the normalized population density~$\rho = n/N$ instead of~$n$ to indicate that the stochastic term is small and scales as~$1/\sqrt{N}$. We also introduce a more compact notation for the noise strength~$\Gamma$ to avoid taking explicit derivatives of~$\sqrt{\gamma_n(\rho)\rho}$.

We seek the solution of Eq.~(\ref{eq:diffusion_problem_formulation}) in the following form

\begin{equation}
\rho(t,x) = \rho_d(x-v_dt-\xi(t)) + \Delta \rho (t, x-v_dt-\xi(t)),
\label{eq:perturbed_profile}
\end{equation}

\noindent where~$\rho_d$ is the deterministic stationary solution satisfying Eq.~(\ref{eq:reduced_ode}), $\xi(t)$~is the shift in the front position due to fluctuations, and~$\Delta \rho(t,\zeta)$ accounts for the effect of the perturbation on the front shape. Because the perturbation,~$\Gamma\eta$, is time dependent,~$\Delta \rho$ explicitly depends on time in addition to the dependence on~$t$ through the comoving coordinate~$\zeta=x-v_dt-\xi(t)$.

To the first order in perturbation theory, there are no terms due to the special rules of It\^o calculus, and we obtain the following expansions for the deterministic terms in Eq.~(\ref{eq:diffusion_problem_formulation})

\begin{equation}
\frac{\partial \rho}{\partial t} \approx - v_d \rho'_d -  \rho'_d \xi'- v_d \frac{\partial \Delta \rho}{\partial \zeta} + \frac{\partial \Delta\rho}{\partial t},
\end{equation}

\noindent the diffusion term as

\begin{equation}
D\frac{\partial^2 \rho}{\partial x^2} \approx D \rho''_d + D\frac{\partial^2 \Delta \rho}{\partial \zeta^2},
\end{equation}

\noindent and the growth term as

\begin{equation}
r(\rho)\rho \approx \rho_d r(\rho_d) + r(\rho_d)\Delta \rho + r'(\rho_d)\rho_d\Delta \rho.
\end{equation}

\noindent For functions with a single argument, primes denote derivatives with respect to that argument.

As before, the zeroth order of the perturbation theory is automatically satisfied, and the first non-trivial equation arises at the first order:

\begin{equation}
\label{eq:perturbation_first_order_diffusion}
\frac{\partial \Delta \rho}{\partial t} - \mathfrak{L}_p \Delta \rho =  \rho'_d \xi' + \frac{1}{\sqrt{N}}\Gamma(\rho_d)\eta ,
\end{equation}

\noindent where~$\mathfrak{L}_p$ is the same as in Eq.~(\ref{eq:perturbation_L}).

To obtain the equation for~$\xi$, we multiply both sides by~$L(\zeta)$, the left eigenvector of~$\mathfrak{L}_p$ with zero eigenvalue, and integrating over~$\zeta$. The result reads

\begin{equation}
\label{eq:projection_on_L}
\frac{\partial}{\partial t}\int_{-\infty}^{+\infty}L\Delta\rho d\zeta  =  \int_{-\infty}^{+\infty} L\rho'_d\xi' d\zeta +  \frac{1}{\sqrt{N}}\int_{-\infty}^{+\infty}L\Gamma(\rho_d)\eta d\zeta.
\end{equation}

\noindent We now use the fact that $\int_{-\infty}^{+\infty}L(\zeta) \Delta \rho(\zeta) d\zeta=0$. The projection of~$\Delta \rho$ on~$L$ vanishes because translations of~$\rho_d$ are excluded from the fluctuations of the front shape and are instead included through~$\xi(t)$. Imposing this condition is also necessary for the perturbation theory to be self-consistent. Otherwise, according to Eq.~(\ref{eq:projection_on_L}), $\int_{-\infty}^{\infty} L\Delta\rho d\zeta$~would perform an unconstrained random walk and grow arbitrarily large, which would violate the assumption that~$\Delta \rho$ is small. After imposing~$\int_{-\infty}^{+\infty} L\Delta\rho d\zeta=0$, we obtain

\begin{equation}
\label{eq:xi_prime}
\xi'(t) = -\frac{1}{\sqrt{N}} \frac{\int_{-\infty}^{+\infty}L(\zeta)\Gamma(\rho_d(\zeta))\eta(t,\zeta) d\zeta}{\int_{-\infty}^{+\infty}L(\zeta)\rho'_d(\zeta)d\zeta},
\end{equation}

From Eq.~(\ref{eq:xi_prime}), it immediately follows that

\begin{equation}
\langle \xi'(t) \rangle = 0.
\end{equation}

Thus, there are no corrections to the wave velocity at this order in the perturbation theory, and the motion of the front position is a random walk. 

The deviation between the position of the front relative to the deterministic expectation is given by~$\xi$, which we obtain by integrating~Eq.~(\ref{eq:xi_prime}):

\begin{equation}
X_{\mathrm{f}} - v_d t = \xi = -\frac{1}{\sqrt{N}}  \int_0^{t}  \frac{\int_{-\infty}^{+\infty}L(\zeta)\Gamma(\rho_d(\zeta))\eta(t,\zeta) d\zeta}{\int_{-\infty}^{+\infty}L(\zeta)\rho'_d(\zeta)d\zeta} dt
\label{eq:xi_solution}
\end{equation}

To determine the diffusion constant of front wandering, we evaluate the mean square displacement of the front position:

\begin{equation}
\label{eq:definition_Df}
\begin{aligned}
D_{\mathrm{f}} &= \frac{\mathrm{Var} \{X_{\mathrm{f}}^2\}}{2t} = \frac{\langle \xi^2\rangle}{2t} \\
& = \frac{1}{2tN} \left\langle \int_0^{t}\frac{\int_{-\infty}^{+\infty}L(\zeta_1)\Gamma(\rho_d(\zeta_1))\eta(t_1,\zeta_1) d\zeta_1}{\int_{-\infty}^{+\infty}L(\zeta)\rho'_d(\zeta)d\zeta}dt_1 \int_0^{t} \frac{\int_{-\infty}^{+\infty}L(\zeta_2)\Gamma(\rho_d(\zeta_2))\eta(t_2,\zeta_2) d\zeta_2}{\int_{-\infty}^{+\infty}L(\zeta)\rho'_d(\zeta)d\zeta}dt_2\right\rangle \\
&= \frac{1}{2tN} \frac{ \int_0^{t}dt_1 \int_0^{t}dt_2\int_{-\infty}^{+\infty}d\zeta_1 \int_{-\infty}^{+\infty}d\zeta_2 L(\zeta_1)\Gamma(\rho_d(\zeta_1))L(\zeta_2)\Gamma(\rho_d(\zeta_2)) \left\langle\eta(t_1,\zeta_1)\eta(t_2,\zeta_2) \right\rangle}{\left(\int_{-\infty}^{+\infty}L(\zeta)\rho'_d(\zeta)d\zeta\right)^2} \\
& = \frac{1}{2N} \frac{ \int_{-\infty}^{+\infty}d\zeta L^2(\zeta)\Gamma^2(\rho_d(\zeta))}{\left(\int_{-\infty}^{+\infty}L(\zeta)\rho'_d(\zeta)d\zeta\right)^2},
\end{aligned}
\end{equation}

\noindent where we used Eq.~(\ref{eq:xi_solution}) to express~$\xi(t)$ and Eq.~(\ref{eq:demographic_noise}) to average over the noise.

Finally, we substitute the expression for~$L(\zeta)$ from Eq.~(\ref{eq:eigen_peturbation_L}) and use the explicit form of~$\Gamma$ from Eq.~(\ref{eq:Gamma}) to obtain Eq.~(5) from the main text:

\begin{equation}
\label{eq:D_result}
D_{\mathrm{f}} = \frac{1}{2N} \frac{\int_{-\infty}^{+\infty}[\rho'_d(\zeta)]^2\rho_d(\zeta)\gamma_n(\rho_d(\zeta))e^{\frac{2v_d\zeta}{D}}d\zeta}{\left(\int^{+\infty}_{-\infty}[\rho_d'(\zeta)]^2e^{\frac{v_d\zeta}{D}}d\zeta\right)^2},
\end{equation}

\noindent which was originally derived in Refs.~\cite{mikhailov:diffusion} and~\cite{rocco:diffusion, meerson:velocity_fluctuations}.

\textit{Perturbation theory for migration fluctuations}\\
The analysis that we performed to compute~$D_\mathrm{f}$ due to demographic noise can be easily generalized to account for the noise in migration; see the discussion below~Eq.~(\ref{eq:drift_noise_space}). This was first done in Ref.~\cite{meerson:velocity_fluctuations} that extend the perturbation theory to the following equation

\begin{equation}
\frac{\partial \rho}{\partial t} = D\frac{\partial^2 \rho}{\partial x^2} + r(\rho)\rho + \frac{1}{\sqrt{N}}\Gamma(\rho)\eta + \frac{1}{\sqrt{N}}\frac{\partial(\Gamma_m(\rho)\chi)}{\partial \zeta},
\label{eq:diffusion_noise_formulation}
\end{equation}

\noindent where~$\Gamma_m$ denotes the strength of the migration fluctuations

\begin{equation}
\label{eq:Gamma}
\Gamma_m(\rho) = \sqrt{\gamma_m(\rho)\rho},
\end{equation}

\noindent and~$\chi =\sum_i \sqrt{n_i}\chi_i/\sqrt{n}$ is a unit-strength, delta-correlated, Gaussian noise that enters the equation for the total population density~$n = \sum_i n_i$ of all neutral genotypes. Note that, for the standard diffusion,~$\gamma_m(\rho) = \mathrm{const}$, but we allow the dependence on~$\rho$ because it does not affect the calculation below. 

The solution for~$\xi$ acquires an additional term due to migration fluctuations:

\begin{equation}
\xi = -\frac{1}{\sqrt{N}}  \int_0^{t}  \frac{\int_{-\infty}^{+\infty}L(\zeta)\Gamma(\rho_d(\zeta))\eta(t,\zeta) d\zeta}{\int_{-\infty}^{+\infty}L(\zeta)\rho'_d(\zeta)d\zeta} dt -\frac{1}{\sqrt{N}}  \int_0^{t}  \frac{\int_{-\infty}^{+\infty}L(\zeta)\frac{\partial \Gamma_m(\rho_d(\zeta))\chi(t,\zeta)}{\partial \zeta} d\zeta}{\int_{-\infty}^{+\infty}L(\zeta)\rho'_d(\zeta)d\zeta} dt.
\label{eq:xi_solution_migration}
\end{equation}

Because~$\chi$ and~$\eta$ are uncorrelated, their contributions to~$D_{\mathrm{f}}$ simply add:

\begin{equation}
\label{eq:D_result_migration}
D_{\mathrm{f}} = \frac{1}{2N} \frac{\int_{-\infty}^{+\infty}[\rho'_d(\zeta)]^2\rho_d(\zeta)\gamma_n(\rho_d(\zeta))e^{\frac{2v_d\zeta}{D}}d\zeta}{\left(\int^{+\infty}_{-\infty}[\rho_d'(\zeta)]^2e^{\frac{v_d\zeta}{D}}d\zeta\right)^2} + \frac{1}{2N} \frac{\int_{-\infty}^{+\infty}\gamma_m(\rho_d(\zeta))\rho_d(\zeta) (\rho''_d(\zeta) + \rho'_d(\zeta)v_d/D)^2e^{\frac{2v_d\zeta}{D}}d\zeta}{\left(\int^{+\infty}_{-\infty}[\rho_d'(\zeta)]^2e^{\frac{v_d\zeta}{D}}d\zeta\right)^2}.
\end{equation}

\noindent The higher order derivatives of~$\rho$ appear in the second term due to the integration by parts that is necessary to remove derivatives from the delta function due to~$\langle \chi(t_1,\zeta_1)\chi(t_2,\zeta_2)\rangle$. See Ref.~\cite{meerson:velocity_fluctuations} for the original derivation and further details.

It is now clear that the qualitative behavior of the two terms in Eq.~(\ref{eq:D_result_migration}) is the same. Indeed, the denominators are identical, and the integrands in the numerators have the same scaling behavior at the front, where divergences could occur. To see this, one can substitute the asymptotic behavior of the population density,~$\rho\sim e^{-k\zeta}$, and confirm that both numerators scale as~$e^{-\zeta(3k-2v_d/D)}$. Thus, the transition from fully-pushed to semi-pushed waves leads to the divergence of both integrals, and the scaling exponent~$\alpha_{\mathrm{\textsc{d}}}$ is the same for both migration and demographic fluctuations. For simplicity, only demographic fluctuations are considered in all other sections of this paper. 

\textit{Results for exactly solvable models}\\
In the regime of fully-pushed waves, we can evaluate~$D_\mathrm{f}$ explicitly for the exactly solvable models introduced in section~I. The details of these calculations are summarized in the subsection on integral evaluation at the end of section~III.

For the model specified by Eq.~(\ref{eq:allee_quadratic}), we find that 

\begin{equation}
	\label{eq:D_quadractic_bd}
D_{\mathrm{f}} = \frac{3}{16\pi}\frac{\gamma_n^0}{N}\sqrt{\frac{2D}{g_0}} \frac{\left( 1 - 4\rho^* \right) \left( 3 - 4\rho^* \right)}{\rho^* \left( 1 - 2\rho^* \right)\left( 1 - \rho^* \right)^2} \tan{2\pi \rho^*},
\end{equation}

\noindent when~$\gamma_m=0$ and~$\gamma_n(n)=\gamma_n^0=\mathrm{const}$.

Note that the choice of~$\gamma_n(n)$ is not specified by the deterministic model of population growth and needs to be determined either from the microscopic dynamics or from empirical observations. For models formulated in terms of independent birth and death rates,~$\gamma_n(n)$ is a constant on the order of~$1/\tau$, where~$\tau$ is the generation time.  However, different~$\gamma_n$ are possible. For example, our simulations that are based on the Wright-Fisher model have~$\gamma_f=\gamma_n^0(1-n/N)$, and the corresponding theoretical prediction for~$D_{\mathrm{f}}$ reads

\begin{equation}
	\label{eq:D_quadractic_wf}
D_{\mathrm{f}} = \frac{3}{20\pi}\frac{\gamma_n^0}{N}\sqrt{\frac{2D}{g_0}} \frac{\left( 1 - 4\rho^* \right) \left( 3 - 4\rho^* \right)}{\rho^* \left( 1 - \rho^* \right) \left( 1 - 2\rho^* \right)} \tan{2\pi \rho^*}.
\end{equation}

For the same model of an Allee effect, the contribution of the noise due to migration with~$\gamma_m=\gamma_m^0=\mathrm{const}$ is given by

\begin{equation}
	\label{eq:D_quadractic_migration}
D_{\mathrm{f}} = \frac{\sqrt{2}}{40\pi}\frac{\gamma_m^0}{N}\sqrt{\frac{g_0}{D}} \frac{\left( 1 + \rho^* \right) \left( 1 + 7\rho^* \right) \left( 1 - 4\rho^* \right) \left( 3 - 4\rho^* \right)}{\rho^* \left( 1 - 2\rho^* \right)\left( 1 - \rho^* \right)^2} \tan{2\pi \rho^*}
\end{equation}

\noindent assuming~$\gamma_n=0$.

For completeness, we also provide the results for other models and different choices of~$\gamma_n$ and~$\gamma_m$.

\begin{equation}
	\label{eq:D_cooperative_bd}
D_{\mathrm{f}} = \frac{3}{16\pi}\frac{\gamma_n^0}{N}\sqrt{\frac{2D}{r_0 B}} \frac{B^2\left( B + 4 \right) \left( 3B + 4 \right)}{\left( B + 2 \right) \left( B + 1 \right)^2} \tan{\frac{2\pi}{B}}
\end{equation}

\noindent for the model of cooperative growth defined in the main text with~$\gamma_n = \gamma_n^0$ and~$\gamma_m=0$.

\begin{equation}
\label{eq:D_cooperative_wf}
D_{\mathrm{f}} = \frac{3}{20\pi}\frac{\gamma_n^0}{N}\sqrt{\frac{2D}{r_0B}} \frac{B \left( B + 4\right) \left( 3B + 4 \right)}{\left( B + 2 \right)\left( B + 1 \right)} \tan{\frac{2\pi}{B}}
\end{equation}

\noindent for the model of cooperative growth defined in the main text with~$\gamma_n = \gamma_n^0(1-n/N)$ and~$\gamma_m=0$.

\begin{equation}
	\label{eq:D_cooperative_migration}
D_{\mathrm{f}} = \frac{\sqrt{2}}{40\pi}\frac{\gamma_m^0}{N}\sqrt{\frac{r_0 B}{D}} \frac{\left( B - 1 \right)\left( B - 7 \right)\left( B + 4 \right) \left( 3B + 4 \right)}{\left( B + 2 \right) \left( B + 1 \right)^2} \tan{\frac{2\pi}{B}}
\end{equation}

\noindent for the model of cooperative growth defined in the main text with~$\gamma_m = \gamma_m^0=\mathrm{const}$ and~$\gamma_n = 0$.

\textbf{\large \hclr{VII. Correction to velocity due to demographic fluctuations}}

In this section, we compute the correction to the wave velocity directly from the stochastic formulation in Eq.~(\ref{eq:diffusion_problem_formulation}) instead of relying on a growth-rate cutoff at low densities. Our main finding is that, for pushed waves, the scaling of~$\Delta v$ with~$N$ coincides\footnote{The scaling behavior of~$\Delta v$ is different for pulled waves because~$\Delta v \sim \ln^{-2}N$; see Ref.~\cite{saarloos:review, panja:review, brunet:phenomenological_pulled, brunet:velocity_cutoff}.} with that of front diffusion constant~$D_{\mathrm{f}}$ and the rate of diversity loss~$\Lambda$. Note that this result cannot be obtain from the cutoff-based calculation of~$\Delta v$ without knowing the correct dependence of~$n_c$ on~$v/v_{\mathrm{\textsc{f}}}$. Thus, the calculation of~$\Delta v$ in the stochastic model provides an additional insight in the dynamics of fluctuating fronts. To the best of our knowledge, the results presented in this section are new.

Because the first order correction to~$v$ is zero (See Eq.~(\ref{eq:xi_prime})), we proceed to the second order in perturbation theory. In this calculation, it is convenient to distinguish the contributions to~$\zeta$ and~$\Delta \rho$ that come from the different orders of the perturbative expansion:

\begin{equation}
\begin{aligned}
&\zeta = x-v_dt -\xi_{(1)}(t) -\xi_{(2)}(t),\\
& \rho(t,\zeta) = \rho_d(\zeta) + \Delta \rho_{(1)}(t,\zeta) + \Delta \rho_{(2)}(t,\zeta),
\end{aligned}
\end{equation}  

\noindent where the order is indicated by a subscript in brackets. For fully-pushed waves, we expect that the first order corrections~$\xi_{(1)}(t)$ and~$\Delta \rho_{(1)}(t,\zeta)$ scale as~$1/\sqrt{N}$, and the second order corrections~$\xi_{(2)}(t)$ and~$\Delta \rho_{(2)}(t,\zeta)$ scale as~$1/N$. Therefore, we expand all terms in Eq.~(\ref{eq:diffusion_problem_formulation}) up to order~$1/N$.

For~$\partial \rho/\partial t$, we obtain

\begin{equation}
\frac{\partial \rho}{\partial t} \approx   \rho'_d(- v_d - \xi_{(1)}' - \xi'_{(2)}) + \frac{1}{2}\rho''_d\left\langle\frac{(d\xi_{(1)})^2}{dt}\right\rangle + \frac{\partial \Delta \rho_{(1)}}{\partial \zeta}(-v_d -\xi'_{(1)}) + \frac{\partial \Delta\rho_{(1)}}{\partial t} + \frac{\partial \Delta \rho_{(2)}}{\partial \zeta}(-v_d ) + \frac{\partial \Delta\rho_{(2)}}{\partial t},
\end{equation}

\noindent where~$\rho''_d/2\left\langle(d\xi_{(1)})^2/dt\right\rangle$ arises due to the It\^o formula of stochastic calculus, which prescribes how to compute derivatives of nonlinear functions; see Refs.~\cite{korolev:review, risken:fpe, gardiner:handbook, oksendal:book}. The unusual derivative~$\left\langle(d\xi_{(1)})^2/dt\right\rangle$ is non-zero because the displacement of a random walk grows as~$\sqrt{dt}$. Using Eqs.~(\ref{eq:xi_prime}) and (\ref{eq:definition_Df}), we express this derivative in terms of~$D_{\mathrm{f}}$, which we know to the order~$1/N$ from the first order of the perturbation theory:

\begin{equation}
\left\langle(d\xi_{(1)})^2/dt\right\rangle = 2 D_{\mathrm{f}}.
\end{equation} 

\noindent The expansion of other terms is more straightforward and does not involve any additional terms due to the special rules of It\^o calculus:

\begin{equation}
\begin{aligned}
& D\frac{\partial^2 \rho}{\partial x^2} \approx D \rho''_d + D\frac{\partial^2 \Delta \rho_{(1)}}{\partial \zeta^2} + D\frac{\partial^2 \Delta \rho_{(2)}}{\partial \zeta^2},\\
	& r(\rho)\rho \approx [r(\rho)\rho]|_{\rho=\rho_d} + [r(\rho)\rho]'|_{\rho=\rho_d}\Delta \rho_{(1)} + \frac{1}{2} [r(\rho)\rho]''|_{\rho=\rho_d}(\Delta \rho_{(1)})^2 + [r(\rho)\rho]'|_{\rho=\rho_d}\Delta \rho_{(2)}, \\
& \Gamma(\rho) \approx \Gamma(\rho_d) +\Gamma'(\rho_d)\Delta \rho_{(1)},
\end{aligned}
\end{equation}

\noindent where we kept only the terms that scale at most as~$1/N$ and used~$|_{\rho=\rho_d}$ to indicate that the expression to the left is evaluated at~$\rho=\rho_d$. Upon choosing~$\Delta \rho_{(1)}$ and~$\xi_{(1)}$ that satisfy the first order equation, i.e. Eq.~(\ref{eq:perturbation_first_order_diffusion}), we obtain the following equation for~$\Delta \rho_{(2)}$ and~$\xi_{(2)}$:

\begin{equation}
\label{eq:perturbation_second_order_diffusion}
	\frac{\partial \Delta \rho_{(2)}}{\partial t} - \mathfrak{L}_p \Delta \rho_{(2)} =  -D_{\mathrm{f}}\rho''_d +\frac{\partial\Delta\rho_{(1)}}{\partial\zeta}\xi'_{(1)} +\frac{1}{2}[r''(\rho_d)\rho_d+2r'(\rho_d)](\Delta \rho_{(1)})^2   + \frac{1}{\sqrt{N}}\Gamma'(\rho_d)\Delta \rho_{(1)}\eta + \rho'_d\xi'_{(2)}.
\end{equation}

The value of~$\xi'_{(2)}$ needs to be chosen to satisfy the solvability condition, which we obtain by multiplying both sides of Eq.~(\ref{eq:perturbation_second_order_diffusion}) by~$L(\zeta)$, integrating over~$\zeta$, and requiring that~$\Delta\rho_{(2)}$ has zero projection on~$L(\zeta)$. The result reads

\begin{equation}
\label{eq:xi2_prime}
	\xi_{(2)}'(t) = - \frac{\int_{-\infty}^{+\infty}L(\zeta) \left [ -D_{\mathrm{f}}\rho''_d + \frac{\partial\Delta\rho_{(1)}}{\partial\zeta}\xi'_{(1)} +\frac{1}{2}[r''(\rho_d)\rho_d+2r'(\rho_d)](\Delta \rho_{(1)})^2   + \frac{1}{\sqrt{N}}\Gamma'(\rho_d)\Delta \rho_{(1)}\eta \right] d\zeta}{\int_{-\infty}^{+\infty}L(\zeta)\rho'_d(\zeta)d\zeta}.
\end{equation}

The correction to the velocity,~$\Delta v$, can now be obtained by averaging Eq.~(\ref{eq:xi2_prime}) over~$\eta$ and substituting the explicit expression for~$L(\zeta)$ from Eq.~(\ref{eq:eigen_peturbation_L}):

\begin{equation}
\label{eq:dv_general_so_preliminary}
\begin{aligned}
&\Delta v = \langle \xi'_{(2)} \rangle = D_{\mathrm{f}}\frac{\int_{-\infty}^{+\infty} e^{v_d \zeta/D} \rho_d'(\zeta) \rho''_d(\zeta) d\zeta}{\int_{-\infty}^{+\infty} e^{v_d \zeta/D} [\rho_d'(\zeta)]^2 d\zeta} - \frac{\int_{-\infty}^{+\infty} e^{v_d \zeta/D} \rho_d'(\zeta) \frac{1}{2}[r''(\rho_d)\rho_d+2r'(\rho_d)] \langle [\Delta \rho_{(1)}(t,\zeta)]^2\rangle d\zeta}{\int_{-\infty}^{+\infty} e^{v_d \zeta/D} [\rho_d'(\zeta)]^2 d\zeta}.
\end{aligned}
\end{equation}

\noindent Note that~$\langle\Delta\rho_{(1)}\eta\rangle=0$ and, therefore,~$\langle \Delta\rho_{(1)}\xi' \rangle = 0$ because~$\Delta\rho_{(1)}(t,\zeta)$ depends only on~$\eta(\tilde{t},\zeta)$ with~$\tilde{t}<t$ and~$\langle\eta(\tilde{t},\zeta)\eta(t,\zeta)\rangle=0$.\footnote{This simplification is specific to the It\^o calculus and does not occur in Stratonovich's formulation. The results of course do not depend on the type of calculus used as long as all calculations are carried using the same calculus and the initial problem statement is correctly formulated. In population dynamics, demographic fluctuations affect only future generations, so It\^o's formulation appears naturally.} The first term could be further simplified through integration by parts in the numerator, assuming that the integrals converge:

\begin{equation}
\label{eq:dv_general_so}
\begin{aligned}
&\Delta v = -v_d\frac{D_{\mathrm{f}}}{2D} - \frac{\int_{-\infty}^{+\infty} e^{v_d \zeta/D} \rho_d'(\zeta) \frac{1}{2}[r''(\rho_d)\rho_d+2r'(\rho_d)] \langle [\Delta \rho_{(1)}(t,\zeta)]^2\rangle d\zeta}{\int_{-\infty}^{+\infty} e^{v_d \zeta/D} [\rho_d'(\zeta)]^2 d\zeta}.
\end{aligned}
\end{equation}

To complete the calculation of~$\Delta v$, we need to obtain~$\Delta \rho_{(1)}$ by solving Eq.~(\ref{eq:perturbation_first_order_diffusion}). Before performing this calculation, let us state the main findings and discuss their implications. For fully-pushed waves, we find that~$\Delta \rho_{(1)}\sim 1/\sqrt{N}$, and all integrals in Eq.~(\ref{eq:dv_general_so}) converge. Thus,~$\Delta v \sim 1/N$ in this regime. For semi-pushed waves, one needs to apply a cutoff at large~$\zeta$ to ensure convergence. We show that the divergence of the last term in Eq.~(\ref{eq:dv_general_so}) does not exceed that of~$D_{\mathrm{f}}$. Thus, the leading behavior is controlled by the first term, and the scaling of~$\Delta v$ coincides with that of~$D_{\mathrm{f}}$. The scaling behavior of~$\Delta v$ and~$D_\mathrm{f}$ is slightly different for pulled waves:~$\Delta v \sim \ln^{-2}N$ and~$D_{\mathrm{f}}\sim\ln^{-3}N$; see Ref.~\cite{saarloos:review, panja:review, brunet:phenomenological_pulled, brunet:velocity_cutoff}.

The calculation of~$\Delta \rho_{(1)}$ can be simplified by transforming Eq.~(\ref{eq:perturbation_first_order_diffusion}) into a Hermitian form. This is accomplished by the following change of variables that eliminates the term linear in~$\partial/\partial\zeta$ from~$\mathfrak{L}_p$:

\begin{equation}
\Delta \rho_{(1)}(t,\zeta) = e^{-\frac{v_d\zeta}{2D}}\Psi(t,\zeta).
\end{equation}

Equation~(\ref{eq:perturbation_first_order_diffusion}) then takes the following form

\begin{equation}
\label{eq:peturbation_psi_diffusion}
\frac{\partial \Psi}{\partial t} - \mathfrak{H}_p \Psi =  e^{\frac{v_d\zeta}{2D}}\left[\rho'_d \xi_{(1)}' + \frac{1}{\sqrt{N}}\Gamma(\rho_d)\eta\right],
\end{equation}

\noindent where~$\mathfrak{H}_p$ is a Hermitian operator:

\begin{equation}
\mathfrak{H}_p = D\frac{\partial^2}{\partial\zeta^2} - \frac{v_d^2}{4D} + r(\rho_d) + r'(\rho_d)\rho_d.
\end{equation}

We solve Eq.~(\ref{eq:peturbation_psi_diffusion}) using the method of separation of variables. Let us denote the eigenvalues and normalized eigenvectors of~$\mathfrak{H}_p$ by~$\lambda_l$ and~$\mathfrak{h}_l$ respectively. The index~$l$ labels both discrete and continuous parts of the spectrum of~$\mathfrak{H}_p$ such that~$\lambda_l$ are in the decreasing order;~$l=0$ corresponds to the zero mode. In the basis of~$\mathfrak{h}_l$, we express~$\Psi$ as:

\begin{equation}
	\label{eq:basis_expansion}
	\Psi(t,\zeta) = \sum_l a_l(t)\mathfrak{h}_l(\zeta).
\end{equation}

\noindent The unknown coefficients~$a_l(t)$ are determined by projecting Eq.~(\ref{eq:peturbation_psi_diffusion}) on~$\mathfrak{h}_l$:

\begin{equation}
	\label{eq:a_l_equation}
	\frac{d a_l}{dt} -\lambda_l a_l = \int_{-\infty}^{+\infty} e^{\frac{v_d\zeta}{2D}}\left[\rho'_d \xi_{(1)}' + \frac{1}{\sqrt{N}}\Gamma(\rho_d)\eta\right] \mathfrak{h}_l d\zeta.
\end{equation}

\noindent and then solving these linear equations: 

\begin{equation}
	\label{eq:a_l_solution}
	a_l(t) = \int_{-\infty}^{t} e^{\lambda_l(t-\tilde{t})} \left\{\int_{-\infty}^{+\infty} e^{\frac{v_d\zeta}{2D}}\left[\rho'_d(\zeta) \xi_{(1)}'(\tilde{t}) + \frac{1}{\sqrt{N}}\Gamma(\rho_d(\zeta))\eta(\tilde{t},\zeta)\right] \mathfrak{h}_l(\zeta) d\zeta\right\}d\tilde{t}.
\end{equation}

\noindent Here, we assumed that the front has been propagating for a very long time and, therefore, set the lower integration limit of the integral over~$\tilde{t}$ to~$-\infty$. The next step is to substitute the solution for~$\xi'_{(1)}$ from Eq.~(\ref{eq:xi_prime}):

\begin{equation}
	\label{eq:a_l_solution_preliminary}
	a_l(t) = \frac{1}{\sqrt{N}}\int_{-\infty}^{t} e^{\lambda_l(t-\tilde{t})} \left\{\int_{-\infty}^{+\infty} \left[ e^{\frac{v_d\zeta}{2D}}\Gamma(\rho_d(\zeta))\eta(\tilde{t},\zeta)-\mathfrak{h}_0(\zeta)\int_{-\infty}^{+\infty} e^{\frac{v_d\tilde{\zeta}}{2D}}\Gamma(\rho_d(\tilde{\zeta}))\eta(\tilde{t},\tilde{\zeta}) \mathfrak{h}_0(\tilde{\zeta})d\tilde{\zeta}\right] \mathfrak{h}_l(\zeta) d\zeta\right\}d\tilde{t},
\end{equation}

\noindent where we used the fact that

\begin{equation}
	\mathfrak{h}_0(\zeta) = \frac{ e^{\frac{v_d\zeta}{2D}}\rho'_d(\zeta)}{\sqrt{\int_{-\infty}^{+\infty}e^{\frac{v_d\zeta}{D}} [\rho'_d(\zeta)]^2d\zeta}},
\end{equation} 

\noindent and, therefore,

\begin{equation}
	\label{eq:xi_h0}
	\xi'_{(1)}(t) = -\frac{1}{\sqrt{N}} \frac{\int_{-\infty}^{+\infty} e^{\frac{v_d\zeta}{2D}}\Gamma(\rho_d(\zeta))\eta(t,\zeta) \mathfrak{h}_0(\zeta)d\zeta}{\sqrt{\int_{-\infty}^{+\infty}e^{\frac{v_d\zeta}{D}} [\rho'_d(\zeta)]^2d\zeta}}.
\end{equation}

Equation~(\ref{eq:a_l_solution_preliminary}) is further simplified by carrying out the integration over~$\zeta$ in the last term and using the orthogonality of~$\mathfrak{h}_0$ and~$\mathfrak{h}_l$ for~$l>0$:

\begin{equation}
	a_l(t) = \frac{1-\delta_{0l}}{\sqrt{N}}\int_{-\infty}^{t} e^{\lambda_l(t-\tilde{t})} \left\{\int_{-\infty}^{+\infty} e^{\frac{v_d\zeta}{2D}}\Gamma(\rho_d(\zeta))\eta(\tilde{t},\zeta) \mathfrak{h}_l(\zeta) d\zeta\right\}d\tilde{t},
	\label{eq:a_l_solution}
\end{equation}

Note that,~$a_0=0$ consistent with the solvability condition that~$\Delta \rho_{(1)}$ has a vanishing projection on the zero mode.

With the solution for~$\Delta \rho_{(1)}$ at hand, we proceed to calculate the average~$[\Delta \rho_{(1)}]^2$ that enters Eq.~(\ref{eq:dv_general_so}):

\begin{equation}
	\begin{aligned}
		& \langle [\Delta \rho_{(1)}(t,\zeta)]^2\rangle = \sum_{l_1>0}\sum_{l_2>0}\mathfrak{h}_{l_{1}}(\zeta)\mathfrak{h}_{l_{2}}(\zeta) \langle a_{l_{1}}(t)a_{l_{2}}(t)\rangle \\&= \frac{1}{N} \sum_{l_1>0}\sum_{l_2>0} \mathfrak{h}_{l_{1}}(\zeta)\mathfrak{h}_{l_{2}}(\zeta)  \frac{-1}{\lambda_{l_1} + \lambda_{l_2}} \int_{-\infty}^{+\infty}e^{\frac{v_d\zeta}{D}}\Gamma^2(\rho_d(\zeta)) \mathfrak{h}_{l_1}(\zeta) \mathfrak{h}_{l_2}(\zeta) d\zeta.
	\end{aligned}
\end{equation}

Upon substituting this result into Eq.~(\ref{eq:dv_general_so}), we obtain

\begin{equation}
\label{eq:dv_so}
\begin{aligned}
	&\Delta v = -v_d\frac{D_{\mathrm{f}}}{2D} - \frac{1}{N} \frac{1}{\int_{-\infty}^{+\infty} e^{v_d \zeta/D} [\rho_d'(\zeta)]^2 d\zeta} \times \\ &\sum_{l_1>0}\sum_{l_2>0} \frac{\left[\int_{-\infty}^{+\infty}e^{\frac{v_d\zeta}{D}}\gamma_n(\rho_d(\zeta))\rho_d(\zeta) \mathfrak{h}_{l_1}(\zeta) \mathfrak{h}_{l_2}(\zeta) d\zeta\right]\left[ \int_{-\infty}^{+\infty} e^{v_d \zeta/D} \rho_d'(\zeta) \frac{1}{2}[r''(\rho_d)\rho_d+2r'(\rho_d)] \mathfrak{h}_{l_{1}}(\zeta)\mathfrak{h}_{l_{2}}(\zeta)d\zeta\right]}{-(\lambda_{l_1} + \lambda_{l_2})} .
\end{aligned}
\end{equation}

\noindent Since the eigenvectors~$\mathfrak{h}_{l}$ decay at least as fast as~$\mathfrak{h}_{0}\sim e^{\frac{v_d\zeta}{2D}}\rho'_d(\zeta)$ as~$\zeta\to+\infty$, all the integrands in Eq.~(\ref{eq:dv_so}) decay faster than~$e^{-(3k-2v_d/D)\zeta}$. For fully-pushed waves, all the integrals converge, and the correction to the velocity scales as~$1/N$. For semi-pushed waves, the term with~$D_{\mathrm{f}}$ shows the fastest divergence with the cutoff and, therefore, determines the scaling of~$\Delta v$ with~$N$.

\textbf{\large \hclr{VIII. Rate of diversity loss,~$\Lambda$}}

In this section, we describe how genetic diversity is lost during a range expansion and provide the derivation of Eq.~(5) from the main text, which was first derived in Ref.~\cite{hallatschek:diversity_wave}. For simplicity, we consider an expansion that started with two neutral genotypes present throughout the population and determine how the probability to sample two different genotypes at the front decreases with time. The calculation of~$\Lambda$ is based on the perturbation theory in~$1/N$ and relies on a mean-field assumption that~$n(t,x)$ can be approximated by~$\langle n(t,x) \rangle$. This analysis is asymptotically exact for fully-pushed waves and could be extended to semi-pushed and pulled waves by applying a cutoff at large~$\zeta$ as we show in section~IX. The current section contain no new results except for the calculation of~$\Lambda$ in exactly solvable models of fully-pushed waves. 

\textit{Forward-in-time analysis of the decay of heterozygosity}\\
We quantify the genetic diversity in the population by the average heterozygosity:

\begin{equation}
\label{eq:H_definition}
H(t,\zeta_1,\zeta_2) = \langle f(t,\zeta_1) [1-f(t,\zeta_2)] + [1-f(t,\zeta_1)]f(t,\zeta_2) \rangle,
\end{equation}

\noindent which is the probability to sample two different genotypes at positions~$\zeta_1$ and~$\zeta_2$ in the comoving reference frame at time~$t$. Here,~$f$ denotes the frequency of one the two genotypes; the frequency of the other genotype is~$1-f$.

To obtain a closed equation for the dynamics of~$H$, we assume that~$n(t,\zeta)$ is given by its non-fluctuating stationary limit,~$n(\zeta)$, from Eq.~(\ref{eq:reduced_ode}). Then, we differentiate Eq.~(\ref{eq:H_definition}) with respect to time and use Eq.~(\ref{eq:stochastic_f_space}) to eliminate the time derivatives of~$f$. The result reads

\begin{equation}
\label{eq:H_dynamics}
\frac{\partial H}{\partial t} = \left( \mathcal{L}_{\zeta_1} + \mathcal{L}_{\zeta_2} \right)H - \delta(\zeta_1-\zeta_2)\frac{\gamma_f(n)}{n}H,
\end{equation}

\noindent where

\begin{equation}
\label{eq:L_definition}
\mathcal{L}_{\zeta} = D\frac{\partial^2}{\partial \zeta^2} + v\frac{\partial}{\partial \zeta} + 2D\frac{\partial \ln n}{\partial \zeta}\frac{\partial}{\partial \zeta}.
\end{equation}

\noindent we note that the first term in Eq.~(\ref{eq:H_dynamics}) follows from the rules of regular calculus, but the last term arises due to the It\^o formula of stochastic calculus, which prescribes how to compute derivatives of nonlinear functions; see Refs.~\cite{korolev:review, risken:fpe, gardiner:handbook, oksendal:book}. This last term encapsulates the effect of genetic drift and ensures that genetic diversity decays to zero due to the fixation of one of the genotypes.

Since~$H$ obeys a linear equation, it will decay to zero exponentially in time with the decay rate given by the solution of the following eigenvalue problem:  

\begin{equation}
\label{eq:lambda_eigen}
- \Lambda H= \left( \mathcal{L}_{\zeta_1} + \mathcal{L}_{\zeta_2} \right)H - \delta(\zeta_1-\zeta_2)\frac{\gamma_f(n)}{n}H,
\end{equation}

\noindent where we seek the smallest~$\Lambda$ or alternatively the largest eigenvalue of the operator on the right hand side. 

We compute~$\Lambda$ perturbatively by treating~$1/N$ as a small parameter. To the zeroth order, we can neglect the last term in Eq.~(\ref{eq:lambda_eigen}) because it scales as~$1/N$. Without the sink term, Eq.~(\ref{eq:H_dynamics}) admits a constant stationary solution~($H(t,\zeta)=\mathrm{const}$), so the smallest decay rate is zero. Thus, the zeroth order solution of Eq.~(\ref{eq:lambda_eigen}) reads

\begin{equation}
	\label{eq:H_zeroth_order_right}
	\begin{aligned}
		\Lambda &= 0,\\
		 R(\zeta_1,\zeta_2) & = 1.
	\end{aligned}
\end{equation}

Because~$\mathcal{L}_{\zeta}$ contains terms linear in~$\frac{\partial}{\partial \zeta}$, the operator in Eq.~(\ref{eq:lambda_eigen}) is not Hermitian. Therefore, we also need~$L(\zeta_1, \zeta_2)$, the left eigenvector of~$ \mathcal{L}_{\zeta_1} + \mathcal{L}_{\zeta_2}$, to compute the first order correction. It is not difficult to guess~$L(\zeta_1, \zeta_2)$ because it corresponds to the right eigenvector of the adjoint operator, and we already obtained the stationary distribution for~$\partial S/ \partial\tau= L^{+}_{\zeta}S$ when we discussed the patterns of ancestry. Since~$\mathcal{L}_{\zeta_1}$ and~$\mathcal{L}_{\zeta_2}$ act on different variables, the sought-after eigenfunction is the product of the eigenfunctions of these two operators:

\begin{equation}
	\label{eq:H_zeroth_order_left}
	L(\zeta_1,\zeta_2) = n^2(\zeta_1)e^{\zeta_1 v /D}n^2(\zeta_2)e^{\zeta_2 v /D}.
\end{equation}

The first order correction to~$\Lambda$ is given by the standard formula~\cite{risken:fpe, gardiner:handbook}:

\begin{equation}
	\label{eq:L_first_order_derivation}
	\begin{aligned}
		\Lambda = \frac{ \int_{-\infty}^{+\infty}d\zeta_1\int_{-\infty}^{+\infty}d\zeta_2  L(\zeta_1,\zeta_2) \delta(\zeta_1-\zeta_2)\frac{\gamma_f(n)}{n}  R(\zeta_1,\zeta_2) } {\int_{-\infty}^{+\infty}d\zeta_1\int_{-\infty}^{+\infty}d\zeta_2  L(\zeta_1,\zeta_2)R(\zeta_1,\zeta_2)}. 
	\end{aligned}
\end{equation}

We now use the expressions of~$L(\zeta_1,\zeta_2)$ and~$R(\zeta_1,\zeta_2)$ from Eqs.~(\ref{eq:H_zeroth_order_right}) and~(\ref{eq:H_zeroth_order_left}) and obtain the final result:

\begin{equation}
	\Lambda = \frac{\int_{-\infty}^{+\infty}\gamma_{f}(n(\zeta))n^{3}(\zeta)e^{\frac{2v\zeta}{D}}d\zeta}{\left(\int^{+\infty}_{-\infty}n^2(\zeta)e^{\frac{v\zeta}{D}}d\zeta\right)^2}.
\label{eq:L_result}
\end{equation}

\noindent which becomes identical to Eq.~(5) in the main text upon substituting~$n = N\rho$. This result was first obtained in Ref.~\cite{hallatschek:diversity_wave}.

For fully-pushed waves, all the integrals in Eq.~(\ref{eq:L_result}) converge and one can obtain the dependence of~$\Lambda$ on model parameters by dimensional analysis. Specifically, each factor of~$n$ contributes a factor of~$N$, and each~$d\zeta$ contributes a width of the front~(the integrands rapidly tend to zero in the bulk and the leading edge). In total,~$\Lambda$ is inversely proportional to the product of~$N$ and front width, i.e. to the number of individuals at the front. This result is quite intuitive because, in well-mixed populations, the rate of diversity loss scales as the total population size, and~$\Lambda^{-1}$ is often denoted as an effective population size~\cite{gillespie:textbook}. Thus, the neutral evolution in a fully-pushed wave could be approximated by that in a well-mixed population consisting of all the organisms at the front. In contrast, only the very tip of the front drives the evolutionary dynamics in semi-pushed and pulled waves.

Equation~(\ref{eq:L_result}) also suggests that the deterministic approximation for~$n(t,\zeta)$ that we made in Eq.~(\ref{eq:H_dynamics}) is asymptotically exact for fully-pushed waves. Indeed, the main contribution to the integrals in Eq.~(\ref{eq:L_result}) comes for the interior regions of the front, where the fluctuations in~$n$ are small compared to the mean population density.

Finally, we note that one can avoid using the perturbation theory for non-Hermitian operators to derive Eq.~(\ref{eq:L_first_order_derivation}). Specifically, one can recast~$(\mathcal{L}_{\zeta_1} + \mathcal{L}_{\zeta_1} )$ in a Hermitian form by finding a function~$\beta(\zeta)$ such that~$(\mathcal{L}_{\zeta_1} + \mathcal{L}_{\zeta_1} ) \beta(\zeta_1)\beta(\zeta_2)\Psi(t,\zeta_1,\zeta_2) = \beta(\zeta_1)\beta(\zeta_2) \mathcal{H}\Psi(t,\zeta_1,\zeta_2)$, where~$\mathcal{H}$ is a Hermitian operator, which contains no terms linear in~$\partial/\partial \zeta_1$ or~$\partial/\partial \zeta_2$. Then, the substitution:~$H(t,\zeta_1,\zeta_2) = \beta(\zeta_1)\beta(\zeta_2)\Psi(t,\zeta_1,\zeta_2)$ converts Eq.~(\ref{eq:lambda_eigen}) into a Hermitian eigenvalue problem.

The following equations summarize the main steps in this approach:

\begin{equation}
\label{eq:hermitian_eigen_L}
\begin{aligned}
& \beta(\zeta) = n^{-1}(\zeta) e^{-\frac{\zeta v}{2D}},\\
& \mathcal{H} = D\frac{\partial^2}{\partial \zeta_1^2} - \left(\frac{D}{n(\zeta_1)}\frac{\partial^2 n(\zeta_1)}{\partial \zeta_1^2} +  \frac{v}{n(\zeta_1)}\frac{\partial n(\zeta_1)} {\partial \zeta_1} + \frac{v^{2}}{4D} \right) + D\frac{\partial^2}{\partial \zeta_2^2} - \left(\frac{D}{n(\zeta_2)}\frac{\partial^2 n(\zeta_2)}{\partial \zeta_2^2} +  \frac{v}{n(\zeta_2)}\frac{\partial n(\zeta_2)} {\partial \zeta_2} + \frac{v^{2}}{4D} \right), \\
& h_0(\zeta_1,\zeta_2) = \frac{1}{\beta(\zeta_1)}\frac{1}{\beta(\zeta_2)} = n(\zeta_1) e^{\frac{\zeta_1 v}{2D}}n(\zeta_2) e^{\frac{\zeta_2 v}{2D}},
\end{aligned}
\end{equation}

\noindent where~$h_0(\zeta)$ is the eigenvector corresponding to the zero eigenvalue. This eigenvector is easily obtained from the reverse transformation from~$H$ to~$\Psi$ and the fact that~$H=\mathrm{const}$ is the right eigenvector of the original operator,~$(\mathcal{L}_{\zeta_1} + \mathcal{L}_{\zeta_1} )$.

Since the eigenvalues of~$\mathcal{H}$ coincide with the eigenvalues of~$(\mathcal{L}_{\zeta_1} + \mathcal{L}_{\zeta_1} )$, one can compute~$\Lambda$ by the standard formula:

\begin{equation}
\label{eq:first_order_hermitian_L}
\Lambda = \frac{\int_{-\infty}^{+\infty}\int_{-\infty}^{+\infty} h_0(\zeta_1,\zeta_2) \delta(\zeta_1-\zeta_2)\frac{\gamma_f(n(\zeta_1))}{n(\zeta_1)} h_0(\zeta_1,\zeta_2) d\zeta_1d\zeta_2}{\int_{-\infty}^{+\infty}h^2_0(\zeta_1,\zeta_2) d\zeta_1d\zeta_2}.
\end{equation}

\noindent It is now easy to see that Eqs.~(\ref{eq:hermitian_eigen_L}) and~(\ref{eq:first_order_hermitian_L}) lead to the same expression for~$\Lambda$ as in Eq.~(\ref{eq:L_result}).

\textit{Backward-in-time analysis of lineage coalescence}\\
To complement the forward-in-time analysis, we show how~$\Lambda$ can be computed by tracing ancestral lineages backward in time. One advantage of this approach is that it provides a more intuitive explanation of Eq.~(\ref{eq:L_result}). The discussion of this approach closely follows Ref.~\cite{hallatschek:diversity_wave}.

We motivate the backward-in-time approach by considering how~$H$ can be estimated from its definition as the probability to sample two different genotypes. To determine whether the genotypes are different, we trace their ancestral lineages backward in time and observe that only two outcomes are possible: Either the lineages never interact with each other until they hit the initial conditions or the lineages coalesce, i.e. converge on the same ancestor, at some point during the range expansion. In the former case, the probability to be different is determined by the initial heterozygosity of the population. In the latter case, the probability to be different is zero because we do not allow mutations. Thus,~$H(t,\zeta_1^d,\zeta_2^d)$ is intimately related to the probability~$S^{(2)}$ that two lineages sampled at time~$t$ at positions~$\zeta_1^d$ and~$\zeta_2^d$ have not coalesced up to time~$\tau$ into the past and were present at~$\zeta_1^a$ and~$\zeta_2^a$ at time~$t-\tau$; the superscripts distinguish between the positions of the descendants and the ancestors. To simplify the notation, we suppress descendent-related variables, drop the subscripts, and write this probability as~$S^{(2)}(\tau, \zeta_1, \zeta_2)$. We keep the superscript to distinguish~$S^{(2)}$ from~$S$, which denotes the position of a single ancestral lineage.

The dynamical equation for~$S^{(2)}$ can be derived either from the forward-in-time formulation for~$H$ or directly from the dynamics of ancestral lineages. The result reads

\begin{equation}
\label{eq:S2_dynamics}
\frac{\partial S^{(2)}}{\partial \tau} = \left( \mathcal{L}^+_{\zeta_1} + \mathcal{L}^+_{\zeta_2} \right)S^{(2)} - \delta(\zeta_1-\zeta_2)\frac{\gamma_f(n)}{n}S^{(2)},
\end{equation}

\noindent where the first term describes the motion of the two ancestral lineages, and the last term accounts for the lineage coalescence. As expected, the rate of coalescence events is inversely proportional to the local effective population size~$n/\gamma_f$; see~\cite{gillespie:textbook, wakeley:textbook, kingman:coalescent}. Because the linear operators on the right hand side of Eqs.~(\ref{eq:S2_dynamics}) and~(\ref{eq:H_dynamics}) are adjoint to each other, their eigenvalues coincide. Therefore, the temporal decay of~$S^{(2)}$ is exponential in~$\tau$ with the decay rate equal to~$\Lambda$. 

The expression for~$\Lambda$ that we obtained previously~(see~Eq.~(\ref{eq:L_result})) is much easier to interpret in the backward-in-time formulation. To show this, let us rewrite Eq.~(\ref{eq:L_result}) as

\begin{equation}
\label{eq:L_bin}
\Lambda = \int_{-\infty}^{+\infty} \frac{\gamma_f(n(\zeta))}{n(\zeta)}S^{2}(\zeta)d\zeta,
\end{equation}

\noindent where we used Eq.~(\ref{eq:S_forward_in_time_stationary}) to express~$\Lambda$ in terms of~$S(\zeta)$, the stationary distribution of the location of a single ancestral lineage. We can now see that the effective coalescence rate,~$\Lambda$, is given by the sum of the local  coalescence rates,~$\gamma_f/n$, weighted by the probability that two lineages are present at the same location,~$S^2$. Thus, the first order perturbation theory is equivalent to assuming that the positions of the two ancestral lineages are uncorrelated with each other and distributed according to their stationary distribution~$S(\zeta)$. 

The last results also clarifies the difference between pulled, semi-pushed, and fully-pushed waves. For pulled waves,~$S(\zeta)$ is peaked at the leading edge and, since the coalescent rate peaks at the same location, the neutral evolution is driven by the very tip of the front. In semi-pushed waves,~$S(\zeta)$ is peaked in the interior of the front, but the~$1/n$ increase in the coalescence rates at the front is sufficiently strong to keep all coalescent events at the front edge. Finally, in fully-pushed waves, the decay of~$S^2(\zeta)$ at the front is stronger than the increase in the coalescence rates, and most coalescence events occur in the interior of the front. Thus, the focus of diversity is located in the interior of the front in fully-pushed waves, but at the front edge in pulled and semi-pushed waves.

\textit{Explicit results for~$\Lambda$ in exactly solvable models and connection}\\
In the regime of fully-pushed waves, we can evaluate~$\Lambda$ explicitly for the exactly solvable models introduced in section~I. Specifically, we find that

\begin{equation}
	\label{eq:L_quadractic_wf}
\Lambda = \frac{\gamma_f^0}{4\pi N}\sqrt{\frac{g_0}{2D}}\frac{1 - 4\rho^*}{\rho^*}\tan{2\pi \rho^*}
\end{equation}

\noindent for the model specified by Eq.~(\ref{eq:allee_quadratic}) with~$\gamma_f(n)$ that does not depend on~$n$ and is equal to~$\gamma_f^0$. Note that the choice of~$\gamma_{f}(n)$ is not specified by the deterministic model of population growth and needs to be determined from the microscopic dynamics, from phenomenological considerations, or empirically. For most commonly used models,~$\gamma_f(n)$ is a constant. In our simulations, this constant is~$1/(a\tau)$, where~$\tau$ is the generation time and~$a$ is the spatial scale over which genetic drift is correlated.  However, different~$\gamma_f$ are possible. For example,~$\gamma_f=\gamma_f^0(1-n/N)$ could be appropriate for models that allow no births or deaths once the population has reached the carrying capacity. In such models~$\gamma_f(N)=0$, and genetic drift operates only at the front.

For completeness, we also provide the results for other models and different choices of~$\gamma_f(n)$:

\begin{equation}
	\label{eq:L_quadractic_go}
\Lambda = \frac{\gamma_f^0}{6\pi N}\sqrt{\frac{g_0}{2D}}\frac{(1 - 2\rho^*)(1 - 4\rho^*)}{\rho^*}\tan{2\pi \rho^*}
\end{equation}

\noindent for the model specified by Eq.~(\ref{eq:allee_quadratic}) with~$\gamma_f= \gamma_f^0(1-n/N)$,

\begin{equation}
	\label{eq:L_cooperative_wf}
	\Lambda = \frac{\gamma_f^0}{4\pi N}\sqrt{\frac{r_0B}{2D}}(B + 4)\tan{\frac{2\pi}{B}}
\end{equation}

\noindent for the model of cooperative growth defined in the main text with~$\gamma_f = \gamma_f^0$;

\begin{equation}
	\label{eq:L_cooperative_go}
	\Lambda = \frac{\gamma_f^0}{6\pi N}\sqrt{\frac{r_0B}{2D}}\frac{(B + 2)(B + 4)}{B}\tan{\frac{2\pi}{B}}
\end{equation}

\noindent for the model of cooperative growth defined in the main text with~$\gamma_f = \gamma_f^0 (1-n/N)$.

\textbf{\large \hclr{IX. Cutoffs for deterministic and fluctuating fronts}}

The integrands that appear in the expressions for~$\Delta v$,~$D_{\mathrm{f}}$, and~$\Lambda$ diverge near the front edge in pulled and semi-pushed waves. Since there are no organisms sufficiently far ahead of the front, these divergences are technical artifacts that do not represent the actual dynamics of the traveling wave. For example, in our calculation of~$\Lambda$, the divergence appears because we approximate the wave front by the stationary, deterministic profile,~$n(\zeta)$, from Eq.~(\ref{eq:reduced_ode}). In this section, we show how to remove these divergences by applying a cutoff at large~$\zeta$. The value of the cutoff,~$\zeta_c$, scales as~$\ln(N)/q$ for fluctuating fronts, but as~$\ln(N)/k$ for deterministic fronts with~$\gamma_n=0$;~$k$ and~$q$ are given in Eq.~(\ref{eq:k_q_front}). The derivation of~$\zeta_c\sim\ln(N)/q$ is the main new result in this section.

\textit{Cutoff for deterministic fronts}\\
A cutoff for the growth rate was first introduced in the context of pulled waves~\cite{saarloos:review}. The primary motivation for the cutoff was to compute the corrections to the wave velocity and to resolve the velocity-selection problem, i.e. to explain why the simulations of discrete entities never exhibit waves with velocities greater than~$v_{\mathrm{\textsc{f}}}$ even though such solutions are possible in the continuum limit. 

The naive argument for a cutoff is that there should be no growth in areas where the average number of individuals falls below one per site in lattice models or one per typical dispersal distance in models with continuous space. We denote the relevant spatial scale, i.e. the distance between lattice sites or the dispersal distance, by~$a$, so the cutoff density is~$1/a$. Since, at such low densities, the front shape is well approximated by the asymptotic solution~$n\sim Ne^{-k\zeta}$, the value of the cutoff is given by

\begin{equation}
\label{eq:deterministic_cutoff}
\zeta_c = \frac{1}{k}\ln(Na).
\end{equation}

While this cutoff regularizes all the integrals and captures the gross effects of the stochastic dynamics, it is not quantitatively accurate for fluctuating fronts. Previous studies showed significant differences between the predictions of Eq.~(\ref{eq:deterministic_cutoff}) and simulations and argued that the factor multiplying~$\ln(N)$ in Eq.~(\ref{eq:deterministic_cutoff}) should be different from~$1/k$~\cite{kessler:velocity_cutoff}. The main goal of this section is to derive the correct cutoff for fluctuating fronts. 

Before proceeding with fluctuating fronts, however, it is important to point out that Eq.~(\ref{eq:deterministic_cutoff}) prescribes the correct cutoff for deterministic models with discrete entities~\cite{moro:numerical_schemes}. In such models, the main effect of discreteness is simply the absence of growth for~$n<1/a$, and, therefore, Eq.~(\ref{eq:deterministic_cutoff}) does apply. Our simulations show clear differences in the scaling of~$\Delta v$ and~$\Lambda$ with~$N$ for deterministic and fluctuating fronts~(Fig.~\ref{fig:fluctuating_deterministic_comparison}). Moreover, these differences are explained entirely by the different cutoffs that one needs to apply for fluctuating and deterministic fronts.

\textit{Cutoff for fluctuating fronts}\\
Analysis of fluctuating fronts is a challenging problem that is typically addressed by matching the nonlinear quasi-deterministic dynamics at the bulk of the front and the linear, but stochastic dynamics at the front edge~\cite{rouzine:wave}. Recently, a more rigorous approach has been developed in Refs.~\cite{hallatschek:tuned_model, hallatschek:tuned_moments}, which relies on modifying the reaction-diffusion model to ensure that the hierarchy of moment equations closes exactly. The details of this approach are sufficiently technical and tangential to the main issues discussed in this paper, so we do not discuss them here. Instead, we refer the readers to~Ref.~\cite{hallatschek:tuned_model} for a self-contained presentation of the new method. 

For our purpose, the most useful result from~Ref.~\cite{hallatschek:tuned_model} is that the deterministic equation for the steady-state density profile needs to be modified as

\begin{equation}
Dn'' + vn' + r(n)n - \frac{\gamma_n(n)n^2e^{\frac{v\zeta}{D}}}{\int_{-\infty}^{+\infty}n^2e^{\frac{v\zeta}{D}}d\zeta}= 0;
\label{eq:tuned_model}
\end{equation}

\noindent see Eq.~(10) in Ref.~\cite{hallatschek:tuned_model}. 

The only difference between Eq.~(\ref{eq:tuned_model}) and Eq.~(\ref{eq:reduced_ode}) is an additional term, which, as we show below, effectively imposes a cutoff on the growth rate. To quantify the magnitude of the new term, it is convenient to define a ratio between the terms due to front fluctuations and population growth: 

\begin{equation}
E(\zeta) = \frac{\frac{\gamma_n(n)n^2e^{\frac{v\zeta}{D}}}{\int_{-\infty}^{+\infty}n^2e^{\frac{v\zeta}{D}}d\zeta}}{r(n)n}= \frac{\gamma_n(n)ne^{\frac{v\zeta}{D}}}{r(n)\int_{-\infty}^{+\infty}n^2e^{\frac{v\zeta}{D}}d\zeta}.
\label{eq:term_ratio}
\end{equation}

\noindent Note that the first three terms in Eq.~(\ref{eq:tuned_model}) are of the same order at the front, so any one of them could be used to define~$E$. 

Since we are only interested in the behavior of~$E$ near the front edge,~$E$ can be further simplified as

\begin{equation}
E(\zeta) \approx \frac{\gamma_n(0)ne^{\frac{v\zeta}{D}}}{r(0)\int_{-\infty}^{+\infty}n^2e^{\frac{v\zeta}{D}}d\zeta} = \frac{\gamma_n(0)v\rho e^{\frac{v\zeta}{D}}}{r(0)DNI},
\label{eq:E}
\end{equation}

\noindent where, in the last equality, we made the dependence on all dimensional quantities explicit by using~$\rho=n/N$ and introducing a non-dimensional integral

\begin{equation}
I = \int_{-\infty}^{+\infty}\rho^2(\zeta)e^{\frac{v\zeta}{D}}d\left(\frac{v\zeta}{D}\right).
\label{eq:I}
\end{equation}

To obtain the scaling behavior of~$E$ for large~$\zeta$, we approximate~$\rho$ as~$e^{-k\zeta}$ and replace~$v/D$ by~$k+q$~(see Eq.~(\ref{eq:k_q_front})) 

\begin{equation}
E\sim e^{-\zeta(k-v/D)}\sim e^{q\zeta}.
\end{equation}

\noindent Note that~$q$ is defined in Eq.~(\ref{eq:solution_front}) as the decay rate for the solution of Eq.~(\ref{eq:reduced_ode_front}) that is inconsistent with the boundary conditions. Therefore,~$q$ corresponds to the unphysical part of the solution for~$\rho$ and does not directly enter the asymptotic behavior of the population density. In the following, $q$~is often used instead of~$v$ to make the formulas more compact.

We now determine the cutoff~$\zeta_c$ for all subtypes of traveling waves. The main idea is to check whether the solution of Eq.~(\ref{eq:reduced_ode}) is consistent with the addition of a term due to front fluctuations in Eq.~(\ref{eq:tuned_model}). If the solution is consistent, then no cutoff is necessary. If the solution is not consistent, then it can be valid only up to some critical~$\zeta$, which acts as an effective cutoff.

\textit{Cutoff for pushed waves expanding into a metastable state}\\
When the invaded state is metastable, the low-density growth rate is negative, and, therefore,~$q<0$; see Eq.~(\ref{eq:k_q_front}). In consequence,~$E\to0$ as~$\zeta\to+\infty$, and front fluctuations have a negligible effect on wave dynamics. Thus, no cutoff is necessary, i.e.~$\zeta_c=+\infty$. 

\textit{Cutoff for pushed waves expanding into an unstable state}\\
When the invaded state is unstable,~$q$ is positive, and~$E$ diverges as~$\zeta\to+\infty$. This contradicts Eq.~(\ref{eq:tuned_model}), where all terms need to cancel out. To satisfy the equation,~$\rho$ must decay faster than~$e^{-k\zeta}$ at the front beyond some critical~$\zeta_c$, so that~$E$ never becomes much greater than one. The value of~$\zeta_c$ is then determined by the solution of~$E(\zeta_c)=1$ with the deterministic approximation for~$\rho$. Hence, we substitute~$\rho\sim e^{-k\zeta}$ in Eq.~(\ref{eq:E}) and find that

\begin{equation}
\label{eq:fluctuating_cutoff}
\zeta_c = \frac{1}{q}\ln\left(N\frac{Dr(0)I}{v\gamma_n(0)}\right) = \frac{1}{q}\ln\left(\frac{N}{k+q}\frac{r(0)I}{\gamma_n(0)}\right) \sim \frac{1}{q}\ln N.
\end{equation}

\noindent This is the most important result of this section because it determines the novel scaling behavior of~$\Delta v$,~$D_{\mathrm{f}}$, and~$\Lambda$ in semi-pushed waves. To the best of our knowledge, Eq.~\ref{eq:fluctuating_cutoff} is a new finding.

Since~$q<k$, fluctuating fronts have a larger~$\zeta_c$ than deterministic fronts and a lower normalized cutoff density~$\rho_c\sim e^{-k\zeta_c}\sim N^{-k/q}<N^{-1}$. The applicability of the continuum theory for~$\rho<1/N$ seems counter-intuitive, not only because the expected number of organisms at a site falls below one, but also because fluctuations should appreciably modify the profile density at least when~$\rho<1/\sqrt{N}$, i.e. well before the naive~$1/N$ cutoff. The key problem with this argument is that it assumes a continual front and neglects the possibility that a sufficiently large group of organisms can occasionally expand well ahead of the deterministic front~\cite{hallatschek:noisy_fisher, brunet:selection_ancestry}. Such front excursions prevent a sharp cutoff in the population density below~$1/N$. More importantly, they significantly amplify both genetic drift and front wandering. In the continuum theory, this increase in fluctuations is captured by greater~$\zeta_c$, which increases both~$\Lambda$ and~$D_{\mathrm{f}}$. The probability of front excursions is controlled not only by the intensity of demographic fluctuations, but also by other parameters of the population dynamics. In particular,~$\rho_c$ depends on cooperativity through~$v/v_{\mathrm{\textsc{f}}}$. Inclusion of this dependence is necessary to accurately describe the dynamics of semi-pushed waves.

\textit{Cutoff for pulled waves}\\
For pulled waves,~$q>0$, and we obtain the value of the cutoff by solving~$E(\zeta_c)=1$ just as for semi-pushed waves:

\begin{equation}
\label{eq:cutoff_pulled_initial}
\gamma_n(0) \rho(\zeta_c) e^{\frac{v\zeta_c}{D}}= r(0)N\int_{-\infty}^{\zeta_c}\rho^2(\zeta)e^{\frac{v\zeta}{D}}d\zeta.
\end{equation}

Note, however, that there are two important differences in the calculation for pulled compared to semi-pushed waves. First, the integral~$I$ diverges and needs to be cut off at~$\zeta_c$. Second, the front shape also acquires corrections due to the cutoff and needs to be determined self-consistently. This sensitivity of the front shape originates from the degeneracy~$k=q$ that occurs in pulled waves. In the continuum limit, this degeneracy modifies the scaling of~$\rho$ from~$e^{-k_{\mathrm{\textsc{f}}}\zeta}$ to~$k_{\mathrm{\textsc{f}}}\zeta e^{-k_{\mathrm{\textsc{f}}}\zeta}$. For a fluctuating front, however, the wave velocity deviates slightly from~$v_{\mathrm{\textsc{f}}}$, and the correction to the front shape is different. 

The shape of the front can be obtained by setting the growth rate to zero for~$\zeta>\zeta_c$ and solving the resulting equation for~$\rho(\zeta)$; see Refs.~\cite{kessler:velocity_cutoff, saarloos:review, brunet:phenomenological_pulled}. The result reads

\begin{equation}
\label{eq:front_shape_pulled}
\rho(\zeta) \sim  \frac{k_{\mathrm{\textsc{f}}}\zeta_c}{\pi}\sin\left(\pi\frac{\zeta_c-\zeta}{\zeta_c}\right)e^{-k_{\mathrm{\textsc{f}}}\zeta}.
\end{equation}

Upon substituting this result in Eq.~(\ref{eq:cutoff_pulled_initial}), we find the following condition on~$\zeta_c$:

\begin{equation}
\label{eq:cutoff_pulled_equation}
e^{k_{\mathrm{\textsc{f}}}\zeta_c} \sim \frac{1}{2\pi^2}\frac{r(0)N}{\gamma_n(0)k_{\mathrm{\textsc{f}}}} (k_{\mathrm{\textsc{f}}}\zeta_c)^3,
\end{equation}

\noindent where we shifted~$\zeta_c$ in the argument of the sine by~$1/k_{\mathrm{\textsc{f}}}$, which does not change the asymptotic scaling of the exponential term, but avoids setting the left hand side to zero. To solve Eq.~(\ref{eq:cutoff_pulled_equation}), we treat~$(k_{\mathrm{\textsc{f}}}\zeta_c)^3$ as a small perturbation compared to~$\frac{r(0)N}{\gamma_n(0)k_{\mathrm{\textsc{f}}}}$ and obtain that, up to additive numerical factors, the leading behavior of~$\zeta_c$ is given by

\begin{equation}
\label{eq:cutoff_pulled_result}
\zeta_c = \frac{1}{k_{\mathrm{\textsc{f}}}}\ln\left(\frac{r(0)N}{\gamma_n(0)k_{\mathrm{\textsc{f}}}}\right) + \frac{3}{k_{\mathrm{\textsc{f}}}}\ln\left[\ln\left(\frac{r(0)N}{\gamma_n(0)k_{\mathrm{\textsc{f}}}}\right)\right]  \sim \frac{1}{k_{\mathrm{\textsc{f}}}}\ln N + \frac{3}{k_{\mathrm{\textsc{f}}}}\ln\ln N .
\end{equation}

\noindent Note that the second order term cannot be neglected in the calculation of~$\Delta v$,~$D_{\mathrm{f}}$, and~$\Lambda$ because these quantities have terms that scale both linearly and exponentially with~$\zeta_c$. Equation~(\ref{eq:cutoff_pulled_result}) was first motivated phenomenologically in Ref.~\cite{brunet:phenomenological_pulled} and then derived more rigorously in Ref.~\cite{hallatschek:tuned_model}.

\textbf{\large \hclr{X. Scaling of~$\Delta v$,~$D_{\mathrm{f}}$, and~$\Lambda$ in pulled, semi-pushed, and fully-pushed waves}}

In this section, we synthesize the results of the perturbation theory for~$\Delta v$,~$D_{\mathrm{f}}$, and~$\Lambda$ and supplement them with an appropriate cutoff~$\zeta_c$ when needed. We show that pushed waves consist of two distinct classes. In fully-pushed waves, the fluctuations scale as~$1/N$ consistent with the central limit theorem, but, in semi-pushed waves, non-trivial power scaling occurs. The exponent of this power law depends only on~$v/v_{\mathrm{\textsc{f}}}$ and is the same for~$\Delta v$,~$D_{\mathrm{f}}$, and~$\Lambda$. For completeness, also provide the corresponding results for pulled waves and models with deterministic fronts. While most results of the perturbation theory are not new, the synthesis of these results, the application of an appropriate cutoff, and the discovery of semi-pushed waves are novel contributions of this paper.  

The main results of the perturbation theory are given by Eq.~(\ref{eq:dv_so}) for the correction to wave velocity, by Eq.~(\ref{eq:D_result}) for the effective diffusion constant of the front, and by Eq.~(\ref{eq:L_result}) for the rate of diversity loss. All of these equations, have a similar form and contain a ratio of two integrals. Both integrals converge for~$\zeta\to-\infty$, but they could diverge for~$\zeta\to+\infty$. At the front, the integrands in the numerator scale as~$e^{-(3k-2v/D)\zeta}$, and the integrands in the denominators scale as~$e^{-(2k-v/D)\zeta}$; therefore, the integrals in the denominators always converge when the integrals in the numerators converge. Since the ratio of~$k$ to~$v/D$ depends on the degree to which the growth is cooperative, the integrals could change their behavior as cooperativity is varied. A change from convergence to divergence in either of the integrals corresponds to a transitions between different classes of waves. Below we consider each class separately.   

\textit{$1/N$~scaling in fully-pushed waves}\\
The class of fully-pushed waves is defined by the requirement that all integrals converge. In this case, the perturbation theory is well-posed without a cutoff and provides not only the scaling, but also the exact values of~$\Delta v$,~$D_{\mathrm{f}}$, and~$\Lambda$. Straightforward dimensional analysis shows that all three quantities scale as~$1/N$, i.e. the central limit theorem holds. 

The convergence of integrals requires that~$kD/v$ is greater than~$2/3$. For waves expanding into an metastable state,~$kD/v>1$~(see Eq.~(\ref{eq:k_q_front})), so these waves are always fully-pushed. For expansions into an unstable state, it is convenient to express the convergence condition only as a function of~$v$ using Eq.~(\ref{eq:k_q_front}):

\begin{equation}
v \ge \frac{3}{2\sqrt{2}}v_{\mathrm{\textsc{f}}}.
\label{eq:v_critical_two}
\end{equation} 

\noindent where~$v_{\mathrm{\textsc{f}}}=2\sqrt{Dr(0)}$ is the linear spreading velocity. We emphasize that~$v_{\mathrm{\textsc{f}}}$ serves only as convenient notation for~$2\sqrt{Dr(0)}$; in particular, the wave is not pulled, and the wave velocity is greater than~$v_{\mathrm{\textsc{f}}}$.

Equation~(\ref{eq:v_critical_two}) immediately implies that not every pushed wave is fully-pushed. Indeed, only~$v>v_{\mathrm{\textsc{f}}}$ is required for a wave to be pushed, which is a weaker condition than Eq.~(\ref{eq:v_critical_two}). Because~$v/v_{\mathrm{\textsc{f}}}$ increases with cooperativity, fully-pushed waves occur once cooperativity exceeds a certain threshold. 

We can also express the condition that~$kD/v>2/3$ in terms of~$k$ and~$q$ using Eq.~(\ref{eq:k_q_front}). Because~$v/D=k+q$, this convergence condition is equivalent to~$k>2q$. For all pushed waves,~$k>q$, but a stronger inequality is required for fully-pushed waves. Note that~$q<0$ for waves propagating into a metastable state, so~$k>q$ is satisfied.

Finally, we discuss the effects of a cutoff derived in the previous section. For expansions into a metastable state,~$\zeta_c=+\infty$, i.e. no cutoff is necessary. For expansions into an unstable state, the theory suggest a finite cutoff:~$\zeta_c\sim \ln(N)/q$. Note, however, that the application of this cutoff in the formulas for~$\Delta v$,~$D_{\mathrm{f}}$, and~$\Lambda$ only produces subleading corrections to the~$1/N$ scaling because convergent integrals are insensitive to small changes in their upper limit of integration.

\textit{$N^{\alpha}$ scaling in semi-pushed waves}\\
We now proceed to the second class of pushed waves, for which the integrals in the numerators diverge, i.e.~$v<\sqrt{9/8}v_{\mathrm{\textsc{f}}}$. We term the waves in this class semi-pushed because the fluctuations at the front make a significant contribution to their dynamics. Note that the integrals in the denominators converge for all pushed waves because~$kD/v>1/2$; see Eq.~(\ref{eq:k_q_front}).  

To estimate the scaling of~$\Delta v$,~$D_{\mathrm{f}}$, and~$\Lambda$, we cut off the integrals in the numerators at~$\zeta_c\sim \ln(N)/q$ and find that three quantities scale as~$N^{\alpha}$ with~$\alpha$ given by

\begin{equation}
\label{eq:alpha_fluctuating}
\alpha = -2\frac{\sqrt{1-v^2_{\mathrm{\textsc{f}}}/v^2}}{1-\sqrt{1-v^2_{\mathrm{\textsc{f}}}/v^2}}.
\end{equation}

The details of this calculation for~$\Lambda$ are summarized below 

\begin{equation}
\begin{aligned}
\Lambda &= \frac{1}{N} \frac{\int_{-\infty}^{\zeta_c}\gamma_{f}(\rho)\rho^{3}(\zeta)e^{\frac{2v\zeta}{D}}d\zeta}{\left(\int^{+\infty}_{-\infty}\rho^2(\zeta)e^{\frac{v\zeta}{D}}d\zeta\right)^2} \sim \frac{1}{N} \frac{\gamma_{f}(0)\int_{-\infty}^{+\zeta_c}e^{-(k-2q)\zeta}d\zeta}{\left(I\frac{D}{v}\right)^2} \sim \frac{(k+q)^2\gamma_f(0)}{I^2N}\frac{e^{-\frac{(k-2q)}{q}\ln\left( \frac{N}{k+q}\frac{r(0)I}{\gamma_n(0)}\right)}}{k-2q}\\
& \sim N^{-\frac{k-q}{q}}I^{-\frac{k}{q}}\gamma_f(0)[\gamma_n(0)]^{\frac{k-2q}{q}}[r(0)]^{-\frac{k-2q}{q}}(k+q)^{\frac{k}{q}}(k-2q)^{-1}\sim N^{-\frac{k-q}{q}}\sim N^{-2\frac{\sqrt{v^2 -v^2_{\mathrm{\textsc{f}}}}}{v-\sqrt{v^2-v^2_{\mathrm{\textsc{f}}}}}}.
\end{aligned}
\end{equation}

\noindent where we used Eqs.~(\ref{eq:k_q_front}), (\ref{eq:L_result}), (\ref{eq:I}), and~(\ref{eq:fluctuating_cutoff}). The calculations for~$D_{\mathrm{f}}$ and~$\Delta v$ are essentially the same.

\textit{Logarithmic scaling in pulled waves}\\
The remaining possibility is that the integrals diverge in both numerators and denominators. This is the case for pulled waves because~$v_{\mathrm{\textsc{f}}}=2k_{\mathrm{\textsc{f}}}D$. To compute the asymptotic scaling of~$D_{\mathrm{f}}$ and~$\Lambda$, we use (\ref{eq:D_result}) and~(\ref{eq:L_result}) together with the cutoff from Eq.~(\ref{eq:cutoff_pulled_result}) and the profile shape from Eq.~(\ref{eq:front_shape_pulled}). The results read

\begin{equation}
\begin{aligned}
& D_{\mathrm{f}} \sim \frac{\gamma_{n}(0)}{k^2_{ \mathrm{\textsc{f}} } }\ln^{-3}\left(\frac{N}{k_{ \mathrm{\textsc{f}} }}\right),\\
& \Lambda \sim \gamma_{f}(0)\ln^{-3}\left(\frac{N}{k_{ \mathrm{\textsc{f}} }}\right).\\
\end{aligned}
\end{equation}

This completes our discussion of different scaling regimes in fluctuating fronts.  

\textit{Scaling of~$\Lambda$ with~$N$ in deterministic fronts}\\
Some of our results for fluctuating fronts depend on the specific choice of the cutoff~$\zeta_c=\ln(N)/q$. This cutoff is different from the naive expectation that~$\zeta_c=\ln(N)/k$ because occasional fluctuations establish a small population far ahead of the deterministic front. To understand the effect of such fluctuations, we now examine the properties of deterministic fronts, where~$\gamma_n=0$, but there is a cutoff on the growth rate below~$\rho_c\sim1/N$. Since deterministic fronts do not fluctuate, their diffusion constant is zero. Genetic drift, however, occurs even without any fluctuations in the total population size, so the rate of diversity loss is well-defined. Therefore, we focus on the scaling of~$\Lambda$ with~$N$ in this subsection.   

Our analysis of fully-pushed waves remains unchanged because all the integrals converge, and a cutoff is not required. Thus,~$\Lambda\sim N^{-1}$ for fully-pushed waves with or without demographic fluctuations. Moreover, the transition point between fully-pushed and semi-pushed waves remains the same because it depends on the behavior of the integrands in Eq.~(\ref{eq:L_result}) rather than on the value of the cutoff.

For semi-pushed waves,~$\zeta_c$ does enter the calculation and changes the value of~$\alpha$. For deterministic fronts, we find that

\begin{equation}
\alpha_{\mathrm{deterministic}} = -2\frac{k-q}{k} = - \frac{4\sqrt{1 - v_{\mathrm{\textsc{F}}}^2/v^2}}{1 + \sqrt{1 - v_{\mathrm{\textsc{F}}}^2/v^2}}. 
\label{eq:alpha_deterministic}
\end{equation}

\noindent Similarly to our results for the fluctuating fronts,~$\alpha_{\mathrm{deterministic}}$ approaches~$0$ and~$-1$ near the transitions to pulled and fully-pushed waves. Within the class of semi-pushed waves, however,~$\alpha_{\mathrm{deterministic}}$ is less than~$\alpha$ for fluctuating fronts~($|\alpha_{\mathrm{deterministic}}|>|\alpha_{\mathrm{fluctuating}}|$), that is genetic drift is amplified by front fluctuations.

For pulled waves, we find that

\begin{equation}
\Lambda\sim\gamma_f(0)\ln^{-6}(N/k_{\mathrm{\textsc{f}}}),
\label{eq:L_deterministic_pulled}
\end{equation}

\noindent which further supports the fact that genetic drift is weaker without front fluctuations. 

The~$\ln^{-6}N$ scaling was previously suggested for the diffusion constant of pulled waves based on the incorrect application of the naive cutoff~\cite{panja:review}. Moreover, simulations that limited the extent of demographic fluctuations indeed observed that~$D_{\mathrm{f}}\sim\ln^{-6}N$~\cite{moro:numerical_schemes}.

\textit{Comparison of~$\Delta v$ in deterministic vs. fluctuating fronts}\\
We close this section by comparing velocity corrections for deterministic and fluctuating fronts. This comparison highlights the conceptual challenges that we resolved in order to describe the stochastic dynamics of range expansions and provides a useful perspective on the potential pitfalls in approximating a fluctuating front by a deterministic front with a cutoff. Because corrections to velocity have been a subject of intense theoretical study~\cite{saarloos:review, panja:review, kessler:velocity_cutoff, brunet:velocity_cutoff, brunet:phenomenological_pulled}, the following discussion also clarifies the connection between our and previous work.

The standard approach to computing~$\Delta v$ is to impose a zero growth rate below a certain population density; typically~$\rho_c=1/N$. The deterministic reaction-diffusion equation is then solved separately for~$\zeta<\zeta_c$ and~$\zeta>\zeta_c$, and the solutions are matched at~$\zeta=\zeta_c$. This approach is thought to be largely correct because it yields the right scaling of~$\Delta v\sim\ln^{-2}N$ for pulled waves~\cite{brunet:velocity_cutoff}, which have been the primary subject of research. Our calculation of the cutoff, however, shows that the agreement between~$\rho_c$ and~$1/N$ for pulled waves is rather accidental because these two quantities are different for all other wave classes. Moreover, further work on pulled waves showed that~$1/N$ cutoff is insufficient to describe all of their properties, and the second term on the right hand side of Eq.~(\ref{eq:cutoff_pulled_result}) is necessary~\cite{brunet:phenomenological_pulled}. This result was first obtained from phenomenological considerations~\cite{brunet:phenomenological_pulled}, but was later derived more rigorously via an approach that also justified the existence of the cutoff~\cite{hallatschek:tuned_model}.

The calculation of~$\Delta v$ based on a fixed growth-rate cutoff at~$\rho_c$ was extended to pushed waves by Kessler~\textit{et al.}~\cite{kessler:velocity_cutoff}, who found that\footnote{Ref.~\protect{\cite{kessler:velocity_cutoff}} computed~$\Delta v$ for an unspecified cutoff at~$\rho=\rho_c$; we substituted~$\rho_c=1/N$ to facilitate the comparison with other results in this paper.} 

\begin{equation}
\label{eq:dv_deterministic}
\Delta v_{\mathrm{deterministic}} \sim N^{-1+\frac{q}{k}}.
\end{equation}

\noindent The same result is obtained from the first order perturbation theory~(Eq.~(\ref{eq:delta_v_cutoff})) with~$\rho_c=1/N$. By numerically solving a reaction-diffusion equation with an imposed growth-rate cutoff, Ref.~\cite{kessler:velocity_cutoff} confirmed that Eq.~(\ref{eq:dv_deterministic}) provides an accurate prediction for~$\Delta v$ in deterministic fronts, but the applicability of Eq.~(\ref{eq:dv_deterministic}) to fluctuating fronts has not been investigated.

Our findings show that there are two qualitative differences between the predictions of Eq.~(\ref{eq:dv_deterministic}) and the actual behavior of fluctuating fronts~(Fig.~\ref{fig:velocity_corrections_comparison}A). First, Eq.~(\ref{eq:dv_deterministic}) predicts that the exponent~$\alpha_{\mathrm{\textsc{v}}}$ changes gradually from~$0$ to~$-2$ as the strength of the Allee effect increases\footnote{At the boundary with pulled waves,~$\alpha_{\mathrm{\textsc{v}}}=0$, and $\alpha_{\mathrm{\textsc{v}}}=-2$ for the maximal strength of the Allee effect at which the invasion can still proceed~($v>0$)}. Between these two limiting cases, there are no transitions that would indicate the existence of distinct classes of pushed waves. Second, Eq.~(\ref{eq:dv_deterministic}) misses the~$1/N$ scaling of~$\Delta v$ in the regime of highly cooperative growth, where the central limit theorem applies because the properties of the wave are determined by the dynamics in the interior of the front rather than at the leading edge.

The origin of these discrepancies is different for semi-pushed and fully-pushed waves. For semi-pushed waves, the different behavior of deterministic and fluctuating front comes from the dependence of the cutoff on the strength of the Allee effect~(Eq.~(\ref{eq:alpha_fluctuating})). Indeed, Eq.~(\ref{eq:delta_v_cutoff}) and the results from Ref.~\cite{kessler:velocity_cutoff} reproduce the correct scaling of~$\Delta v$ with~$N$~(Eq.~(\ref{eq:alpha_fluctuating})) once we substitute~$\rho_c=N^{-k/q}$ instead of the~$\rho_c=N^{-1}$. For fully-pushed waves, this approach still produces unrealistic scaling with~$\alpha_{\mathrm{\textsc{v}}}<-1$ because both the first order perturbation theory and the approach in Ref.~\cite{kessler:velocity_cutoff} assume that the main contribution to~$\Delta v$ comes from the stochastic dynamics of the tip of front. The dynamics of fully-pushed waves are, however, controlled by the fluctuations throughout the front, and, therefore, cannot be described by an effective cutoff. This is clearly demonstrated by the second order perturbation theory~(Eq.~(\ref{eq:dv_so})), which shows how the~$1/N$ scaling, expected from the central limit theorem, emerges from the stochastic dynamics at the entire front.

Thus, replacing the full stochastic dynamics by a deterministic front with a cutoff can fail to describe population dynamics both because the value of the cutoff has a nontrivial dependence on model parameters and because the dominant contribution of fluctuations may not be restricted to the leading edge of the reaction-diffusion wave.

\textbf{\large \hclr{XI. Precise definitions of the foci of growth, ancestry, and diversity}}

In this section, we consolidate the results obtained above on the spatial distribution of growth, ancestry, diversity processes within the wave front. We also provide the precise definitions of the foci of growth, ancestry, and diversity.

The spatial distribution of the per capita growth rate is given by

\begin{equation}
\label{eq:growth_focus}
\mathrm{growth\;\;distribution} \sim r(n(\zeta)).
\end{equation}

\noindent The mode of this distribution is the focus of growth. For monotonically decreasing~$r(n)$, the focus of growth is at the very edge of the front, i.e. at~$\zeta=\zeta_c$, but it is in the interior of the front otherwise. 

The spatial distribution of the most recent common ancestor of the entire population at the front is given by Eq.~(\ref{eq:S_forward_in_time_stationary}):

\begin{equation}
\label{eq:ancestry_focus}
\mathrm{ancestry\;\;distribution}=S(\zeta) \sim n^{2}(\zeta)e^{v\zeta/D}.
\end{equation}
 
\noindent The mode of this distribution is the focus of ancestry, which is the most likely location of the most recent common ancestor. The focus of ancestry is located at~$\zeta=\zeta_c$ for pulled waves and in the interior of the front for pushed waves.

To characterize the contribution of the different regions of the front to genetic diversity, we consider the spatial distribution of the locations where two ancestral lineages coalesce. That is we consider the spatial location of the most recent common ancestor of two randomly sampled individuals. From Eq.~(\ref{eq:L_bin}), it follows that this distribution is given by

\begin{equation}
\label{eq:diversity_focus}
\mathrm{diversity\;\;distribution}=C(\zeta)  \sim \frac{\gamma_f(n(\zeta))}{n(\zeta)}S^{2}(\zeta) \sim \gamma_f(n(\zeta))n^{3}(\zeta)e^{2v\zeta/D}.
\end{equation}

\noindent The mode of this distribution is the focus of ancestry. The focus of diversity is located at~$\zeta=\zeta_c$ for pulled and semi-pushed waves and in the interior of the front for fully-pushed waves.

The definitions above are somewhat arbitrary as one could have used the mean or median of the corresponding distributions rather than the mode in defining the foci of ancestry and diversity. The precise definitions of foci are, however, irrelevant for understanding the differences in wave properties because the spatial distributions fundamentally change at the transitions between different wave classes. For pulled waves, the distribution of ancestor~$S(\zeta)$ becomes independent of~$\zeta$ for large positive~$\zeta$. Therefore, the distribution is not normalizable, and effectively all the weight of the distribution is concentrated on large positive values of~$\zeta$. In consequence, both the mean, median, and the mode are at large positive~$\zeta$. Thus, the transition from pulled to pushed waves is marked by a fundamental change in the distribution and an infinite jump in the focus of ancestry. A similar transition occurs for the focus of diversity as waves transition from semi-pushed to fully-pushed. For fully-pushed waves,~$C(\zeta)$ is normalizable and peaked at a well-defined value of~$\zeta$. For semi-pushed waves,~$C(\zeta)$ diverges at large~$\zeta$ and is therefore not normalizable. The weight of the distribution shifts to very large~$\zeta$, so we described this transition as the shift in the focus of diversity from the bulk to the edge of the front.

The focus of growth is less informative because waves could still be pulled even when the growth is not maximal at the very edge of the front. Nevertheless, the transition from pulled to pushed waves is marked by a nonzero contribution of growth throughout the front to the wave velocity, so one can loosely speak of a shift in growth from the edge to the bulk of the front.

Figure~\ref{fig:foci} graphically summarizes how the locations of different processes change as waves transition from pulled, to semi-pushed, and to fully-pushed waves.

\textbf{\large \hclr{XII Prevalence of semi-pushed waves}}

The range of velocities of semi-pushed waves appears to be small from~$1.00$ to about~$1.06$ times the Fisher velocity. Therefore, one might be tempted to conclude that semi-pushed waves are rare. Below we show that this conclusion is not justified.

While the ratio of wave velocity to Fisher velocity is a universal metric of cooperativity, it does not faithfully represent the size of the parameter space. Indeed, the entire region of pulled waves collapses to a single point~$v/v_{\mathrm{\textsc{f}}}=1$. Pulled waves of course occur for more than a single parameter: The growth rate could include an arbitrary density-dependence as long as it decreases with population density, and the growth rate could even be mildly cooperative. Because semi-pushed waves are bordering pulled waves, the parameter space also undergoes compression when mapped into the space of~$v/v_{\mathrm{\textsc{f}}}$. To illustrate this, we consider three models of the growth rate: the cooperative model from the main text (Eq.~(3)), a completely different model with predator satiation, and a model of an experimental system that was recently used to show a transition from pulled to pushed waves~\cite{gandhi:pulled_pushed}.

For the model in the manuscript, the growth rate is given by~$r(n)=r_0(1-n/N)(1+Bn/N)$. Here, parameter~$B$ represents cooperativity in the growth rate and controls the transition from pulled to pushed waves. For this model, pulled waves occurs for~$B$ between~$0$ and~$2$, semi-pushed waves for~$B$ between~$2$ and~$4$, and fully-pushed waves for~$B$ greater than~$4$. Thus, the extensively-studied pulled waves and the newly-discovered semi-pushed waves occupy regions in the parameter space of exactly the same size. This model can be parameterized differently, see Eq.~\ref{eq:allee_quadratic}. For this parameterization, the region of pulled waves occurs for~$n^*/N <-0.5$, semi-pushed waves for~$n^*/N$ between~$-0.5$ and~$-0.25$, and fully-pushed waves for~$n^*/N$ between~$-0.25$ and~$0.5$. From this comparison of essentially the same models, it is clear that the size of a region in the parameter spaces depend on the type of parameterization, but, generically, semi-pushed waves occupy about as much parameter space as pulled and fully-pushed waves.

To demonstrate, that the above conclusion is not specific to the cooperative model studied in the manuscript, we considered a completely different mechanism behind pushed waves: namely, predator satiation. This type of an Allee effect can be modeled by

\begin{equation}
 r(n)=r_0\left(1-\frac{n}{N}\right) - d \frac{n^*}{n+n^*}.
 \label{eq:satiation}
\end{equation}
 
\begin{figure}[!ht]
\begin{center}
\includegraphics[width=17.8cm]{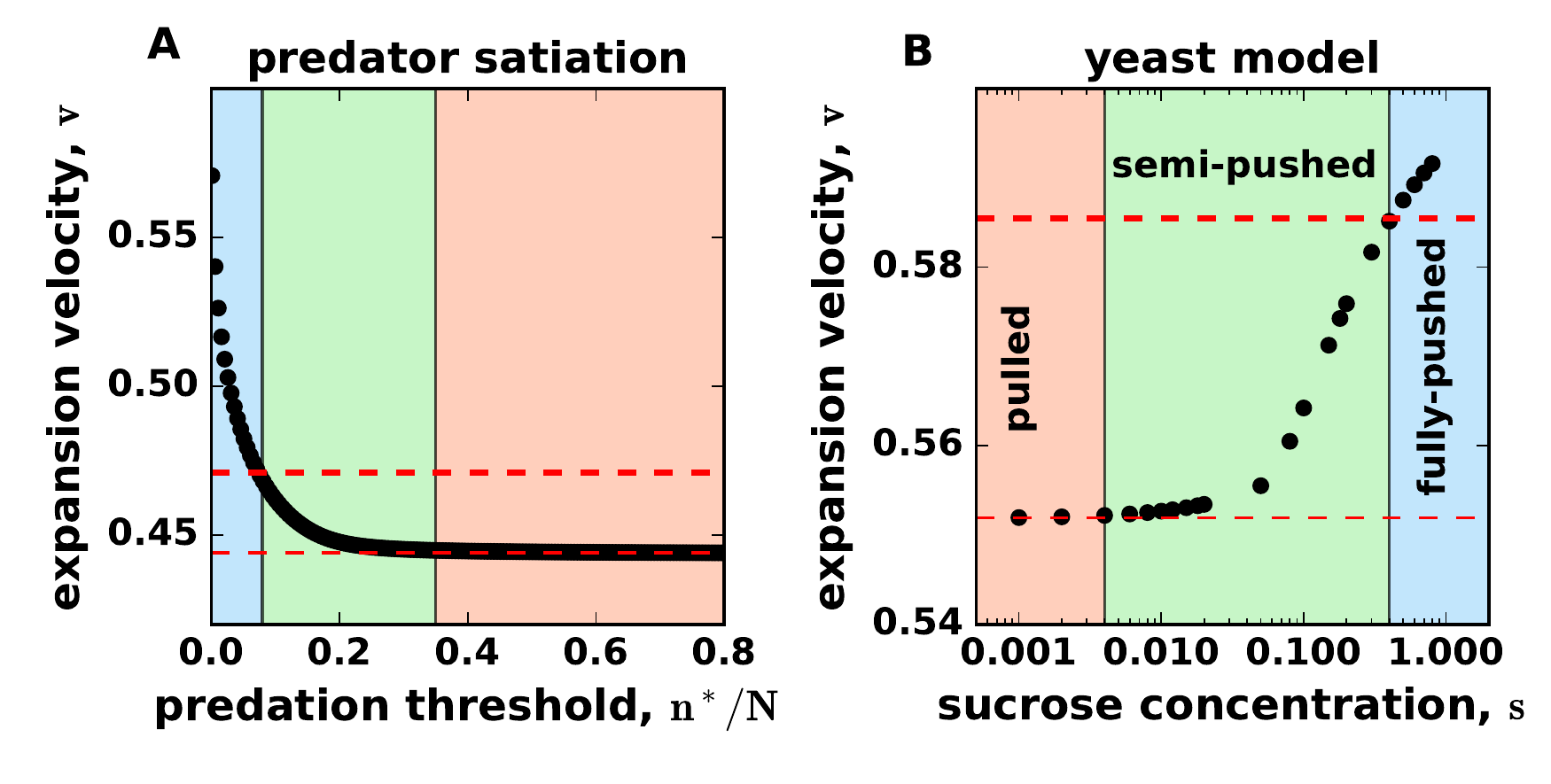}
\caption{\textbf{Semi-pushed waves occupy a sizable region in the parameter space.} The panels show the expansion velocity as a function of cooperativity for two alternative models of Allee effect. Even though the transition from pulled to fully-pushed waves requires a modest change in wave velocity, it requires a substantial change in the parameter controlling the cooperativity of growth. \textbf{(A)}~shows the velocities obtained by numerically solving Eq.~(\ref{eq:deterministic_n}) with the growth rate from the predator satiation model defined by Eq.~(\ref{eq:satiation}). \textbf{(B)}~shows the results for a model of cooperative extracellular digestion defined by Eq.~(\protect{\ref{eq:yeast}}), which was solved as described in Ref.~\cite{gandhi:pulled_pushed}, using $m=0.1$ and $df=2$.}
\label{fig:phase_space}
\end{center}  
\end{figure}

\begin{figure}[!h]
\begin{center}
\includegraphics[width=17.8cm]{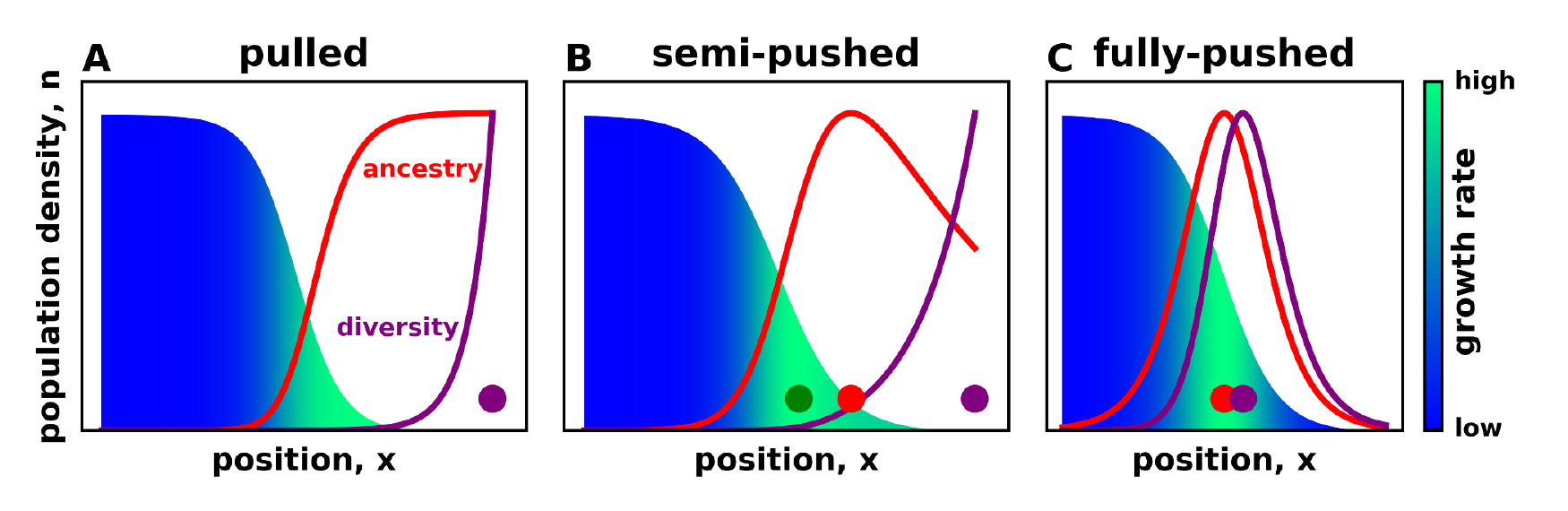}
\caption{\textbf{Foci of growth, ancestry, and diversity spatially segregate in semi-pushed waves.} The three panels compare the spatial distribution of growth, ancestry, and diversity among pulled, semi-pushed, and fully-pushed waves. The color gradient shows how the per-capita growth rate changes along the wave front. The ancestry curve shows the distribution of the most recent common ancestor of the entire population at the front, which is the same as the probability of a neutral mutation arising at a particular location and then reaching fixation. The diversity curve shows the spatial distribution of the most recent common ancestor of two individuals sampled randomly from the front. In other words, this curve shows the probability that two ancestral lineages coalesce at a specific location at the front. Thus, the maximum of the diversity curve corresponds to the location that contributes most to the rate of diversity loss. The colored dots show the positions of foci of growth~(green), ancestry~(red), and diversity~(purple). In both pulled and pushed waves these foci are colocalized, but they are spatially separated in semi-pushed waves. As a result of this, semi-pushed waves posses characteristics of both pulled and pushed waves. The color gradient and the curves are theoretical predictions from Eqs.~(\ref{eq:allee_quadratic}),~(\ref{eq:ancestry_focus}), and~(\ref{eq:diversity_focus}) using the growth model defined by Eq.~(\ref{eq:allee_quadratic});~$D=0.125$,~$g_0=0.01$,~$\zeta_c=30$. The values of~$n^*/N$ used were -1.0, -0.4 and 0 for pulled, semi-pushed, and fully-pushed, respectively.}
\label{fig:foci}
\end{center}  
\end{figure}

\noindent Below the Allee threshold~$n^*$, the population experiences a high per capita death rate~$d$ from predation, but, above~$n^*$, the limited number of predators cannot keep up with the prey, and the per capita death rate declines. We found that pulled waves occur for~$n^*/N$ greater than~$0.35$, semi-pushed waves for~$n^*/N$ between~$0.08$ and~$0.35$, and fully-pushed waves for~$n^*/N$ less than~$0.08$; see Fig.~\ref{fig:phase_space}A. In this model, semi-pushed waves occupy a larger region in the parameter space than fully-pushed waves, supporting the conclusion that all three types of waves are likely to occur in nature. In drawing this conclusion, we assumed that probability distribution of parameters such as~$B$ or~$n^*/N$ is uniform in the parameter space. While this is certainly a gross approximation, it could be more accurate than the assumption that the values of~$v/v_{\mathrm{\textsc{f}}}$ are uniformly distributed.

Finally, we analyzed the model of cooperative yeast growth in sucrose from Ref.~\cite{gandhi:pulled_pushed}. As far as we know, this is the only study that both measured the wave velocity and parameters necessary to determine~$v/v_{\mathrm{\textsc{f}}}$ and also varied the environmental parameter~(sucrose concentration) to change the mode of propagation from pulled to pushed. Because the computational growth model in Ref.~\cite{gandhi:pulled_pushed} showed excellent agreement with the experimental data, we used the model instead of the actual data to compare the regions in the parameter space occupied by the three classes of waves. This model is described in detail in Ref.~\cite{gandhi:pulled_pushed}, but is briefly summarized below. The expansions occur in a one-dimensional metapopulation with discrete cycles of migration and growth. The dynamics during the growth cycle is described by the following set of differential equations for the population density~$n$, the glucose concentration~$g$ and the sucrose concentration~$s$:

\begin{equation}
\begin{aligned}
\frac{dn}{dt} & = \gamma_{\mathrm{max}}\frac{g + g_{\mathrm{loc}}}{g + g_{\mathrm{loc}} + k_g}n\\
\frac{dg}{dt} & = - Y \frac{dn}{dt} + nv_s\frac{s}{s + k_s}\\
\frac{ds}{dt} & = - nv_s\frac{s}{s + k_s},\\
\end{aligned}
\label{eq:yeast}
\end{equation}

\noindent where~$g_{\mathrm{loc}}$ is given by
\begin{equation}
g_{\mathrm{loc}} = g_{\mathrm{eff}}v_s\frac{s}{s + k_s}.
\label{eq:g_loc}
\end{equation}

The behavior of this model is illustrated in Fig.~\ref{fig:phase_space}B. We found that pulled waves occur for a sucrose concentration between 0 and~$0.004\%$, semi-pushed waves for a sucrose concentration~$0.004$ to $0.4\%$, and fully-pushed waves for a sucrose concentration between~$0.4\%$ and~$2\%$, which was the upper value of the sugar explored in the study; presumably very high concentrations of sucrose become toxic. Thus, semi-pushed waves occur in a substantial part of the parameter space for this experimental population.

Overall, we believe all three wave classes could be readily observed in nature, but further empirical work is necessary to test this hypothesis. We also think that this conclusion should hold for physical systems. Indeed, the quadratic~$r(n)$ from Eq.~(3) corresponds to the quartic potential function~$V(n) = -\frac{d}{dn}(r(n)n)$, which is a common model for phase transitions. External parameters such as temperature or chemical potential can change~$B$ and drive the transition between different wave classes. Since the ranges of~$B$ for pulled and semi-pushed waves are the same, so should be the ranges of the external parameter. Therefore, one should be able to observe both types of waves.

\textbf{\large \hclr{XIII. Computer simulations}}

In this section, we explain the details of our computer simulations and the subsequent data analysis.

\textit{Interpretation of the simulations as the Wright-Fisher model with vacancies}\\
Deterministic migration between patches followed by the Wright-Fisher sampling provides one of the most efficient ways to simulate population dynamics. In its standard formulation, the Wright-Fisher model cannot simulate population growth because it assumes that the population size is fixed at the carrying capacity. To overcome this difficulty, we generalized the Wright-Fisher model by considering the number of vacancies, i.e. the difference between the carrying capacity~$N$ and the total population density~$n$, as the abundance of an additional species. With this modification, the total abundance of the two genotypes can increase at the expense of the number of vacancies. 

Following Ref.~\cite{hallatschek:diversity_wave}, the growth of the population was modeled by introducing a fitness difference between the vacancies and the actual species. Specifically, the fitness of the two genotypes was set to~$w_i=1$ and the fitness of the vacancies was set to~$w_{\mathrm{v}}=1-r(n)/(1-n/N)$. The probability to sample genotype~$i$ was then proportional to the ratio of~$w_i$ to the mean fitness of the population~$\bar{w}=n/N + w_{\mathrm{v}}(N-n)/N=1-r(n)$, which explains why we used~$1/(1-r(n)\tau)$ instead of~$1+r(n)\tau$ in Eq.~(11).

\textit{Simulations of deterministic fronts}\\
We also simulated range expansions without demographic fluctuations~($\gamma_n = 0$), but with genetic drift. In these simulations, the total population density was updated deterministically:

\begin{equation}
n(t+\tau,x) = \left \lfloor (p_1+p_2)N \right \rfloor, 
\end{equation}

\noindent where~$p_i$ are the same as in Eq.~(11), and~$\left \lfloor y \right \rfloor$ denotes the floor function, which is equal to the greast integer less than~$y$. The abundances of the two neutral genotypes were then determined by Binomial sampling with~$n(t+\tau,x)$ trials and~$p_i/(p_1+p_2)$ probability of choosing genotype~$i$.

For all simulations~$m=0.25$ and~$r_0=g_0=0.01$ were used, unless noted otherwise.

\textit{Boundary and initial conditions}\\
The most direct approach to simulating a range expansion is to use a stationary habitat, in which the range expansion proceeds from one end to the other. This approach is however expensive because the computational times grows quadratically with the duration of the simulations. Instead, we took advantage of the fact that all population dynamics are localized to the vicinity of the expansion front and simulated only a region of 300 patches comoving with the expansion. Specifically, every simulation time step, we shifted the front backward if the total population inside the simulation array~$n_{\mathrm{array}}$ exceeded~$150N$, i.e. half of the maximally possible population size. The magnitude of the shift was equal to~$\lfloor (n_{\mathrm{array}}-150N)/N \rfloor+1$. The population density in the patches that were added ahead of the front was set to zero, and the number of the individuals moved outside the box was stored, so that we could compute the total number of individuals~$n_{\mathrm{tot}}$ in the entire population including both inside and outside of the simulation array. Our choice of~$300$ patches in the simulation array was sufficient to ensure that at least one patch remained always unoccupied ahead of the expansion front and that the patches shifted outside the array were always at the carrying capacity. 

We initialized all simulations by leaving the right half of the array unoccupied and filling the left half to the carrying capacity. In each occupied patch, we determined the relative abundance of the two neutral genotypes by sampling the binomial distribution with~$N$ trials and equal probabilities of choosing each of the genotypes.

\textit{Duration of simulations and data collection}\\
To ensure that we can access the exponential decay of the average heterozygosity, simulations were carried out for~$2\sqrt{N}$ generations for pulled waves and for~$N$ generations for pushed waves. Although, for pulled waves, the expected timescale of heterozygosity decay is~$\ln^3{N}$, we chose a longer duration of simulations to account for possible deviations from this asymptotic scaling. In all simulations, the minimal simulation time was set to~$10^4$ time steps. 

For each simulation, we saved~the total population size~$n_{\mathrm{tot}}$ and the population heterozygosity~$h$ at~$1000$ time points evenly distributed across the simulation time. These were used to compute~$\mathrm{Var}X_{\mathrm{f}}$ and~$H$ by averaging over~$1000$ independent simulation runs.

\textit{Computing front velocity}\\
The velocity of the front was measured by fitting~$n_{\mathrm{tot}}/N$ to~$vt+\mathrm{const}$. For this fit, we discarded the first~$10\%$ of the total simulation time (1000 generations for the shortest runs) 
to account for the transient dynamics. The length of the transient is the largest for pulled waves and is specified by the following result from Ref.~\cite{saarloos:review}:

\begin{equation}
v(t) = v_{\mathrm{\textsc{F}}} \left[ 1 - \frac{3}{4r_0 t} + \mathcal{O}\left( t^{-3/2} \right) \right].
\end{equation}

\noindent Thus, discarding time points prior to~$t\sim1/r_0$ was sufficient to eliminate the transient dynamics in all of our simulations.

\textit{Computing the diffusion constant of the front}\\
To measure~$D_{\mathrm{f}}$, we discarded early time points as described above and then fitted~$\mathrm{Var}\{n_{\mathrm{tot}}/N\}$ to~$2D_{\mathrm{f}}t+\mathrm{const}$.

\textit{Computing heterozygosity and the rate of its decay}\label{DA:diversity}\\
For each time point, the heterozygosity~$h$ was computed as follows

\begin{equation}
\label{eq:h_definition}
h(t) = \frac{1}{300}\sum_{x}\frac{2n_1(t,x)n_2(t,x)}{[n_1(t,x)+n_2(t,x)]^2},
\end{equation} 

\noindent where the sum is over~$x$ within the simulation array. The average heterozygosity~$H$ was then obtained by averaging over independent simulation runs. To compute~$\Lambda$ we fitted~$\ln H$ to~$-\Lambda t +\mathrm{const}$.

The transient, non-exponential, decay of~$H$ lasted much longer compared to the transient dynamics of~$v$ and~$\mathrm{Var}X_{\mathrm{f}}$; in addition, our estimates of~$H$ had large uncertainty for large~$t$ because only a few simulation runs had non-zero heterozygosity at the final time point. To avoid these sources of error, we restricted the analysis~$H(t)$ to~$t\in(t_{\mathrm{i}},t_{\mathrm{f}})$. The value of~$t_{\mathrm{f}}$ was chosen such that at least~$50$ simulations had non-zero heterozygosity at~$t=t_{\mathrm{f}}$. The value of~$t_{\mathrm{i}}$ was chosen to maximize the goodness of fit~($R^2$) between the fit to~$H\sim e^{-\Lambda t}$ and the data subject to the constraint that~$t_{\mathrm{f}}-t_{\mathrm{i}}>1000$. The latter constraint ensured that we had a sufficient number of uncorrelated data points to carry out the fitting procedure.

\textit{Computing the scaling exponents for~$D_{\mathrm{f}}$,~$\Lambda$, and~$v-v_{\mathrm{d}}$}\\
To quantify the dependence of~$D_{\mathrm{f}}$,~$\Lambda$, and~$v-v_{\mathrm{d}}$ on~$N$, we fitted a power-law dependence using linear regression on log-log scale. Because the power-law behavior is only asymptotic and did not match the results for low~$N$, the exponents were calculated using the data only for~$N>10^4$. 

For the velocity corrections, we also needed to determine the value of~$v_{\mathrm{d}}$. This was done by maximizing the goodness of fit~($R^2$) between the simulation results and theoretical predictions.

\textbf{\large \hclr{XIV. Supplemental results and figures}}

In this section, we present additional simulation data that further supports and clarifies the conclusions made in the main text. Of particular interest is the comparison between deterministic and fluctuating fronts and the results for an alternative model of an Allee effect that can describe propagation into a metastable state~(strong Allee effect).

Figure~\ref{fig:phase_space} shows that the semi-pushed waves occupy a sizable region in the parameter space for two additional models of an Allee effect: one with predator satiation and one with cooperative breakdown of sucrose by yeast.

Figure~\ref{fig:foci} graphically summarizes how the locations of different processes change as waves transition from pulled, to semi-pushed, and to fully-pushed waves.

\begin{figure}[!ht]
\begin{center}
\includegraphics[width=17.8cm]{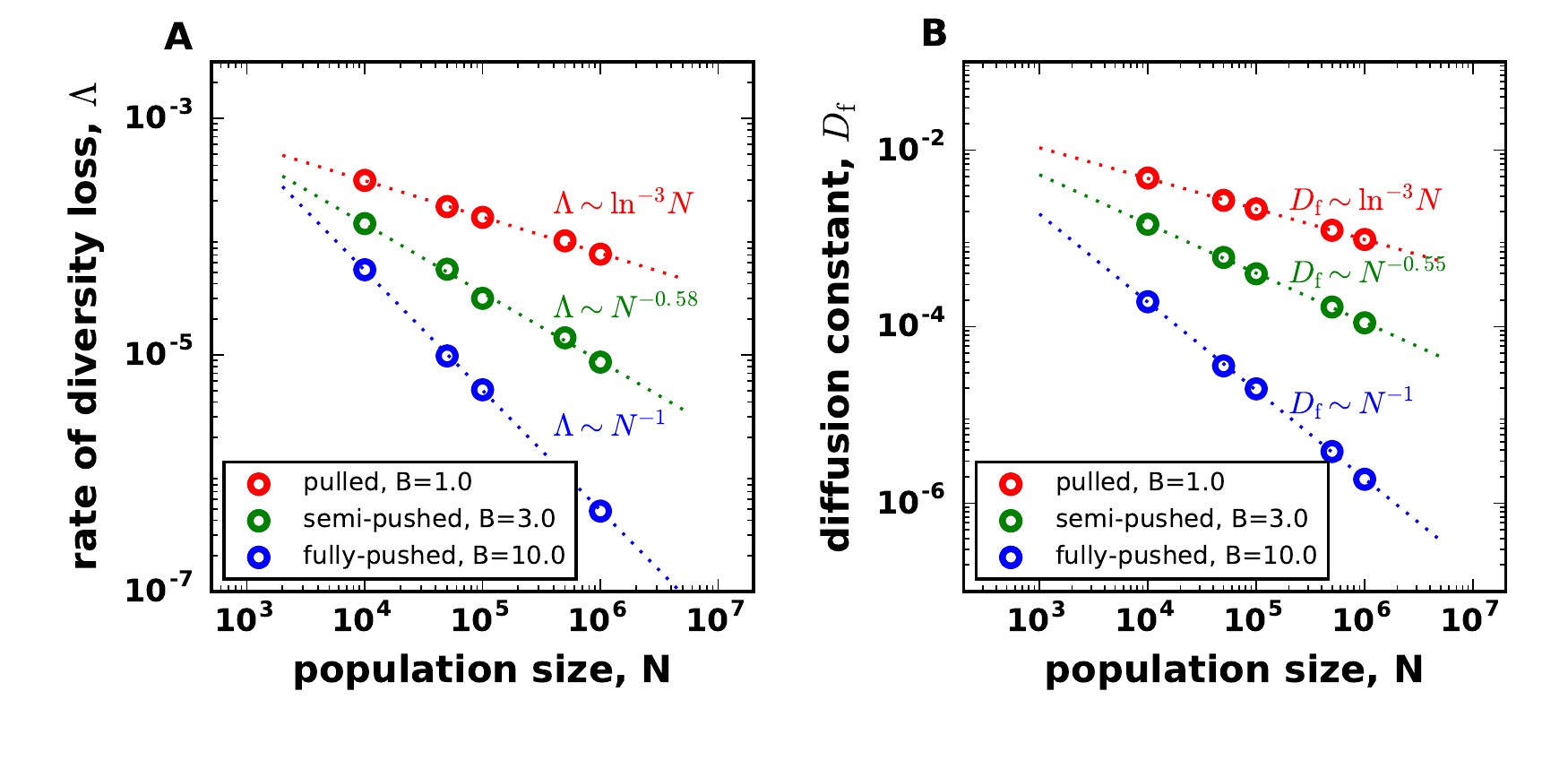}
	\caption{\textbf{Dependence of~$D_{\mathrm{f}}$ and~$\Lambda$ on~$N$ for all three classes of waves.} Circles show the results from the simulations, and dashed lines show the fits of the expected asymptotic scaling: power law for pushed waves and~$\ln^{-3} N$ for pulled waves. The simulation results are for the growth rate specified by Eq.~(\protect{3}).}
\label{fig:semipushed_examples} 
\end{center}
\end{figure}

\begin{figure}[!ht]
\includegraphics[width=17.8cm]{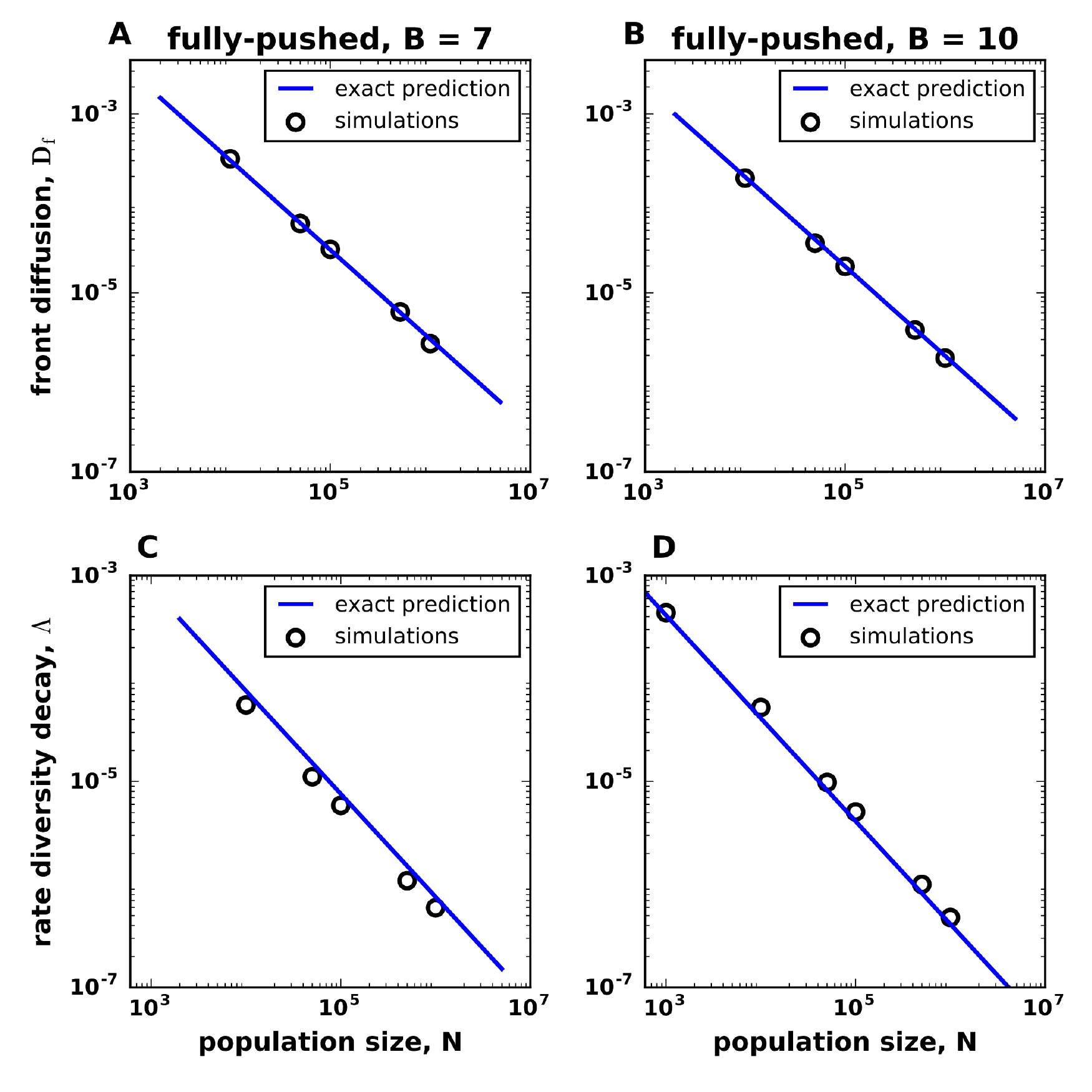}
\caption{\textbf{For fully-pushed waves, theoretical predictions agree with the simulation results without any fitting parameters.} The diffusion constant of the front~$D_{\mathrm{f}}$ and the rate of diversity loss~$\Lambda$ are shown with circles for two values of cooperativity~$B$. Both values of~$B$ are greater than the minimal cooperativity required for fully-pushed waves. The solid lines are the predictions of the perturbation theory: Eq.~(\protect{\ref{eq:D_cooperative_wf}}) for~$D_{\mathrm{f}}$ and Eq.~(\ref{eq:L_cooperative_wf}) for~$\Lambda$. The simulation results are for the growth rate specified by Eq.~(\protect{3}), with~$r_0=0.01$ and~$m=0.25$.}
\label{fig:exact_theory} 
\end{figure}

Figure~\ref{fig:semipushed_examples} shows the data that we used to conclude that fluctuations in semi-pushed waves exhibit different scaling behavior compared to pulled and fully-pushed waves. Figure~\ref{fig:exact_theory} demonstrates that the perturbation theory accurately predicts not only the scaling with~$N$, but also the exact values of~$D_{\mathrm{f}}$ and~$\Lambda$ for fully-pushed waves.

\begin{figure}[!ht]
\begin{center}
\includegraphics[width=17.8cm]{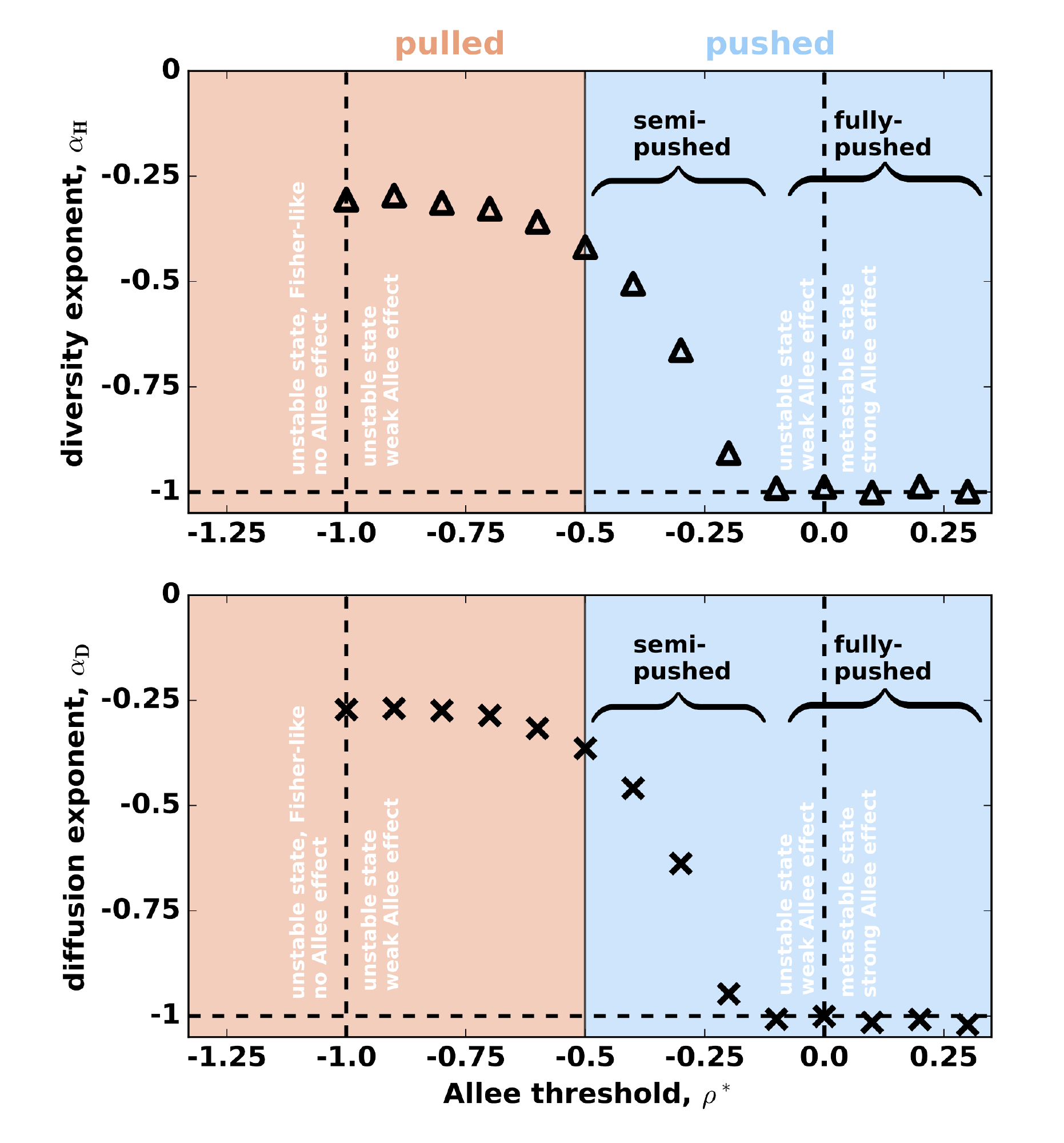}
\caption{\textbf{Waves propagating into a metastable state are fully-pushed.} This figures shows the scaling exponents~$\alpha_{\mathrm{\textsc{D}}}$ and~$\alpha_{\mathrm{\textsc{H}}}$ for an alternative model of an Allee effect specified by Eq.~(\protect{\ref{eq:allee_quadratic}}). A strong Allee effect is possible in this model, so the waves can propagate both into an unstable and metastable states depending on whether the Allee threshold is negative or positive. Note that, both the diffusion constant of the front and the rate of diversity loss scale as~$1/N$ for fully-pushed waves irrespective of the stability of the invaded state. The transition between pulled and semi-pushed waves occurs at~$\rho^*=-0.5$ and between semi-pushed and fully-pushed waves at~$\rho^*=-0.25$. }
\label{fig:metastable} 
\end{center}
\end{figure}

The scaling properties of fully-pushed waves that propagate into a metastable state are shown in Fig.~\ref{fig:metastable}. This figure also illustrates the transition from pulled to semi-pushed and then to fully-pushed waves in an alternative model of an Allee effect.

\begin{figure}[!ht]
\begin{center}
\includegraphics[width=17.8cm]{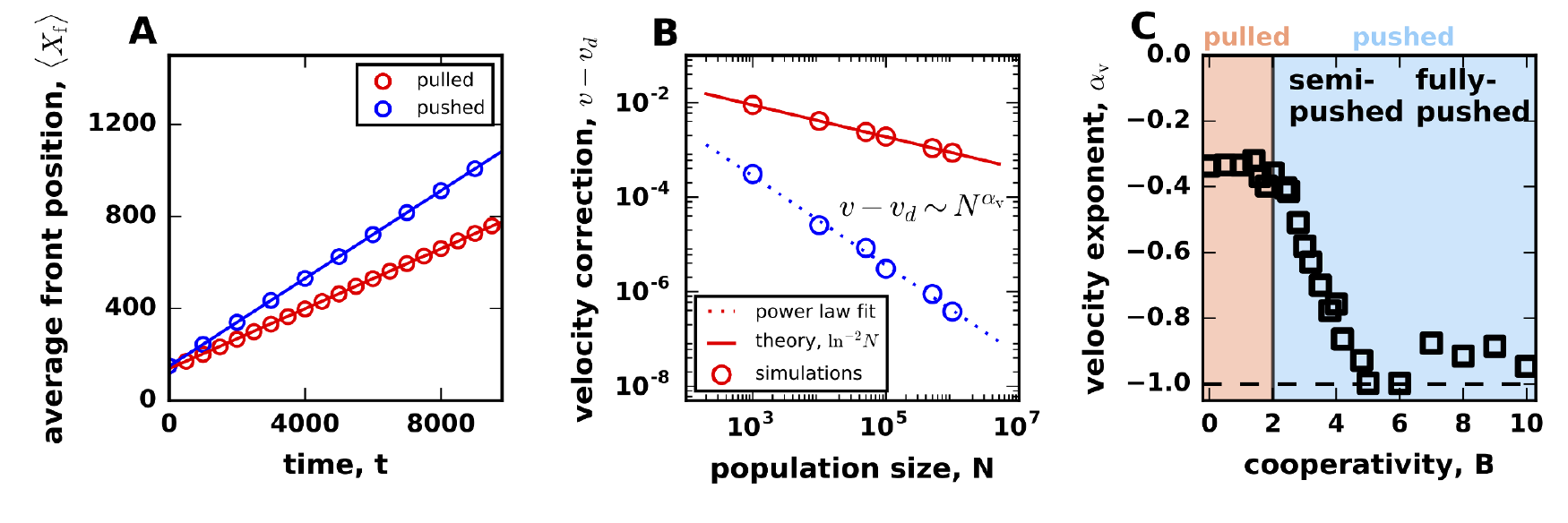}
\caption{\textbf{Correction to wave velocity also shows a transition between semi-pushed and fully-pushed waves.} \textbf{(A)}~The average position of the front increases linearly with time for both pulled and pushed expansions. \textbf{(B)}~The velocity of the front is reduced by demographic fluctuations below its deterministic value~$v_d$. For pulled waves,~$v-v_d \sim \ln^{-2}N$, while, for pushed waves,~$v-v_d$ decreases as a power law~$N^{-\alpha_{\mathrm{\textsc{v}}}}$. \textbf{(C)}~The dependence of~$\alpha_{\mathrm{\textsc{v}}}$ on cooperativity is the same as for~$\alpha_{\mathrm{\textsc{D}}}$ and~$\alpha_{\mathrm{\textsc{H}}}$. In particular,~$v(N)$ clearly shows that the class of pushed waves consists of two subclasses: fully-pushed waves with~$v-v_d\sim 1/N$ and semi-pushed waves with~$\alpha_{\mathrm{\textsc{D}}}\in(-1,0)$. Similar to other figures, the limited range of~$N$ in simulations makes logarithmic scaling for pulled waves also consistent with a power-law with a small negative exponent. The simulation results are for the growth rate specified by Eq.~(\protect{3}).} 
\label{fig:velocities} 
\end{center}
\end{figure}

Figure~\ref{fig:velocities} shows that the transition between different wave classes can also be detected from the small corrections to the wave velocity due to demographic fluctuations.

\begin{figure}[!ht]
\begin{center}
\includegraphics[width=17.8cm]{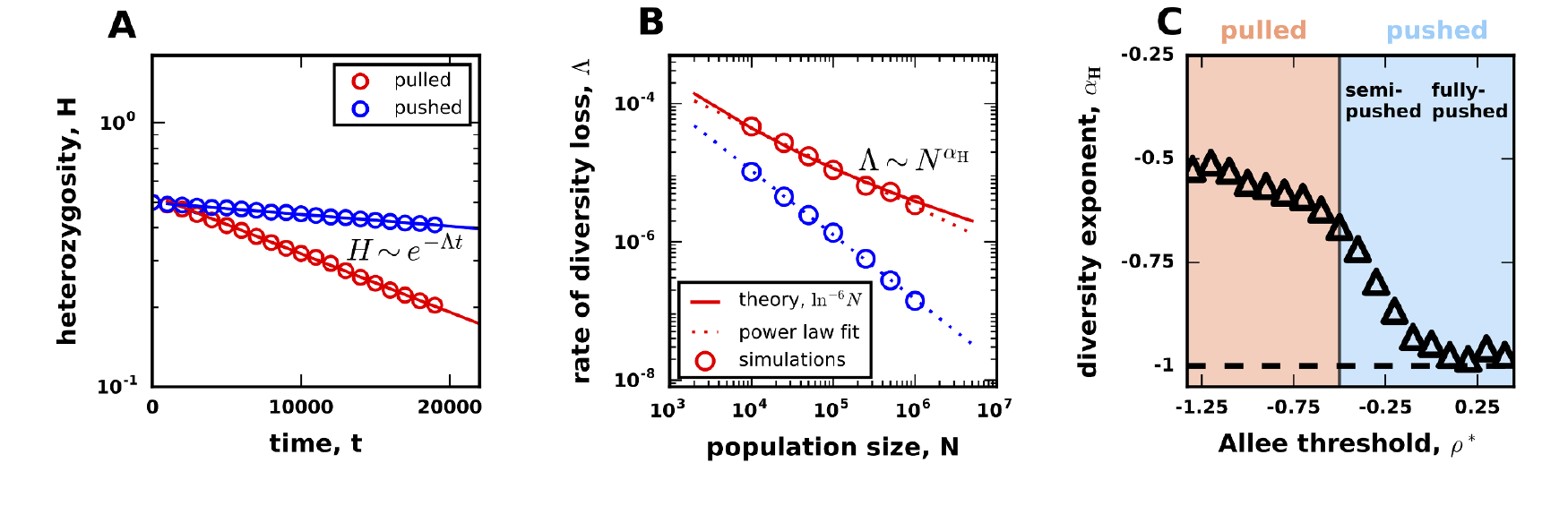}
\caption{\textbf{The rate of diversity decay for deterministic fronts.} \textbf{(A)}~The average heterozygosity,~$H$, is a measure of diversity equal to the probability to sample two distinct genotypes in the population and is well-defined for both fluctuating and deterministic fronts. The decay of genetic diversity is exponential in time:~$H\sim e^{-\Lambda t}$ for both pulled and pushed waves. \textbf{(B)} For deterministic pulled waves,~$\Lambda\sim\ln^{-6}N$ from Eq.~(\protect{\ref{eq:L_deterministic_pulled}}), while, for fully-pushed waves,~$\Lambda\sim N^{-1}$ from Eq.~(\protect{5}). To quantify the dependence of~$\Lambda$ on~$N$, we fit~$\Lambda \sim N^{\alpha_{\mathrm{\textsc{h}}}}$. The dashed red line shows that even though~$\alpha_{\mathrm{\textsc{h}}}$ should equal~$0$ for pulled waves, the limited range of~$N$ results in a different value of~$\alpha_{\mathrm{\textsc{h}}}\approx-0.51$. \textbf{(C)}~The dependence of the scaling exponent on cooperativity identifies the same three classes of waves as in Fig.~\protect{5}C for fluctuating fronts. The transitions between the wave classes occur at the same values of~$B$, but the values of the exponents for semi-pushed waves are slightly different.}
\label{fig:diversity_deterministic} 
\end{center}
\end{figure}

\begin{figure}[!ht]
\begin{center}
\includegraphics[width=11.4cm]{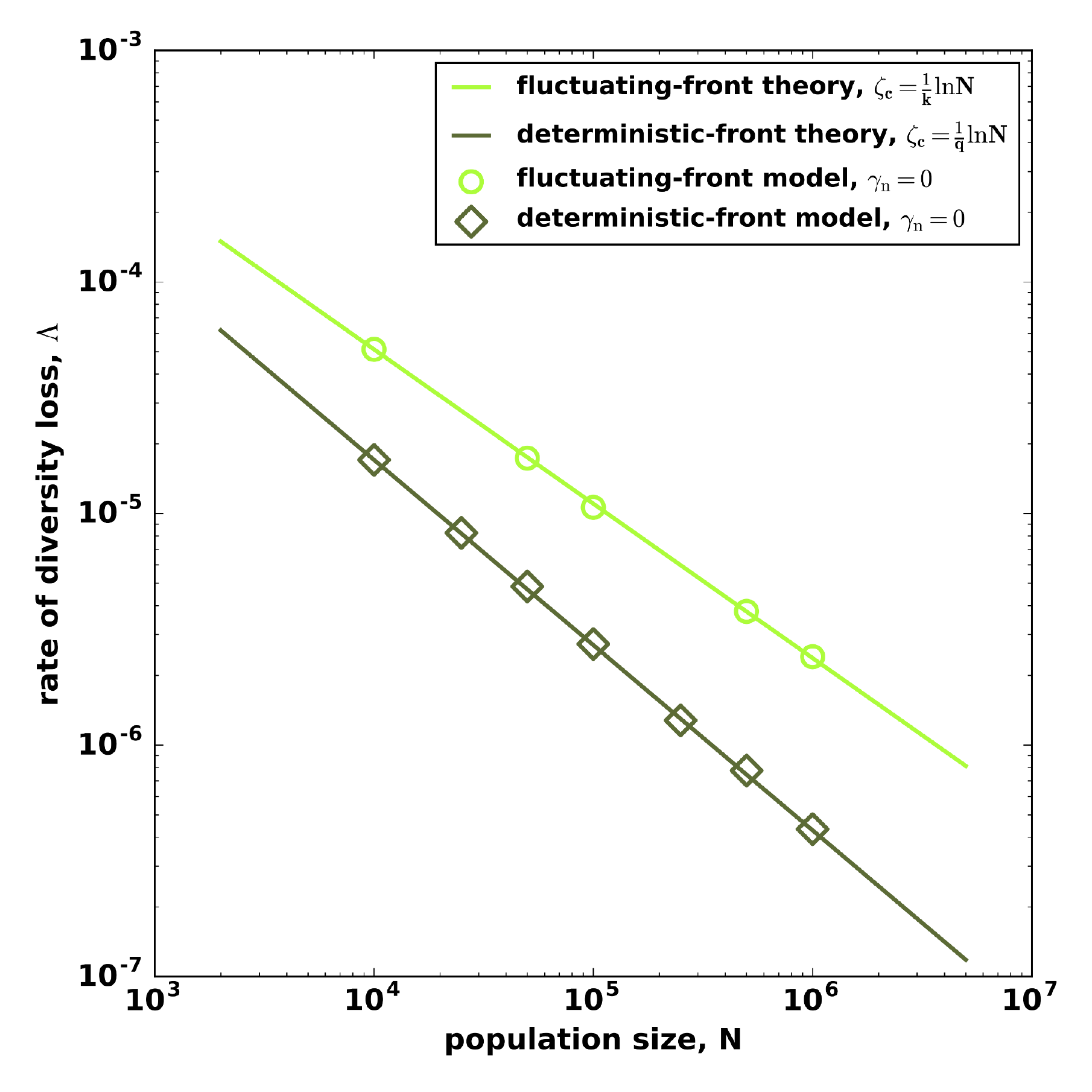}
\caption{\textbf{Comparison between deterministic and fluctuating fronts.} The circles show the dependence of~$\Lambda$ on~$N$ for both stochastic and deterministic simulations of population fronts, and the lines show the corresponding theoretical predictions from Eqs.~(\protect{\ref{eq:alpha_fluctuating}}) and~(\protect{\ref{eq:alpha_deterministic}}). It is clear that the theory correctly captures the contribution of the front fluctuations to the rate of genetic drift. For this figure, we chose the value of cooperativity well within the class of semi-pushed waves to avoid the contribution of the crossover behavior near the transition to pulled and fully-pushed waves. The simulation results are for the growth rate specified by Eq.~(\protect{\ref{eq:allee_quadratic}}) with~$\rho^*=-0.3$. }
\label{fig:fluctuating_deterministic_comparison} 
\end{center}
\end{figure}

\begin{figure}[!ht]
\begin{center}
\includegraphics[width=17.8cm]{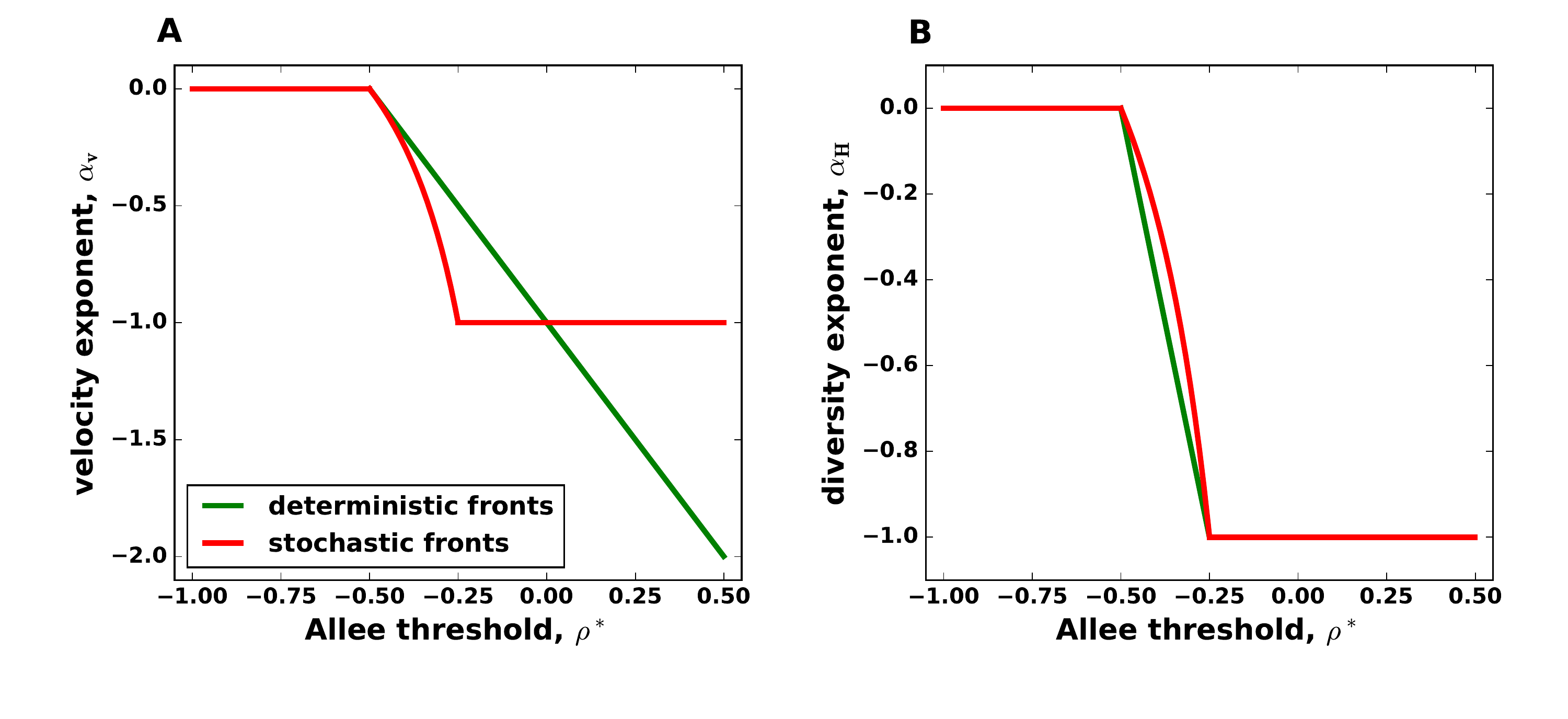}
\caption{\textbf{Corrections to velocity in deterministic and stochastic fronts.} \textbf{(A)} Theoretical predictions for the scaling of~$\Delta v$ with~$N$ for deterministic~(Eq.~(\protect{\ref{eq:dv_deterministic}})) and fluctuating fronts~(Eq.~(\protect{\ref{eq:alpha_fluctuating}})). Note that~$\alpha_{\mathrm{\textsc{v}}}$ gradually changes between~$0$ and~$-2$ for deterministic fronts and does not exhibit a transition between two distinct behaviors. In contrast,~$\alpha_{\mathrm{\textsc{v}}}$ for stochastic fronts clearly shows that the scaling is very different for fully-pushed and semi-pushed waves. \textbf{(B)} Theoretical predictions for the scaling of~$\Lambda$ with~$N$ for deterministic~(Eq.~(\protect{\ref{eq:alpha_deterministic}})) and fluctuating fronts~(Eq.~(\protect{\ref{eq:alpha_fluctuating}})). In contrast to (A), the results for~$\Lambda$ show the existence of two subclasses of pushed waves for both deterministic and fluctuating fronts. Within the semi-pushed class, there is a quantitative difference between~$\alpha_{\mathrm{\textsc{h}}}$ for deterministic and fluctuating fronts. In both panels, we used the growth model from Eq.~(\protect{\ref{eq:allee_quadratic}}) to show the behavior of the exponents across all possible strengths of an Allee effect, which include propagation into unstable as well as into metastable state.}
\label{fig:velocity_corrections_comparison} 
\end{center}
\end{figure}

Genetic drift in deterministic fronts is examined in Fig.~\ref{fig:diversity_deterministic}, and Fig.~\ref{fig:fluctuating_deterministic_comparison} compares the scaling behavior of~$\Lambda$ with~$N$ in deterministic vs. fluctuating fronts.

Finally, Fig.~\ref{fig:velocity_corrections_comparison} contrasts the behavior of~$\Delta v$ and~$\Lambda$ in fluctuating \textit{vs.} deterministic fronts. For~$\Lambda$, both deterministic and stochastic fronts show a transition between large fluctuations in semi-pushed waves and regular~$1/N$ fluctuations in fully-pushed waves. Moreover, both deterministic and stochastic fronts have quite similar values~$\alpha$ for semi-pushed waves. In contrast, the behavior of~$\Delta v$ is qualitatively different. Only stochastic fronts exhibit a transition between large fluctuations and~$1/N$ scaling. For deterministic fronts,~$\alpha_{\mathrm{\textsc{v}}}$ smoothly decreases with the Allee threshold and does not signal the existence of two types of pushed waves. Thus, neglecting front fluctuations has a fundamentally different effect on~$\Lambda$ and~$\Delta v$. For~$\Lambda$, the transition between fully-pushed and semi-pushed waves is indicated by the divergence of the integrals in the perturbation theory. Front fluctuations simply modify the cutoff necessary to regularize these integrals and change~$\alpha$ only quantitatively. For~$\Delta v$, on the other hand, the cutoff is the sole cause of slower expansion velocity of deterministic fronts. For semi-pushed waves, which are sensitive to the dynamics at the front edge, the cutoff qualitatively captures the nontrivial power law dependence of~$\Delta v$ on N. The cutoff, however, cannot account for velocity corrections in fully-pushed waves because~$\Delta v$ arise due to fluctuations throughout the whole front and the contribution from the front edge is negligible.

\clearpage

\clearpage
\bibliography{references}
\bibliographystyle{naturemag}

\end{document}